%\input harvmac

%%%%%%%%%%%%%%%%%%  tex macros for preprints, cm version %%%%%%%%%%%%%%
%         (P. Ginsparg <ginsparg@lanl.gov>, last updated 7/94)
%                if confused, type `b' in response to query
%           hypertex extensions (still provisional), 7/26/94
%
%---------------------------------------------------------------------%
%\input hyperbasics %comment out this line to restore non-hyper functionality
%
%% site dependent options:
%% \unredoffs and \redoffs define horizontal and vertical offsets
%% respectively for unreduced and reduced modes. \speclscape defines
%% the \special{} call that sets printer to landscape (sideways) mode.
%% from standard set below, leave uncommented as appropriate or redefine
%
%%% next 400dpi
\def\unredoffs{} \def\redoffs{\voffset=-.31truein\hoffset=-.48truein}
\def\speclscape{}
%\def\speclscape{\special{papersize=11in,8.5in}}
%
%%% apple lw
%\def\unredoffs{} \def\redoffs{\voffset=-.31truein\hoffset=-.59truein}
%\def\speclscape{\special{ps: landscape}}
%
%%% qms lasergrafix:
%\def\unredoffs{} \def\redoffs{\voffset=-.4truein\hoffset=.125truein}
%\def\speclscape{\special{qms: landscape}}
%
%%% saclay A4 paper:
%\def\unredoffs{\hoffset-.14truein\voffset-.2truein}
%\def\redoffs{\voffset=-.45truein\hoffset=-.21truein}
%\def\speclscape{\special{landscape}}
%
%---------------------------------------------------------------------%
%
\newbox\leftpage \newdimen\fullhsize \newdimen\hstitle \newdimen\hsbody
\tolerance=1000\hfuzz=2pt
\catcode`\@=11 % This allows us to modify PLAIN macros.
\ifx\hyperdef\UNd@FiNeD\def\hyperdef#1#2#3#4{#4}\def\hyperref#1#2#3#4{#4}\fi
\def\bigans{b }
\def\answ{b }
%\message{ big or little (b/l)? }\read-1 to\answ
%
\ifx\answ\bigans\message{(This will come out unreduced.}
\magnification=1200\unredoffs\baselineskip=16pt plus 2pt minus 1pt
\hsbody=\hsize \hstitle=\hsize %take default values for unreduced format
\else\message{(This will be reduced.} \let\l@r=L
\magnification=1000\baselineskip=16pt plus 2pt minus 1pt \vsize=7truein
\redoffs \hstitle=8truein\hsbody=4.75truein\fullhsize=10truein\hsize=\hsbody
\output={\ifnum\pageno=0 %%% This is the HUTP version
  \shipout\vbox{\speclscape{\hsize\fullhsize\makeheadline}
    \hbox to \fullhsize{\hfill\pagebody\hfill}}\advancepageno
  \else
  \almostshipout{\leftline{\vbox{\pagebody\makefootline}}}\advancepageno
  \fi}
\def\almostshipout#1{\if L\l@r \count1=1 \message{[\the\count0.\the\count1]}
      \global\setbox\leftpage=#1 \global\let\l@r=R
 \else \count1=2
  \shipout\vbox{\speclscape{\hsize\fullhsize\makeheadline}
      \hbox to\fullhsize{\box\leftpage\hfil#1}}  \global\let\l@r=L\fi}
\fi
%---------------------------------------------------------------------
%
\newcount\yearltd\yearltd=\year\advance\yearltd by -2000

\def\Title#1#2{\nopagenumbers\abstractfont\hsize=\hstitle\rightline{#1}%
\vskip 1in\centerline{\titlefont #2}\abstractfont\vskip .5in\pageno=0}
\def\Date#1{\vfill\leftline{#1}\tenpoint\supereject\global\hsize=\hsbody%
\footline={\hss\tenrm\hyperdef\hypernoname{page}\folio\folio\hss}}%
% (restores pagenumbers)
%
%       use following instead of \Date on the preliminary draft,
%       puts date/time on each page in big mode, writes labels in margins

\def\draftmode{\message{ DRAFTMODE }\def\draftdate{{\rm preliminary draft:
\number\month/\number\day/\number\yearltd\ \ \hourmin}}%
\headline={\hfil\draftdate}\writelabels\baselineskip=20pt plus 2pt minus 2pt
 {\count255=\time\divide\count255 by 60 \xdef\hourmin{\number\count255}
  \multiply\count255 by-60\advance\count255 by\time
  \xdef\hourmin{\hourmin:\ifnum\count255<10 0\fi\the\count255}}}
%       use \nolabels to get rid of eqn, ref, and fig labels in draft mode
\def\nolabels{\def\wrlabeL##1{}\def\eqlabeL##1{}\def\reflabeL##1{}}
\def\writelabels{\def\wrlabeL##1{\leavevmode\vadjust{\rlap{\smash%
{\line{{\escapechar=` \hfill\rlap{\sevenrm\hskip.03in\string##1}}}}}}}%
\def\eqlabeL##1{{\escapechar-1\rlap{\sevenrm\hskip.05in\string##1}}}%
\def\reflabeL##1{\noexpand\llap{\noexpand\sevenrm\string\string\string##1}}}
\nolabels
%
% tagged sec numbers
\global\newcount\secno \global\secno=0
\global\newcount\meqno \global\meqno=1
\def\s@csym{}
\def\newsec#1{\global\advance\secno by1%
{\toks0{#1}\message{(\the\secno. \the\toks0)}}%
%\ifx\answ\bigans \vfill\eject \else \bigbreak\bigskip \fi  %if desired
\global\subsecno=0\eqnres@t\let\s@csym\secsym\xdef\secn@m{\the\secno}\noindent
{\bf\hyperdef\hypernoname{section}{\the\secno}{\the\secno.} #1}%
\writetoca{{\string\hyperref{}{section}{\the\secno}{\the\secno.}} {#1}}%
\par\nobreak\medskip\nobreak}
\def\eqnres@t{\xdef\secsym{\the\secno.}\global\meqno=1\bigbreak\bigskip}
\def\sequentialequations{\def\eqnres@t{\bigbreak}}\xdef\secsym{}
\global\newcount\subsecno \global\subsecno=0
\def\subsec#1{\global\advance\subsecno by1%
{\toks0{#1}\message{(\s@csym\the\subsecno. \the\toks0)}}%
\ifnum\lastpenalty>9000\else\bigbreak\fi
\noindent{\it\hyperdef\hypernoname{subsection}{\secn@m.\the\subsecno}%
{\secn@m.\the\subsecno.} #1}\writetoca{\string\quad
{\string\hyperref{}{subsection}{\secn@m.\the\subsecno}{\secn@m.\the\subsecno.}}
{#1}}\par\nobreak\medskip\nobreak}
\def\appendix#1#2{\global\meqno=1\global\subsecno=0\xdef\secsym{\hbox{#1.}}%
\bigbreak\bigskip\noindent{\bf Appendix \hyperdef\hypernoname{appendix}{#1}%
{#1.} #2}{\toks0{(#1. #2)}\message{\the\toks0}}%
\xdef\s@csym{#1.}\xdef\secn@m{#1}%
\writetoca{\string\hyperref{}{appendix}{#1}{Appendix {#1.}} {#2}}%
\par\nobreak\medskip\nobreak}
%
%       \eqn\label{a+b=c}	gives displayed equation, numbered
%				consecutively within sections.
%     \eqnn and \eqna define labels in advance (of eqalign?)
%
\def\checkm@de#1#2{\ifmmode{\def\f@rst##1{##1}\hyperdef\hypernoname{equation}%
{#1}{#2}}\else\hyperref{}{equation}{#1}{#2}\fi}
\def\eqnn#1{\DefWarn#1\xdef #1{(\noexpand\relax\noexpand\checkm@de%
{\s@csym\the\meqno}{\secsym\the\meqno})}%
\wrlabeL#1\writedef{#1\leftbracket#1}\global\advance\meqno by1}
\def\f@rst#1{\c@t#1a\em@ark}\def\c@t#1#2\em@ark{#1}
\def\eqna#1{\DefWarn#1\wrlabeL{#1$\{\}$}%
\xdef #1##1{(\noexpand\relax\noexpand\checkm@de%
{\s@csym\the\meqno\noexpand\f@rst{##1}}{\hbox{$\secsym\the\meqno##1$}})}
\writedef{#1\numbersign1\leftbracket#1{\numbersign1}}\global\advance\meqno by1}
\def\eqn#1#2{\DefWarn#1%
\xdef #1{(\noexpand\hyperref{}{equation}{\s@csym\the\meqno}%
{\secsym\the\meqno})}$$#2\eqno(\hyperdef\hypernoname{equation}%
{\s@csym\the\meqno}{\secsym\the\meqno})\eqlabeL#1$$%
\writedef{#1\leftbracket#1}\global\advance\meqno by1}
\def\xeqn{\expandafter\xe@n}\def\xe@n(#1){#1}
\def\xeqna#1{\expandafter\xe@n#1}
\def\eqns#1{(\e@ns #1{\hbox{}})}
\def\e@ns#1{\ifx\UNd@FiNeD#1\message{eqnlabel \string#1 is undefined.}%
\xdef#1{(?.?)}\fi{\let\hyperref=\relax\xdef\next{#1}}%
\ifx\next\em@rk\def\next{}\else%
\ifx\next#1\xeqn#1\else\def\n@xt{#1}\ifx\n@xt\next#1\else\xeqna#1\fi
\fi\let\next=\e@ns\fi\next}

\def\DefWarn#1{\ifx\UNd@FiNeD#1\else
\immediate\write16{*** WARNING: the label \string#1 is already defined ***}\fi}
%
%			 footnotes
\newskip\footskip\footskip14pt plus 1pt minus 1pt %sets footnote baselineskip
\def\footnotefont{\ninepoint}\def\f@t#1{\footnotefont #1\@foot}
\def\f@@t{\baselineskip\footskip\bgroup\footnotefont\aftergroup\@foot\let\next}
\setbox\strutbox=\hbox{\vrule height9.5pt depth4.5pt width0pt}
\global\newcount\ftno \global\ftno=0
\def\foot{\global\advance\ftno by1\def\foot@rg{\hyperref{}{footnote}%
{\the\ftno}{\the\ftno}\xdef\foot@rg{\noexpand\hyperdef\noexpand\hypernoname%
{footnote}{\the\ftno}{\the\ftno}}}\footnote{$^{\foot@rg}$}}
%
%say \footend to put footnotes at end
%will cause problems if \ref used inside \foot, instead use \nref before
\newwrite\ftfile
\def\footend{\def\foot{\global\advance\ftno by1\chardef\wfile=\ftfile
%%$^{\the\ftno}$\ifnum\ftno=1\immediate\openout\ftfile=\jobname.fts\fi%
\hyperref{}{footnote}{\the\ftno}{$^{\the\ftno}$}%
\ifnum\ftno=1\immediate\openout\ftfile=\jobname.fts\fi%
\immediate\write\ftfile{\noexpand\smallskip%
%%\noexpand\item{f\the\ftno:\ }\pctsign}\findarg}%
\noexpand\item{\noexpand\hyperdef\noexpand\hypernoname{footnote}
{\the\ftno}{f\the\ftno}:\ }\pctsign}\findarg}%
\def\footatend{\vfill\eject\immediate\closeout\ftfile{\parindent=20pt
\centerline{\bf Footnotes}\nobreak\bigskip\input \jobname.fts }}}
\def\footatend{}
%
%     \ref\label{text}
% generates a number, assigns it to \label, generates an entry.
% To list the refs on a separate page,  \listrefs
%
\global\newcount\refno \global\refno=1
\newwrite\rfile
\def\ref{[\hyperref{}{reference}{\the\refno}{\the\refno}]\nref}
\def\nref#1{\DefWarn#1%
\xdef#1{[\noexpand\hyperref{}{reference}{\the\refno}{\the\refno}]}%
\writedef{#1\leftbracket#1}%
\ifnum\refno=1\immediate\openout\rfile=\jobname.refs\fi
\chardef\wfile=\rfile\immediate\write\rfile{\noexpand\item{[\noexpand\hyperdef%
\noexpand\hypernoname{reference}{\the\refno}{\the\refno}]\ }%
\reflabeL{#1\hskip.31in}\pctsign}\global\advance\refno by1\findarg}
%	horrible hack to sidestep tex \write limitation
\def\findarg#1#{\begingroup\obeylines\newlinechar=`\^^M\pass@rg}
{\obeylines\gdef\pass@rg#1{\writ@line\relax #1^^M\hbox{}^^M}%
\gdef\writ@line#1^^M{\expandafter\toks0\expandafter{\striprel@x #1}%
\edef\next{\the\toks0}\ifx\next\em@rk\let\next=\endgroup\else\ifx\next\empty%
\else\immediate\write\wfile{\the\toks0}\fi\let\next=\writ@line\fi\next\relax}}
\def\striprel@x#1{} \def\em@rk{\hbox{}}
\def\lref{\begingroup\obeylines\lr@f}
\def\lr@f#1#2{\DefWarn#1\gdef#1{\let#1=\UNd@FiNeD\ref#1{#2}}\endgroup\unskip}

\def\addref#1{\immediate\write\rfile{\noexpand\item{}#1}} %now unnecessary
\def\listrefs{\footatend\vfill\supereject\immediate\closeout\rfile\writestoppt
\baselineskip=\footskip\centerline{{\bf References}}\bigskip{\parindent=20pt%
\frenchspacing\escapechar=` \input \jobname.refs\vfill\eject}\nonfrenchspacing}
\def\startrefs#1{\immediate\openout\rfile=\jobname.refs\refno=#1}
\def\xref{\expandafter\xr@f}\def\xr@f[#1]{#1}
\def\refs#1{\count255=1[\r@fs #1{\hbox{}}]}
\def\r@fs#1{\ifx\UNd@FiNeD#1\message{reflabel \string#1 is undefined.}%
\nref#1{need to supply reference \string#1.}\fi%
\vphantom{\hphantom{#1}}{\let\hyperref=\relax\xdef\next{#1}}%
\ifx\next\em@rk\def\next{}%
\else\ifx\next#1\ifodd\count255\relax\xref#1\count255=0\fi%
\else#1\count255=1\fi\let\next=\r@fs\fi\next}
%

%
% this is ugly, but moore insists
\newwrite\ffile\global\newcount\figno \global\figno=1
\def\fig{fig.~\hyperref{}{figure}{\the\figno}{\the\figno}\nfig}
\def\nfig#1{\DefWarn#1%
\xdef#1{fig.~\noexpand\hyperref{}{figure}{\the\figno}{\the\figno}}%
\writedef{#1\leftbracket fig.\noexpand~\xfig#1}%
\ifnum\figno=1\immediate\openout\ffile=\jobname.figs\fi\chardef\wfile=\ffile%
{\let\hyperref=\relax
\immediate\write\ffile{\noexpand\medskip\noexpand\item{Fig.\ %
\noexpand\hyperdef\noexpand\hypernoname{figure}{\the\figno}{\the\figno}. }
\reflabeL{#1\hskip.55in}\pctsign}}\global\advance\figno by1\findarg}
\def\listfigs{\vfill\eject\immediate\closeout\ffile{\parindent40pt
\baselineskip14pt\centerline{{\bf Figure Captions}}\nobreak\medskip
\escapechar=` \input \jobname.figs\vfill\eject}}
\def\xfig{\expandafter\xf@g}\def\xf@g fig.\penalty\@M\ {}
\def\figs#1{figs.~\f@gs #1{\hbox{}}}
\def\f@gs#1{{\let\hyperref=\relax\xdef\next{#1}}\ifx\next\em@rk\def\next{}\else
\ifx\next#1\xfig #1\else#1\fi\let\next=\f@gs\fi\next}
\def\figin{\epsfcheck\figin}\def\figins{\epsfcheck\figins}
\def\epsfcheck{\ifx\epsfbox\UNd@FiNeD
\message{(NO epsf.tex, FIGURES WILL BE IGNORED)}
\gdef\figin##1{\vskip2in}\gdef\figins##1{\hskip.5in}% blank space instead
\else\message{(FIGURES WILL BE INCLUDED)}%
\gdef\figin##1{##1}\gdef\figins##1{##1}\fi}
\def\DefWarn#1{}
\def\figinsert{\goodbreak\midinsert}
\def\ifig#1#2#3{\DefWarn#1\xdef#1{fig.~\noexpand\hyperref{}{figure}%
{\the\figno}{\the\figno}}\writedef{#1\leftbracket fig.\noexpand~\xfig#1}%
\figinsert\figin{\centerline{#3}}\medskip\centerline{\vbox{\baselineskip12pt
\advance\hsize by -1truein\noindent\wrlabeL{#1=#1}\footnotefont%
{\bf Fig.~\hyperdef\hypernoname{figure}{\the\figno}{\the\figno}:} #2}}
\bigskip\endinsert\global\advance\figno by1}
\newwrite\lfile
{\escapechar-1\xdef\pctsign{\string\%}\xdef\leftbracket{\string\{}
\xdef\rightbracket{\string\}}\xdef\numbersign{\string\#}}
\def\writedefs{\immediate\openout\lfile=\jobname.defs \def\writedef##1{%
{\let\hyperref=\relax\let\hyperdef=\relax\let\hypernoname=\relax
 \immediate\write\lfile{\string\def\string##1\rightbracket}}}}%
\def\writestop{\def\writestoppt{\immediate\write\lfile{\string\pageno
 \the\pageno\string\startrefs\leftbracket\the\refno\rightbracket
 \string\def\string\secsym\leftbracket\secsym\rightbracket
 \string\secno\the\secno\string\meqno\the\meqno}\immediate\closeout\lfile}}
\def\writestoppt{}\def\writedef#1{}
\def\seclab#1{\DefWarn#1%
\xdef #1{\noexpand\hyperref{}{section}{\the\secno}{\the\secno}}%
\writedef{#1\leftbracket#1}\wrlabeL{#1=#1}}
\def\subseclab#1{\DefWarn#1%
\xdef #1{\noexpand\hyperref{}{subsection}{\secn@m.\the\subsecno}%
{\secn@m.\the\subsecno}}\writedef{#1\leftbracket#1}\wrlabeL{#1=#1}}
\def\applab#1{\DefWarn#1%
\xdef #1{\noexpand\hyperref{}{appendix}{\secn@m}{\secn@m}}%
\writedef{#1\leftbracket#1}\wrlabeL{#1=#1}}
\newwrite\tfile \def\writetoca#1{}
\def\leaderfill{\leaders\hbox to 1em{\hss.\hss}\hfill}
%	use this to write file with table of contents
\def\writetoc{\immediate\openout\tfile=\jobname.toc
   \def\writetoca##1{{\edef\next{\write\tfile{\noindent ##1
   \string\leaderfill {\string\hyperref{}{page}{\noexpand\number\pageno}%
                       {\noexpand\number\pageno}} \par}}\next}}}
%       and this lists table of contents on second pass
\newread\ch@ckfile
\def\listtoc{\immediate\closeout\tfile\immediate\openin\ch@ckfile=\jobname.toc
\ifeof\ch@ckfile\message{no file \jobname.toc, no table of contents this pass}%
\else\closein\ch@ckfile\centerline{\bf Contents}\nobreak\medskip%
{\baselineskip=12pt\footnotefont\parskip=0pt\catcode`\@=11\input\jobname.toc
\catcode`\@=12\bigbreak\bigskip}\fi}
\catcode`\@=12 % at signs are no longer letters
%
%	Unpleasantness in calling in abstract and title fonts
\edef\tfontsize{\ifx\answ\bigans scaled\magstep3\else scaled\magstep4\fi}
\font\titlerm=cmr10 \tfontsize \font\titlerms=cmr7 \tfontsize
\font\titlermss=cmr5 \tfontsize \font\titlei=cmmi10 \tfontsize
\font\titleis=cmmi7 \tfontsize \font\titleiss=cmmi5 \tfontsize
\font\titlesy=cmsy10 \tfontsize \font\titlesys=cmsy7 \tfontsize
\font\titlesyss=cmsy5 \tfontsize \font\titleit=cmti10 \tfontsize
\skewchar\titlei='177 \skewchar\titleis='177 \skewchar\titleiss='177
\skewchar\titlesy='60 \skewchar\titlesys='60 \skewchar\titlesyss='60
\def\titlefont{\def\rm{\fam0\titlerm}% switch to title font
\textfont0=\titlerm \scriptfont0=\titlerms \scriptscriptfont0=\titlermss
\textfont1=\titlei \scriptfont1=\titleis \scriptscriptfont1=\titleiss
\textfont2=\titlesy \scriptfont2=\titlesys \scriptscriptfont2=\titlesyss
\textfont\itfam=\titleit \def\it{\fam\itfam\titleit}\rm}
 \ifx\answ\bigans\else scaled\magstep1\fi
\ifx\answ\bigans\def\abstractfont{\tenpoint}\else
\font\absit=cmti10 scaled \magstep1
\font\abssl=cmsl10 scaled \magstep1
\font\absrm=cmr10 scaled\magstep1 \font\absrms=cmr7 scaled\magstep1
\font\absrmss=cmr5 scaled\magstep1 \font\absi=cmmi10 scaled\magstep1
\font\absis=cmmi7 scaled\magstep1 \font\absiss=cmmi5 scaled\magstep1
\font\abssy=cmsy10 scaled\magstep1 \font\abssys=cmsy7 scaled\magstep1
\font\abssyss=cmsy5 scaled\magstep1 \font\absbf=cmbx10 scaled\magstep1
\skewchar\absi='177 \skewchar\absis='177 \skewchar\absiss='177
\skewchar\abssy='60 \skewchar\abssys='60 \skewchar\abssyss='60
\def\abstractfont{\def\rm{\fam0\absrm}% switch to abstract font
\textfont0=\absrm \scriptfont0=\absrms \scriptscriptfont0=\absrmss
\textfont1=\absi \scriptfont1=\absis \scriptscriptfont1=\absiss
\textfont2=\abssy \scriptfont2=\abssys \scriptscriptfont2=\abssyss
\textfont\itfam=\absit \def\it{\fam\itfam\absit}\def\footnotefont{\tenpoint}%
\textfont\slfam=\abssl \def\sl{\fam\slfam\abssl}%
\textfont\bffam=\absbf \def\bf{\fam\bffam\absbf}\rm}\fi
\def\tenpoint{\def\rm{\fam0\tenrm}% switch back to 10-point type
\textfont0=\tenrm \scriptfont0=\sevenrm \scriptscriptfont0=\fiverm
\textfont1=\teni  \scriptfont1=\seveni  \scriptscriptfont1=\fivei
\textfont2=\tensy \scriptfont2=\sevensy \scriptscriptfont2=\fivesy
\textfont\itfam=\tenit \def\it{\fam\itfam\tenit}\def\footnotefont{\ninepoint}%
\textfont\bffam=\tenbf \def\bf{\fam\bffam\tenbf}\def\sl{\fam\slfam\tensl}\rm}
\font\ninerm=cmr9 \font\sixrm=cmr6 \font\ninei=cmmi9 \font\sixi=cmmi6
\font\ninesy=cmsy9 \font\sixsy=cmsy6 \font\ninebf=cmbx9
\font\nineit=cmti9 \font\ninesl=cmsl9 \skewchar\ninei='177
\skewchar\sixi='177 \skewchar\ninesy='60 \skewchar\sixsy='60
\def\ninepoint{\def\rm{\fam0\ninerm}% switch to footnote font
\textfont0=\ninerm \scriptfont0=\sixrm \scriptscriptfont0=\fiverm
\textfont1=\ninei \scriptfont1=\sixi \scriptscriptfont1=\fivei
\textfont2=\ninesy \scriptfont2=\sixsy \scriptscriptfont2=\fivesy
\textfont\itfam=\ninei \def\it{\fam\itfam\nineit}\def\sl{\fam\slfam\ninesl}%
\textfont\bffam=\ninebf \def\bf{\fam\bffam\ninebf}\rm}
%
%---------------------------------------------------------------------
%

\hyphenation{anom-aly anom-alies coun-ter-term coun-ter-terms}
\def\inv{^{\raise.15ex\hbox{${\scriptscriptstyle -}$}\kern-.05em 1}}

\def\Dsl{\,\raise.15ex\hbox{/}\mkern-13.5mu D} %this one can be subscripted
\def\dsl{\raise.15ex\hbox{/}\kern-.57em\partial}

\def\tr{{\rm tr}} \def\Tr{{\rm Tr}}
 %pound sterling
\def\lspace{\ifx\answ\bigans{}\else\qquad\fi}
\def\lbspace{\ifx\answ\bigans{}\else\hskip-.2in\fi} % $$\lbspace...$$

\def\boxeqn#1{\vcenter{\vbox{\hrule\hbox{\vrule\kern3pt\vbox{\kern3pt
	\hbox{${\displaystyle #1}$}\kern3pt}\kern3pt\vrule}\hrule}}}
\def\mbox#1#2{\vcenter{\hrule \hbox{\vrule height#2in
		\kern#1in \vrule} \hrule}}  %e.g. \mbox{.1}{.1}
%	matters of taste
%\def\tilde{\widetilde} \def\bar{\overline} \def\hat{\widehat}
%
% some sample definitions
  %     curly letters

\def\e#1{{\rm e}^{^{\textstyle#1}}}

\def\vev#1{\langle #1 \rangle}

\def\darr#1{\raise1.5ex\hbox{$\leftrightarrow$}\mkern-16.5mu #1}
 %pound sterling

 %puts a small half in a displayed eqn
\def\roughly#1{\raise.3ex\hbox{$#1$\kern-.75em\lower1ex\hbox{$\sim$}}}

\def\bb{
\font\tenmsb=msbm10
\font\sevenmsb=msbm7
\font\fivemsb=msbm5
\textfont1=\tenmsb
\scriptfont1=\sevenmsb
\scriptscriptfont1=\fivemsb
}

\input amssym

\input epsf

\def\IZ{\relax\ifmmode\mathchoice
{\hbox{\cmss Z\kern-.4em Z}}{\hbox{\cmss Z\kern-.4em Z}} {\lower.9pt\hbox{\cmsss Z\kern-.4em Z}}
{\lower1.2pt\hbox{\cmsss Z\kern-.4em Z}}\else{\cmss Z\kern-.4em Z}\fi}

\newif\ifdraft\draftfalse
%\drafttrue
\newif\ifinter\interfalse
%\intertrue
\ifdraft\draftmode\else\interfalse\fi
\def\journal#1&#2(#3){\unskip, \sl #1\ \bf #2 \rm(19#3) }
\def\andjournal#1&#2(#3){\sl #1~\bf #2 \rm (19#3) }

\def\frac#1#2{{#1\over#2}}

\def\vev#1{\langle#1\rangle}

\def\inbar{\,\vrule height1.5ex width.4pt depth0pt}
\def\IC{\relax\hbox{$\inbar\kern-.3em{\rm C}$}}
\def\IR{\relax{\rm I\kern-.18em R}}
\def\IP{\relax{\rm I\kern-.18em P}}
%\def\Z{{\bf Z}}

%
%%%%%%%%%%%%%%%%%%%%%%%%%%%%%%%%%%%%
%

%\def\ap#1#2#3{Ann. Phys. {\bf #1} (#2) #3}

%
\catcode`\@=11
\def\slash#1{\mathord{\mathpalette\c@ncel{#1}}}
\overfullrule=0pt

\def\II{{\cal I}}

\def\S{\hbox{$\bb S$}}
\def\ZZ{\hbox{$\bb Z$}}
\def\Z{\hbox{$\bb Z$}}
\def\R{\hbox{$\bb R$}}

\def\underrel#1\over#2{\mathrel{\mathop{\kern\z@#1}\limits_{#2}}}

\catcode`\@=12

%%%%%%%%%%%%%%%%%%%%%%%%%%%%%%%%%%%%%%%%%%%%%%%%%%%%%%%%%%%%%%

%

\def\vev#1{\left\langle #1 \right\rangle}
\def\det{{\rm det}}
\def\tr{{\rm tr}}

\def\det{{\rm det}}
\def\exp{{\rm exp}}

%%%%%%%%%%%%%%%%%%%%%%%%%%%%%%%%%%%%%%%%%%%%%%%%%%%%%%%%%%%%%%
% new defs:

\def\e{\epsilon}

\def\[{[}
\def\]{]}

\def\comment#1{ }

%%%%%%%%%%%%%%%%%%%%%%%%%%%%%%%%%%%%%%%%%%%%%%%%%%%%%%%%%%%%%%
%%% Oskar's definitions:
%
%% A box for a short draft note
\def\draftnote#1{\ifdraft{\baselineskip2ex
                 \vbox{\kern1em\hrule\hbox{\vrule\kern1em\vbox{\kern1ex
                 \noindent \underbar{NOTE}: #1
             \vskip1ex}\kern1em\vrule}\hrule}}\fi}
%% A box for a short internal note
\def\internote#1{\ifinter{\baselineskip2ex
                 \vbox{\kern1em\hrule\hbox{\vrule\kern1em\vbox{\kern1ex
                 \noindent \underbar{Internal Note}: #1
             \vskip1ex}\kern1em\vrule}\hrule}}\fi}
%% A few internal words

%
%% Greek letters
%
%\def\al{\alpha}
%\def\bt{\beta}
%\def\gm{\gamma}                \def\Gm{\Gamma}
%\def\dl{\delta}                \def\Dl{\Delta}
%\def\ep{\epsilon}
%\def\vep{\varepsilon}

%\def\io{\iota}
%\def\kp{\kappa}
%\def\lm{\lambda}               \def\Lm{\Lambda}
%%\mu,\nu unchanged

%\def\th{\theta}               \def\Th{\Theta}
%\def\vth{\vartheta}
%%\phi unchanged               \Phi unchanged
%\def\vph{\varphi}
%%\psi unchanged               \Psi unchanged
%%\chi unchanged

%\def\om{\omega}               \def\Om{\Omega}
%%\pi unchanged                \Pi unchanged
%\def\vpi{\varpi}
%%\rho unchanged
%\def\vro{\varrho}
%\def\sg{\sigma}               \def\Sg{\Sigma}
%\def\vsg{\varsigma}
%%\tau unchanged
%\def\up{\upsilon}             \Up{\Upsilon}
%%\xi unchanged                \Xi unchanged
%%\eta unchanged
%\def\zt{\zeta}
%
%% Capital roman double letters (blackboard font)
%
%\def\inbar{\hskip.3em\vrule height1.5ex width.4pt depth0pt}
%\def\IC{\relax{\inbar\kern-.3em{\rm C}}}
%\def\IN{\relax{\rm I\kern-.16em N}}
%\def\IP{\relax{\rm I\kern-.18em P}}
%\def\IQ{\relax\hbox{$\inbar$\kern-.3em{\rm Q}}}
%\def\IR{\relax{\rm I\kern-.18em R}}
%\def\IZ{\relax{\rm Z\kern-.8em Z}}
%
%% Other Defs
%

%

\def\inv{^{-1}}

%mydef

\def\ts{{\tilde \sigma}}
\def\tz{{\tilde z}}
\def\tr{{\rm tr}}
\def\Tr{{\rm Tr}}

\def\cN{{\cal N}}

%\KutasovVE
\lref\KutasovVE{
  D.~Kutasov,
  ``A Comment on duality in N=1 supersymmetric nonAbelian gauge theories,''
Phys.\ Lett.\ B {\bf 351}, 230 (1995).
[hep-th/9503086].
%%CITATION = hep-th/9503086%%
}

\lref\KinneyEJ{
  J.~Kinney, J.~M.~Maldacena, S.~Minwalla and S.~Raju,
  ``An Index for 4 dimensional super conformal theories,''
Commun.\ Math.\ Phys.\  {\bf 275}, 209 (2007).
[hep-th/0510251].
%%CITATION = hep-th/0510251%%
}

%\NiarchosAA
\lref\NiarchosAA{
  V.~Niarchos,
  ``R-charges, Chiral Rings and RG Flows in Supersymmetric Chern-Simons-Matter Theories,''
JHEP {\bf 0905}, 054 (2009).
[arXiv:0903.0435 [hep-th]].
%%CITATION = arXiv:0903.0435%%
}

%\NiarchosJB
\lref\NiarchosJB{
  V.~Niarchos,
  ``Seiberg Duality in Chern-Simons Theories with Fundamental and Adjoint Matter,''
JHEP {\bf 0811}, 001 (2008).
[arXiv:0808.2771 [hep-th]].
%%CITATION = arXiv:0808.2771%%
}

%\BorokhovIB
\lref\BorokhovIB{
  V.~Borokhov, A.~Kapustin and X.~-k.~Wu,
  ``Topological disorder operators in three-dimensional conformal field theory,''
JHEP {\bf 0211}, 049 (2002).
[hep-th/0206054].
%%CITATION = hep-th/0206054%%
}

%\BorokhovCG
\lref\BorokhovCG{
  V.~Borokhov, A.~Kapustin and X.~-k.~Wu,
  ``Monopole operators and mirror symmetry in three-dimensions,''
JHEP {\bf 0212}, 044 (2002).
[hep-th/0207074].
%%CITATION = hep-th/0207074%%
}

%\NakanishiHJ
\lref\NakanishiHJ{
  T.~Nakanishi and A.~Tsuchiya,
  ``Level rank duality of WZW models in conformal field theory,''
Commun.\ Math.\ Phys.\  {\bf 144}, 351 (1992).
%%CITATION = NU-MATH-002%%
}
\lref\NiarchosAH{
  V.~Niarchos,
  ``Seiberg dualities and the 3d/4d connection,''
JHEP {\bf 1207}, 075 (2012).
[arXiv:1205.2086 [hep-th]].
%%CITATION = arXiv:1205.2086%%
}

%\SeibergBZ
\lref\SeibergBZ{
  N.~Seiberg,
  ``Exact results on the space of vacua of four-dimensional SUSY gauge theories,''
Phys.\ Rev.\ D {\bf 49}, 6857 (1994).
[hep-th/9402044].
%%CITATION = hep-th/9402044%%
}

\lref\BhattacharyaZY{
  J.~Bhattacharya, S.~Bhattacharyya, S.~Minwalla and S.~Raju,
  ``Indices for Superconformal Field Theories in 3,5 and 6 Dimensions,''
JHEP {\bf 0802}, 064 (2008).
[arXiv:0801.1435 [hep-th]].
%%CITATION = arXiv:0801.1435%%
}

%\ZupnikRY
\lref\ZupnikRY{
   B.~M.~Zupnik and D.~G.~Pak,
   ``Topologically Massive Gauge Theories In Superspace,''
Sov.\ Phys.\ J.\  {\bf 31}, 962 (1988).
}

%\IvanovFN
\lref\IvanovFN{
   E.~A.~Ivanov,
   ``Chern-Simons matter systems with manifest N=2 supersymmetry,''
Phys.\ Lett.\ B {\bf 268}, 203 (1991).
}

%\NiemiRQ
\lref\NiemiRQ{
  A.~J.~Niemi and G.~W.~Semenoff,
  ``Axial Anomaly Induced Fermion Fractionization and Effective Gauge Theory Actions in Odd Dimensional Space-Times,''
Phys.\ Rev.\ Lett.\  {\bf 51}, 2077 (1983).
%%CITATION = Print-83-0988 (IAS,PRINCETON)%%
}

%\RedlichDV
\lref\RedlichDV{
  A.~N.~Redlich,
  ``Parity Violation and Gauge Noninvariance of the Effective Gauge Field Action in Three-Dimensions,''
Phys.\ Rev.\ D {\bf 29}, 2366 (1984).
%%CITATION = MIT-CTP-1128%%
}

\lref\AharonyGP{
  O.~Aharony,
  ``IR duality in d = 3 N=2 supersymmetric USp(2N(c)) and U(N(c)) gauge theories,''
Phys.\ Lett.\ B {\bf 404}, 71 (1997).
[hep-th/9703215].
%%CITATION = hep-th/9703215%%
}

%\AffleckAS
\lref\AffleckAS{
  I.~Affleck, J.~A.~Harvey and E.~Witten,
  ``Instantons and (Super)Symmetry Breaking in (2+1)-Dimensions,''
Nucl.\ Phys.\ B {\bf 206}, 413 (1982).
%%CITATION = PRINT-82-0478 (PRINCETON)%%
}

\lref\BeemMB{
  C.~Beem, T.~Dimofte and S.~Pasquetti,
  ``Holomorphic Blocks in Three Dimensions,''
[arXiv:1211.1986 [hep-th]].
%%CITATION = arXiv:1211.1986%%
}

\lref\HwangJH{
  C.~Hwang, H.~-C.~Kim and J.~Park,
  ``Factorization of the 3d superconformal index,''
[arXiv:1211.6023 [hep-th]].
%%CITATION = arXiv:1211.6023%%
}

\lref\KrattenthalerDA{
  C.~Krattenthaler, V.~P.~Spiridonov and G.~S.~Vartanov,
  ``Superconformal indices of three-dimensional theories related by mirror symmetry,''
JHEP {\bf 1106}, 008 (2011).
[arXiv:1103.4075 [hep-th]].
%%CITATION = arXiv:1103.4075%%
}

\lref\GaddeEN{
  A.~Gadde, L.~Rastelli, S.~S.~Razamat and W.~Yan,
  ``On the Superconformal Index of N=1 IR Fixed Points: A Holographic Check,''
JHEP {\bf 1103}, 041 (2011).
[arXiv:1011.5278 [hep-th]].
%%CITATION = arXiv:1011.5278%%
}

\lref\ImamuraWG{
  Y.~Imamura and D.~Yokoyama,
 ``N=2 supersymmetric theories on squashed three-sphere,''
Phys.\ Rev.\ D {\bf 85}, 025015 (2012).
[arXiv:1109.4734 [hep-th]].
%%CITATION = arXiv:1109.4734%%
}

%\deBoerKA
\lref\deBoerKA{
  J.~de Boer, K.~Hori, Y.~Oz and Z.~Yin,
  ``Branes and mirror symmetry in N=2 supersymmetric gauge theories in three-dimensions,''
Nucl.\ Phys.\ B {\bf 502}, 107 (1997).
[hep-th/9702154].
%%CITATION = hep-th/9702154%%
}

\lref\ClossetRU{
  C.~Closset, T.~T.~Dumitrescu, G.~Festuccia and Z.~Komargodski,
  ``Supersymmetric Field Theories on Three-Manifolds,''
JHEP {\bf 1305}, 017 (2013).
[arXiv:1212.3388 [hep-th]].
%%CITATION = PUPT-2432%%
}

\lref\ImamuraRQ{
  Y.~Imamura and D.~Yokoyama,
 ``$S^3/Z_n$ partition function and dualities,''
JHEP {\bf 1211}, 122 (2012).
[arXiv:1208.1404 [hep-th]].
%%CITATION = arXiv:1208.1404%%
}

\lref\KapustinSim{
A.~Kapustin,  2010 Simons Workshop talk, a video of this talk can be found at
{\tt
http://media.scgp.stonybrook.edu/video/video.php?f=20110810\_1\_qtp.mp4}
}

%\PolyakovFU
\lref\PolyakovFU{
  A.~M.~Polyakov,
  ``Quark Confinement and Topology of Gauge Groups,''
Nucl.\ Phys.\ B {\bf 120}, 429 (1977).
%%CITATION = NORDITA-76/33%%
}

\lref\newIS{
K.~Intriligator and N.~Seiberg,
  ``Aspects of 3d N=2 Chern-Simons-Matter Theories,''
[arXiv:1305.1633 [hep-th]].
%%CITATION = UCSD-PTH-12-17%%
}

%\JafferisUN
\lref\JafferisUN{
  D.~L.~Jafferis,
  ``The Exact Superconformal R-Symmetry Extremizes Z,''
JHEP {\bf 1205}, 159 (2012).
[arXiv:1012.3210 [hep-th]].
%%CITATION = arXiv:1012.3210%%
}

%\JafferisZI
\lref\JafferisZI{
  D.~L.~Jafferis, I.~R.~Klebanov, S.~S.~Pufu and B.~R.~Safdi,
  ``Towards the F-Theorem: N=2 Field Theories on the Three-Sphere,''
JHEP {\bf 1106}, 102 (2011).
[arXiv:1103.1181 [hep-th]].
%%CITATION = arXiv:1103.1181%%
}

%\IntriligatorID
\lref\IntriligatorID{
  K.~A.~Intriligator and N.~Seiberg,
  ``Duality, monopoles, dyons, confinement and oblique confinement in supersymmetric SO(N(c)) gauge theories,''
Nucl.\ Phys.\ B {\bf 444}, 125 (1995).
[hep-th/9503179].
%%CITATION = hep-th/9503179%%
}

%\SeibergPQ
\lref\SeibergPQ{
  N.~Seiberg,
  ``Electric - magnetic duality in supersymmetric nonAbelian gauge theories,''
Nucl.\ Phys.\ B {\bf 435}, 129 (1995).
[hep-th/9411149].
%%CITATION = hep-th/9411149%%
}

%\AharonyBX
\lref\AharonyBX{
  O.~Aharony, A.~Hanany, K.~A.~Intriligator, N.~Seiberg and M.~J.~Strassler,
  ``Aspects of N=2 supersymmetric gauge theories in three-dimensions,''
Nucl.\ Phys.\ B {\bf 499}, 67 (1997).
[hep-th/9703110].
%%CITATION = hep-th/9703110%%
}

%\IntriligatorNE
\lref\IntriligatorNE{
  K.~A.~Intriligator and P.~Pouliot,
  ``Exact superpotentials, quantum vacua and duality in supersymmetric SP(N(c)) gauge theories,''
Phys.\ Lett.\ B {\bf 353}, 471 (1995).
[hep-th/9505006].
%%CITATION = hep-th/9505006%%
}

%\IntriligatorAU
\lref\IntriligatorAU{
  K.~A.~Intriligator and N.~Seiberg,
  ``Lectures on supersymmetric gauge theories and electric - magnetic duality,''
Nucl.\ Phys.\ Proc.\ Suppl.\  {\bf 45BC}, 1 (1996).
[hep-th/9509066].
%%CITATION = hep-th/9509066%%
}

%\KarchUX
\lref\KarchUX{
  A.~Karch,
  ``Seiberg duality in three-dimensions,''
Phys.\ Lett.\ B {\bf 405}, 79 (1997).
[hep-th/9703172].
%%CITATION = hep-th/9703172%%
}

%\SafdiRE
\lref\SafdiRE{
  B.~R.~Safdi, I.~R.~Klebanov and J.~Lee,
  ``A Crack in the Conformal Window,''
[arXiv:1212.4502 [hep-th]].
%%CITATION = arXiv:1212.4502%%
}

%\KutasovNP
\lref\KutasovNP{
  D.~Kutasov and A.~Schwimmer,
  ``On duality in supersymmetric Yang-Mills theory,''
Phys.\ Lett.\ B {\bf 354}, 315 (1995).
[hep-th/9505004].
%%CITATION = hep-th/9505004%%
}

%\KutasovSS
\lref\KutasovSS{
  D.~Kutasov, A.~Schwimmer and N.~Seiberg,
  ``Chiral rings, singularity theory and electric - magnetic duality,''
Nucl.\ Phys.\ B {\bf 459}, 455 (1996).
[hep-th/9510222].
%%CITATION = hep-th/9510222%%
}

%\GiveonZN
\lref\GiveonZN{
  A.~Giveon and D.~Kutasov,
  ``Seiberg Duality in Chern-Simons Theory,''
Nucl.\ Phys.\ B {\bf 812}, 1 (2009).
[arXiv:0808.0360 [hep-th]].
%%CITATION = arXiv:0808.0360%%
}

\lref\HoriDK{
  K.~Hori and D.~Tong,
  ``Aspects of Non-Abelian Gauge Dynamics in Two-Dimensional N=(2,2) Theories,''
JHEP {\bf 0705}, 079 (2007).
[hep-th/0609032].
%%CITATION = hep-th/0609032%%
}

\lref\NiarchosAH{
  V.~Niarchos,
  ``Seiberg dualities and the 3d/4d connection,''
JHEP {\bf 1207}, 075 (2012).
[arXiv:1205.2086 [hep-th]].
%%CITATION = arXiv:1205.2086%%
}

\lref\KapustinJM{
  A.~Kapustin and B.~Willett,
  ``Generalized Superconformal Index for Three Dimensional Field Theories,''
[arXiv:1106.2484 [hep-th]].
%%CITATION = arXiv:1106.2484%%
}

%\KimCMA
\lref\KimCMA{
  H.~Kim and J.~Park,
  ``Aharony Dualities for 3d Theories with Adjoint Matter,''
[arXiv:1302.3645 [hep-th]].
%%CITATION = arXiv:1302.3645%%
}

%\KapustinVZ
\lref\KapustinVZ{
  A.~Kapustin, H.~Kim and J.~Park,
  ``Dualities for 3d Theories with Tensor Matter,''
JHEP {\bf 1112}, 087 (2011).
[arXiv:1110.2547 [hep-th]].
%%CITATION = arXiv:1110.2547%%
}

\lref\AharonyGP{
  O.~Aharony,
  ``IR duality in d = 3 N=2 supersymmetric USp(2N(c)) and U(N(c)) gauge theories,''
Phys.\ Lett.\ B {\bf 404}, 71 (1997).
[hep-th/9703215].
%%CITATION = hep-th/9703215%%
}

%\WittenDS
\lref\WittenDS{
  E.~Witten,
  ``Supersymmetric index of three-dimensional gauge theory,''
In *Shifman, M.A. (ed.): The many faces of the superworld* 156-184.
[hep-th/9903005].
%%CITATION = hep-th/9903005%%
}

\lref\FestucciaWS{
  G.~Festuccia and N.~Seiberg,
  ``Rigid Supersymmetric Theories in Curved Superspace,''
JHEP {\bf 1106}, 114 (2011).
[arXiv:1105.0689 [hep-th]].
%%CITATION = arXiv:1105.0689%%
}

\lref\SpiridonovHF{
  V.~P.~Spiridonov and G.~S.~Vartanov,
  ``Elliptic hypergeometry of supersymmetric dualities II. Orthogonal groups, knots, and vortices,''
[arXiv:1107.5788 [hep-th]].
%%CITATION = arXiv:1107.5788%%
}
\lref\SpiridonovZR{
  V.~P.~Spiridonov and G.~S.~Vartanov,
  ``Superconformal indices for N = 1 theories with multiple duals,''
Nucl.\ Phys.\ B {\bf 824}, 192 (2010).
[arXiv:0811.1909 [hep-th]].
%%CITATION = arXiv:0811.1909%%
}

\lref\HoriPD{
  K.~Hori,
  ``Duality In Two-Dimensional (2,2) Supersymmetric Non-Abelian Gauge Theories,''
[arXiv:1104.2853 [hep-th]].
%%CITATION = arXiv:1104.2853%%
}

\lref\RomelsbergerEG{
  C.~Romelsberger,
  ``Counting chiral primaries in N = 1, d=4 superconformal field theories,''
Nucl.\ Phys.\ B {\bf 747}, 329 (2006).
[hep-th/0510060].
%%CITATION = hep-th/0510060%%
}

\lref\KapustinKZ{
  A.~Kapustin, B.~Willett and I.~Yaakov,
  ``Exact Results for Wilson Loops in Superconformal Chern-Simons Theories with Matter,''
JHEP {\bf 1003}, 089 (2010).
[arXiv:0909.4559 [hep-th]].
%%CITATION = arXiv:0909.4559%%
}

%\NaculichNC
\lref\NaculichNC{
  S.~G.~Naculich and H.~J.~Schnitzer,
  ``Level-rank duality of the U(N) WZW model, Chern-Simons theory, and 2-D qYM theory,''
JHEP {\bf 0706}, 023 (2007).
[hep-th/0703089 [HEP-TH]].
%%CITATION = hep-th/0703089%%
}

\lref\Naculich{
  S.~G.~Naculich, H.~A.~Riggs and H.~J.~Schnitzer,
  ``Group Level Duality In Wzw Models And Chern-simons Theory,''
  Phys.\ Lett.\ B {\bf 246} (1990) 417.
  %%CITATION = PHLTA,B246,417;%%
}

\lref\KimAVA{
  H.~-C.~Kim and S.~Kim,
  ``M5-branes from gauge theories on the 5-sphere,''
[arXiv:1206.6339 [hep-th]].
%%CITATION = SNUTP12-002%%
}

\lref\RazamatUV{
  S.~S.~Razamat,
  ``On a modular property of N=2 superconformal theories in four dimensions,''
JHEP {\bf 1210}, 191 (2012).
[arXiv:1208.5056 [hep-th]].
%%CITATION = arXiv:1208.5056%%
}
%\cite{Camperi:1990dk}
\lref\Camperi{
  M.~Camperi, F.~Levstein and G.~Zemba,
  ``The Large N Limit Of Chern-simons Gauge Theory,''
  Phys.\ Lett.\ B {\bf 247} (1990) 549.
  %%CITATION = PHLTA,B247,549;%%
}

%\cite{Mlawer:1990uv}
\lref\Mlawer{
  E.~J.~Mlawer, S.~G.~Naculich, H.~A.~Riggs and H.~J.~Schnitzer,
  ``Group level duality of WZW fusion coefficients and Chern-Simons link observables,''
  Nucl.\ Phys.\ B {\bf 352} (1991) 863.
  %%CITATION = NUPHA,B352,863;%%
}

%\DolanQI
\lref\DolanQI{
  F.~A.~Dolan and H.~Osborn,
  ``Applications of the Superconformal Index for Protected Operators and q-Hypergeometric Identities to N=1 Dual Theories,''
Nucl.\ Phys.\ B {\bf 818}, 137 (2009).
[arXiv:0801.4947 [hep-th]].
%%CITATION = arXiv:0801.4947%%
}

\lref\EagerHX{
  R.~Eager, J.~Schmude and Y.~Tachikawa,
  ``Superconformal Indices, Sasaki-Einstein Manifolds, and Cyclic Homologies,''
[arXiv:1207.0573 [hep-th]].
%%CITATION = arXiv:1207.0573%%
}

\lref\GaddeIA{
  A.~Gadde and W.~Yan,
  ``Reducing the 4d Index to the $S^3$ Partition Function,''
JHEP {\bf 1212}, 003 (2012).
[arXiv:1104.2592 [hep-th]].
%%CITATION = arXiv:1104.2592%%
}

\lref\DolanRP{
  F.~A.~H.~Dolan, V.~P.~Spiridonov and G.~S.~Vartanov,
  ``From 4d superconformal indices to 3d partition functions,''
Phys.\ Lett.\ B {\bf 704}, 234 (2011).
[arXiv:1104.1787 [hep-th]].
%%CITATION = arXiv:1104.1787%%
}

%\IntriligatorEX
\lref\IntriligatorEX{
  K.~A.~Intriligator and N.~Seiberg,
  ``Mirror symmetry in three-dimensional gauge theories,''
Phys.\ Lett.\ B {\bf 387}, 513 (1996).
[hep-th/9607207].
%%CITATION = hep-th/9607207%%
}

%\deBoerMP
\lref\deBoerMP{
  J.~de Boer, K.~Hori, H.~Ooguri and Y.~Oz,
  ``Mirror symmetry in three-dimensional gauge theories, quivers and D-branes,''
Nucl.\ Phys.\ B {\bf 493}, 101 (1997).
[hep-th/9611063].
%%CITATION = hep-th/9611063%%
}

\lref\ImamuraUW{
  Y.~Imamura,
 ``Relation between the 4d superconformal index and the $S^3$ partition function,''
JHEP {\bf 1109}, 133 (2011).
[arXiv:1104.4482 [hep-th]].
%%CITATION = arXiv:1104.4482%%
}

%\SeibergBZ
\lref\SeibergBZ{
  N.~Seiberg,
  ``Exact results on the space of vacua of four-dimensional SUSY gauge theories,''
Phys.\ Rev.\ D {\bf 49}, 6857 (1994).
[hep-th/9402044].
%%CITATION = hep-th/9402044%%
}

\lref\HamaEA{
  N.~Hama, K.~Hosomichi and S.~Lee,
  ``SUSY Gauge Theories on Squashed Three-Spheres,''
JHEP {\bf 1105}, 014 (2011).
[arXiv:1102.4716 [hep-th]].
%%CITATION = arXiv:1102.4716%%
}

%\AffleckAS
\lref\AffleckAS{
  I.~Affleck, J.~A.~Harvey and E.~Witten,
  ``Instantons and (Super)Symmetry Breaking in (2+1)-Dimensions,''
Nucl.\ Phys.\ B {\bf 206}, 413 (1982).
%%CITATION = PRINT-82-0478 (PRINCETON)%%
}

%\SeibergPQ
\lref\SeibergPQ{
  N.~Seiberg,
  ``Electric - magnetic duality in supersymmetric nonAbelian gauge theories,''
Nucl.\ Phys.\ B {\bf 435}, 129 (1995).
[hep-th/9411149].
%%CITATION = hep-th/9411149%%
}

%\CveticXN
\lref\CveticXN{
  M.~Cvetic, T.~W.~Grimm and D.~Klevers,
  ``Anomaly Cancellation And Abelian Gauge Symmetries In F-theory,''
JHEP {\bf 1302}, 101 (2013).
[arXiv:1210.6034 [hep-th]].
%%CITATION = arXiv:1210.6034%%
}

\lref\debult{
  F.~van~de~Bult,
  ``Hyperbolic Hypergeometric Functions,''
University of Amsterdam Ph.D. thesis
}

%\MoritaCS
\lref\MoritaCS{
  T.~Morita and V.~Niarchos,
  ``F-theorem, duality and SUSY breaking in one-adjoint Chern-Simons-Matter theories,''
Nucl.\ Phys.\ B {\bf 858}, 84 (2012).
[arXiv:1108.4963 [hep-th]].
%%CITATION = arXiv:1108.4963%%
}

\lref\Shamirthesis{
  I.~Shamir,
  ``Aspects of three dimensional Seiberg duality,''
  M. Sc. thesis submitted to the Weizmann Institute of Science, April 2010.
  }

\lref\slthreeZ{
  J.~Felder, A.~Varchenko,
  ``The elliptic gamma function and $SL(3,Z) \times Z^3$,'' $\;\;$
[arXiv:math/0001184].
}

\lref\SpiridonovZA{
  V.~P.~Spiridonov and G.~S.~Vartanov,
  ``Elliptic Hypergeometry of Supersymmetric Dualities,''
Commun.\ Math.\ Phys.\  {\bf 304}, 797 (2011).
[arXiv:0910.5944 [hep-th]].
%%CITATION = arXiv:0910.5944%%
}

%\BeniniMF
\lref\BeniniMF{
  F.~Benini, C.~Closset and S.~Cremonesi,
  ``Comments on 3d Seiberg-like dualities,''
JHEP {\bf 1110}, 075 (2011).
[arXiv:1108.5373 [hep-th]].
%%CITATION = arXiv:1108.5373%%
}

%\ClossetVG
\lref\ClossetVG{
  C.~Closset, T.~T.~Dumitrescu, G.~Festuccia, Z.~Komargodski and N.~Seiberg,
  ``Contact Terms, Unitarity, and F-Maximization in Three-Dimensional Superconformal Theories,''
JHEP {\bf 1210}, 053 (2012).
[arXiv:1205.4142 [hep-th]].
%%CITATION = arXiv:1205.4142%%
}

%\ClossetVP
\lref\ClossetVP{
  C.~Closset, T.~T.~Dumitrescu, G.~Festuccia, Z.~Komargodski and N.~Seiberg,
  ``Comments on Chern-Simons Contact Terms in Three Dimensions,''
JHEP {\bf 1209}, 091 (2012).
[arXiv:1206.5218 [hep-th]].
%%CITATION = arXiv:1206.5218%%
}

\lref\SpiridonovHF{
  V.~P.~Spiridonov and G.~S.~Vartanov,
  ``Elliptic hypergeometry of supersymmetric dualities II. Orthogonal groups, knots, and vortices,''
[arXiv:1107.5788 [hep-th]].
%%CITATION = arXiv:1107.5788%%
}

%\ElitzurFH
\lref\ElitzurFH{
  S.~Elitzur, A.~Giveon and D.~Kutasov,
  ``Branes and N=1 duality in string theory,''
Phys.\ Lett.\ B {\bf 400}, 269 (1997).
[hep-th/9702014].
%%CITATION = hep-th/9702014%%
}

%\ElitzurHC
\lref\ElitzurHC{
  S.~Elitzur, A.~Giveon, D.~Kutasov, E.~Rabinovici and A.~Schwimmer,
  ``Brane dynamics and N=1 supersymmetric gauge theory,''
Nucl.\ Phys.\ B {\bf 505}, 202 (1997).
[hep-th/9704104].
%%CITATION = hep-th/9704104%%
}

%\KapustinGH
\lref\KapustinGH{
  A.~Kapustin,
  ``Seiberg-like duality in three dimensions for orthogonal gauge groups,''
[arXiv:1104.0466 [hep-th]].
%%CITATION = arXiv:1104.0466%%
}

%\HwangHT
\lref\HwangHT{
  C.~Hwang, K.~-J.~Park and J.~Park,
  ``Evidence for Aharony duality for orthogonal gauge groups,''
JHEP {\bf 1111}, 011 (2011).
[arXiv:1109.2828 [hep-th]].
%%CITATION = arXiv:1109.2828%%
}

\lref\SpiridonovWW{
  V.~P.~Spiridonov and G.~S.~Vartanov,
  ``Elliptic hypergeometric integrals and 't Hooft anomaly matching conditions,''
JHEP {\bf 1206}, 016 (2012).
[arXiv:1203.5677 [hep-th]].
%%CITATION = arXiv:1203.5677%%
}

\lref\DimoftePY{
  T.~Dimofte, D.~Gaiotto and S.~Gukov,
  ``3-Manifolds and 3d Indices,''
[arXiv:1112.5179 [hep-th]].
%%CITATION = arXiv:1112.5179%%
}

\lref\KimWB{
  S.~Kim,
  ``The Complete superconformal index for N=6 Chern-Simons theory,''
Nucl.\ Phys.\ B {\bf 821}, 241 (2009), [Erratum-ibid.\ B {\bf 864}, 884 (2012)].
[arXiv:0903.4172 [hep-th]].
%%CITATION = arXiv:0903.4172%%
}

%\WillettGP
\lref\WillettGP{
  B.~Willett and I.~Yaakov,
  ``N=2 Dualities and Z Extremization in Three Dimensions,''
[arXiv:1104.0487 [hep-th]].
%%CITATION = arXiv:1104.0487%%
}

%\ParkWTA
\lref\ParkWTA{
  J.~Park and K.~-J.~Park,
  ``Seiberg-like Dualities for 3d N=2 Theories with SU(N) gauge group,''
[arXiv:1305.6280 [hep-th]].
%%CITATION = arXiv:1305.6280%%
}

\lref\KapustinXQ{
  A.~Kapustin, B.~Willett and I.~Yaakov,
  ``Nonperturbative Tests of Three-Dimensional Dualities,''
JHEP {\bf 1010}, 013 (2010).
[arXiv:1003.5694 [hep-th]].
%%CITATION = arXiv:1003.5694%%
}

\lref\ImamuraSU{
  Y.~Imamura and S.~Yokoyama,
  ``Index for three dimensional superconformal field theories with general R-charge assignments,''
JHEP {\bf 1104}, 007 (2011).
[arXiv:1101.0557 [hep-th]].
%%CITATION = arXiv:1101.0557%%
}

%\GaddeDDA
\lref\GaddeDDA{
  A.~Gadde and S.~Gukov,
  ``2d Index and Surface operators,''
[arXiv:1305.0266 [hep-th]].
%%CITATION = CALT-68.2932%%
}

\lref\HwangQT{
  C.~Hwang, H.~Kim, K.~-J.~Park and J.~Park,
  ``Index computation for 3d Chern-Simons matter theory: test of Seiberg-like duality,''
JHEP {\bf 1109}, 037 (2011).
[arXiv:1107.4942 [hep-th]].
%%CITATION = arXiv:1107.4942%%
}

\lref\GreenDA{
  D.~Green, Z.~Komargodski, N.~Seiberg, Y.~Tachikawa and B.~Wecht,
  ``Exactly Marginal Deformations and Global Symmetries,''
JHEP {\bf 1006}, 106 (2010).
[arXiv:1005.3546 [hep-th]].
%%CITATION = arXiv:1005.3546%%
}

%\IntriligatorID
\lref\IntriligatorID{
  K.~A.~Intriligator and N.~Seiberg,
  ``Duality, monopoles, dyons, confinement and oblique confinement in supersymmetric SO(N(c)) gauge theories,''
Nucl.\ Phys.\ B {\bf 444}, 125 (1995).
[hep-th/9503179].
%%CITATION = hep-th/9503179%%
}

  %\SeibergQD
\lref\SeibergQD{
  N.~Seiberg,
  ``Modifying the Sum Over Topological Sectors and Constraints on Supergravity,''
JHEP {\bf 1007}, 070 (2010).
[arXiv:1005.0002 [hep-th]].
%%CITATION = arXiv:1005.0002%%
}
%\BanksZN
\lref\BanksZN{
  T.~Banks and N.~Seiberg,
  ``Symmetries and Strings in Field Theory and Gravity,''
Phys.\ Rev.\ D {\bf 83}, 084019 (2011).
[arXiv:1011.5120 [hep-th]].
%%CITATION = arXiv:1011.5120%%
}

%\SeibergNZ
\lref\SeibergNZ{
  N.~Seiberg and E.~Witten,
  ``Gauge dynamics and compactification to three-dimensions,''
In *Saclay 1996, The mathematical beauty of physics* 333-366.
[hep-th/9607163].
%%CITATION = hep-th/9607163%%
}

%\AharonyCI
\lref\AharonyCI{
  O.~Aharony and I.~Shamir,
  ``On $O(N_c)$ d=3 N=2 supersymmetric QCD Theories,''
JHEP {\bf 1112}, 043 (2011).
[arXiv:1109.5081 [hep-th]].
%%CITATION = arXiv:1109.5081%%
}

%\GiveonSR
\lref\GiveonSR{
  A.~Giveon and D.~Kutasov,
  ``Brane dynamics and gauge theory,''
Rev.\ Mod.\ Phys.\  {\bf 71}, 983 (1999).
[hep-th/9802067].
%%CITATION = hep-th/9802067%%
}

%\KapustinPY
\lref\KapustinPY{
  A.~Kapustin,
  ``Wilson-'t Hooft operators in four-dimensional gauge theories and S-duality,''
Phys.\ Rev.\ D {\bf 74}, 025005 (2006).
[hep-th/0501015].
%%CITATION = hep-th/0501015%%
}

%\KapustinGH
\lref\KapustinGH{
  A.~Kapustin,
  ``Seiberg-like duality in three dimensions for orthogonal gauge groups,''
[arXiv:1104.0466 [hep-th]].
%%CITATION = arXiv:1104.0466%%
}

\lref\toappearthree{O. Aharony, S. S. Razamat, N.~Seiberg and B.~Willett, ``$3d$ dualities from $4d$ dualities for orthogonal groups,'' to appear.}

\lref\workinprogresstwo{O. Aharony, S. S. Razamat, N.~Seiberg and
B.~Willett, work in progress.}

\lref\AST{
O.~Aharony, N.~Seiberg and Y.~Tachikawa,
  ``Reading between the lines of four-dimensional gauge theories,''
[arXiv:1305.0318 [hep-th]].
%%CITATION = WIS-03-13-APR-DPPA%%
}

\Title{\vbox{\baselineskip12pt
\hbox{WIS/04/13-APR-DPPA}
}}
{\vbox{\centerline{$3d$ dualities from $4d$ dualities}
%\centerline{}
}}
\centerline{Ofer Aharony$^{a,b}$, Shlomo S. Razamat$^a$, Nathan Seiberg$^a$,
and Brian Willett$^a$}
\bigskip
\centerline{$^a${\it School of Natural Sciences, Institute for Advanced Study, Princeton, NJ 08540, USA}}
\centerline{}
\centerline{${}^b${\it Department of Particle Physics and Astrophysics}}
\centerline{{\it Weizmann Institute of Science, Rehovot 76100, Israel}}
\vskip.1in \vskip.1in \centerline{\bf Abstract}

\noindent
Many examples of low-energy dualities have been found in supersymmetric gauge theories with four supercharges, both in four and in three space-time dimensions. In these
dualities, two theories that are different at high energies have the same
low-energy limit. In this paper we clarify the relation between the dualities
in four and in three dimensions. We show that every four dimensional duality
gives rise to a three dimensional duality between theories that are similar,
but not identical, to the dimensional reductions of the four dimensional dual gauge
theories to three dimensions. From these specific three dimensional dualities one can flow to many other low-energy dualities, including known  three dimensional dualities and many new ones. We discuss in detail the case of three dimensional $SU(N_c)$ supersymmetric QCD theories, showing how to derive new duals for these theories from the four dimensional duality.

\vfill

\Date{May 2013}

%\draftmode

\newsec{Introduction and Summary}

It is interesting to study gauge theories with four supercharges in various dimensions,
since on one hand, their dynamics is quite similar to that of non-supersymmetric gauge
theories (exhibiting phenomena like confinement and chiral symmetry breaking), and on the
other hand, supersymmetry allows some control over their strong coupling behavior (some
things, like effective superpotentials and expectation values of chiral operators, can often
be computed exactly).

An interesting dynamical phenomenon that was discovered in these gauge
theories almost twenty years ago is IR (infrared) dualities -- two gauge theories that are
different at high energies can have the same low-energy limit. The original and typical
example of this is the duality \refs{\SeibergPQ,\IntriligatorAU} between theory A which is an $SU(N_c)$ ${\cal N}=1$ supersymmetric
$4d$ gauge theory with $N_f$ flavors (chiral multiplets $Q_i$ ($i=1,\cdots,N_f$) in the fundamental representation and ${\tilde Q}^{\tilde i}$ (${\tilde i}=1,\cdots,N_f$) in the anti-fundamental representation) and no superpotential ($W_A=0$), and theory B that is an $SU(N_f-N_c)$ gauge theory,
also with $N_f$ flavors $q^i$ and ${\tilde q}_{\tilde i}$ ($i,{\tilde i}=1,\cdots,N_f$), but also with
$N_f^2$ singlets $M_{i}^{{\tilde i}}$, and with a superpotential $W_B = M_{i}^{{\tilde i}} q^i {\tilde q}_{\tilde i}$. Whenever both of these theories are asymptotically free, they are believed to flow to the
same low-energy interacting superconformal field theory, while if one of them is asymptotically free
and the other IR-free, then the IR-free theory gives the low-energy effective theory for the
asymptotically free theory. By now many examples of such dualities have been found, and a lot of
evidence has been collected for their validity. However, there is still no general understanding of
the origin of these dualities, nor a prescription to find the dual for a given gauge theory.

Examples of similar dualities have been found also in lower space-time dimensions, and in particular many dualities are known for $3d$ gauge theories. The main question we would like to address in this paper is whether there is any relation between such $4d$ dualities and $3d$ dualities.\foot{This question was recently discussed in \NiarchosAH.} For instance, one may wonder whether, given a pair of dual $4d$ gauge theories like theories
A and B above, the dimensional reductions of these theories to three dimensions also exhibit an IR
duality or not.

At first sight it seems unlikely that there would be any relation between $4d$ and $3d$ dualities. Strong
coupling gauge dynamics is very different in $4d$ and in $3d$: $4d$ gauge theories exhibit confinement for non-Abelian gauge theories, which is related to monopole condensation, and non-perturbative effects
related to instantons, while in $3d$ there is confinement also for Abelian gauge theories, where
completely different non-perturbative effects sometimes make the gauge fields massive. Moreover, the $4d$
dualities are related to electric-magnetic duality (this is most evident in the example of the duality
for $SO(N_c)$ gauge theories when they are broken to an Abelian $SO(2)$ gauge theory \refs{\SeibergPQ,\IntriligatorID}), which is obviously unique to four space-time dimensions. So, one would not expect a pair of theories that is related by a strong coupling duality in $4d$ to exhibit an IR duality also in $3d$, and indeed one can check (for instance) that theories A and B above, when dimensionally reduced to $3d$, are not equivalent at low energies (they do not even have the same moduli space of supersymmetric vacua). As another trivial example
of this, in $4d$ the pure $SU(N_c)$ gauge theory has $N_c$ supersymmetric vacua exhibiting a mass gap
(so it is IR-dual to a trivial theory), while in $3d$ the dimensional reduction of this theory generates an effective superpotential which leads to a runaway behavior with no supersymmetric vacua \AffleckAS.

Nevertheless, $3d$ dualities have been found between gauge theories that are quite similar to the ones appearing in $4d$ dualities. For instance, the $3d$ $U(N)$ gauge theories with the same $N$'s and matter content as that of theories A and B above (with an extra singlet for theory B), and with a slightly modified superpotential (including also monopole operators), are dual to each other \AharonyGP. Similarly, in $4d$ there is a duality between $USp(2N_c)$ gauge theories with $2N_f$ flavors, and $USp(2N_f-2N_c-4)$ theories with $2N_f$ flavors and extra singlets \refs{\SeibergPQ,\IntriligatorNE}, and a similar duality between $USp(2N_c)$ and $USp(2N_f-2N_c-2)$ theories was discovered in $3d$ \refs{\KarchUX,\AharonyGP}. One explanation for this similarity is that all of the dualities mentioned so far (though not all known IR dualities) can be realized by brane constructions in string theory \refs{\ElitzurFH,\ElitzurHC} (see \GiveonSR\ for a review), and the brane configurations related to $4d$ dualities (involving D4-branes stretched between NS5-branes) are similar to the ones related to $3d$ dualities (involving D3-branes stretched between NS5-branes). But the brane constructions do not suggest any direct relation between the $4d$ and the $3d$ dualities (one can go between then by compactifying the $4d$ brane configurations on a circle and performing a T-duality transformation, but this goes beyond the low-energy field theory limit). So, one may think that the similarity between
the $4d$ and the $3d$ dualities is accidental, because of the similar brane configurations that are related to them.

In this paper we would like to argue that there is a direct way to relate $4d$ dualities to $3d$ dualities,
and we will explain exactly how it works. In fact, there are two simple arguments that suggest that $4d$
dualities should reduce directly to $3d$ dualities. First, let's take two dual theories like theories A and B above, in the range of parameters where both theories are asymptotically free, and compactify them on a circle of radius $r$. The IR duality implies that the two theories exhibit the same physics at energies below their strong coupling scales $\Lambda$, ${\tilde \Lambda}$. This is still true after the compactification; indeed we can just take the low-energy superconformal field theory (SCFT) that both theories flow to and compactify this theory on a circle of radius $r$. But now, if we look at energies below the scale $1/r$, then the Kaluza-Klein modes on the circle decouple, and if we take $r\to 0$ it seems that
we should obtain an IR duality between the corresponding $3d$ gauge theories, which would say that the $3d$ dimensional reductions of theories A and B are the same at low energies (and equivalent to the dimensional reduction of the $4d$ IR SCFT on a circle).

We can also give a second, more formal argument, along the same lines. Supersymmetric $4d$ ${\cal N}=1$ theories have a supersymmetric partition function
on a Euclidean $\S^3\times \S^1$ which is renormalization-group invariant \refs{\RomelsbergerEG,\FestucciaWS}, and should thus be the same in any pair of theories that are IR-dual; this was verified explicitly in many examples \refs{\DolanQI\SpiridonovHF\SpiridonovZA\GaddeEN\SpiridonovZR-\EagerHX}. This supersymmetric partition function depends on the ratio of the radii of the $\S^3$ and the $\S^1$, and in the limit of small $\S^1$ radius $r\to 0$ it reduces to the Euclidean partition function on $\S^3$ of the dimensionally reduced $3d$ theories~\refs{\DolanRP,\GaddeIA,\ImamuraUW,\NiarchosAH}. Thus, the $4d$ duality implies that the Euclidean partition function on $\S^3$ of the $3d$ versions of theories A and B should be the same, and usually this is considered as very strong evidence for the equivalence of these theories~\KapustinXQ. This type of relation between the $4d$ and $3d$ dualities was recently investigated in \NiarchosAH.

But we mentioned above that the $3d$ versions of theories A and B are not dual to each other, so something must be wrong with the arguments of the two previous paragraphs. Let us discuss more carefully what we mean by the $r\to 0$ limit. As we described above, the two $4d$ theories on a circle are identical at energies obeying $E \ll \Lambda, {\tilde \Lambda}, 1/r$. However, what we mean by a $3d$ duality is that we start from the $3d$ gauge theories with fixed $3d$ gauge couplings $g_3^2, {\tilde g}_3^2$, and that the theories should be equivalent for $E \ll g_3^2, {\tilde g}_3^2$. The strong coupling scale $\Lambda$ is related to the $4d$
gauge coupling through
\eqn\lambdadef{\Lambda^b = \exp(-8\pi^2 / g_4^2),}
where $b > 0$ is the one-loop beta function
coefficient, and we have set the renormalization scale to one and the theta angle to zero for simplicity. ${\tilde \Lambda}$ obeys a similar relation. When we compactify the theories on a circle we have $g_4^2 = 2 \pi r g_3^2$, so that
\eqn\lambdainthree{\Lambda^b = \exp(-4\pi / r g_3^2).}
This means that if we take the limit $r\to 0$ keeping $g_3^2$ fixed, then
$\Lambda \to 0$ very fast (and also the dimensionless parameter $\Lambda r \to 0$), and the statement that the two theories are dual below the scales $\Lambda, {\tilde \Lambda}$ becomes meaningless. In other words, the low-energy limit $E \ll \Lambda, {\tilde \Lambda}$ in which theories A and B become equivalent does not commute with the $3d$ limit $r \to 0$ keeping the $3d$ gauge couplings fixed, which is relevant for $3d$ dualities. In particular, the parameters of the $4d$ dual theories obey a relation of the form $\Lambda^b {\tilde \Lambda}^{\tilde b} = (-1)^{N_f-N_c}$ \IntriligatorAU, which is not consistent with the $3d$ gauge theory limit in which
$\Lambda, {\tilde \Lambda} \to 0$.

However, all is not lost, since we can discuss a different limit, in which we keep $\Lambda$, ${\tilde \Lambda}$ and $r$ fixed, and look at energies $E \ll \Lambda, {\tilde \Lambda}, 1/r$. In this limit theories A and B are the same, and since we are below the Kaluza-Klein scale, their effective low-energy behavior is three dimensional. The point is that the effective $3d$ theories that we get by this procedure differ from the low-energy limits of the naive dimensional reductions of theories A and B in two important ways:
\item{1)} Four dimensional gauge theories compactified on $\S^1$ have extra scalars coming from the holonomy of the gauge field around the circle. A generic vacuum expectation value (VEV) for this holonomy breaks the gauge group $G$ to $U(1)^{r_G}$ (where $r_G$ is the rank of $G$). In the low-energy effective action one can then dualize the $r_G$ photons into scalars, and obtain (at least classically) a low-energy theory involving $2r_G$ scalar fields which are classically massless. In supersymmetric theories the VEVs of these scalars parameterize the classical ``Coulomb branch'' of the moduli space. In the theory on a circle, the scalar fields coming from the holonomy are periodic, with a periodicity $\sim 1/r$, since only the eigenvalues of $P\exp(i\oint A_3)$ are gauge-invariant, while in the $3d$ theory the scalars are not periodic. So the effective low-energy theory that we get for finite values of $r$ has a compact Coulomb branch, while the $3d$ theories have a non-compact Coulomb branch.
\item{2)} As we will review below, the $4d$ theories on a circle have non-perturbative superpotentials that are generated by instantons on their Coulomb branch \SeibergNZ, of the schematic form
\eqn\weff{W = \eta Y_{low},}
where $\eta \equiv \Lambda^b$ and $Y_{low}$ is one of the Coulomb branch coordinates. These superpotentials are not present in the $3d$ gauge theories that we get by dimensional reduction (since in the $3d$ limit, $\eta \to 0$).

\noindent
Our arguments above imply that the low-energy theories with these two extra features are dual to
each other, but they are not the same as the naive $3d$ theories, explaining why the latter do not
match.

This also helps to clarify the fallacy in our argument that the partition functions on $\S^3$ of the dimensionally reduced $3d$ theories match. The $3d$ versions of theories A and B have a large global symmetry group, but in $4d$ one combination of the global $U(1)$ symmetries is anomalous. The supersymmetric partition functions of theories A and B on $\S^3\times \S^1$ are parameterized by the expectation values of background fields coupled to the $4d$ global symmetry current multiplets, and they reduce as $r\to 0$ to the Euclidean partition functions on $\S^3$ parameterized by background fields coupled to the $3d$ reductions of these current multiplets. But the $3d$ partition functions on $\S^3$ have an extra parameter related to the $U(1)$ symmetry that was anomalous in $4d$, and the argument above tells us that the partition functions must agree when this parameter vanishes, but they do not have to agree when this parameter is non-zero. And indeed, looking at the Euclidean partition functions on $\S^3$ of the $3d$ theories A and B, one finds that they are the same on this codimension-one subspace of their parameter space, but not in general~\refs{\DolanRP,\NiarchosAH}. In the effective $3d$ theory that we get at finite $r$ as described above, the extra superpotential $W = \eta Y_{low}$ breaks exactly the same $U(1)$ symmetry that is anomalous in $4d$. Since the only effect of superpotentials on the $\S^3$ partition function is through the global symmetries they break, the Euclidean $\S^3$ partition functions of the effective $3d$ theories are exactly the same as the ones we get by dimensional reduction from $4d$, so they agree (as implied by the discussion above). But this does not imply a duality between the $3d$ gauge theories without the superpotential \weff. Note, in particular, that if we first flow to the IR in four dimensions, and then compactify the resulting SCFT on a circle, then the low-energy limit that we will find will be equivalent to the one we find above with finite $\eta$ (or finite $\tilde \eta$), but it is not the same as the low-energy limit of either of the corresponding $3d$ gauge theories.

\midinsert\bigskip{\vbox{{\epsfxsize=4.2in
        \nobreak
    \centerline{\epsfbox{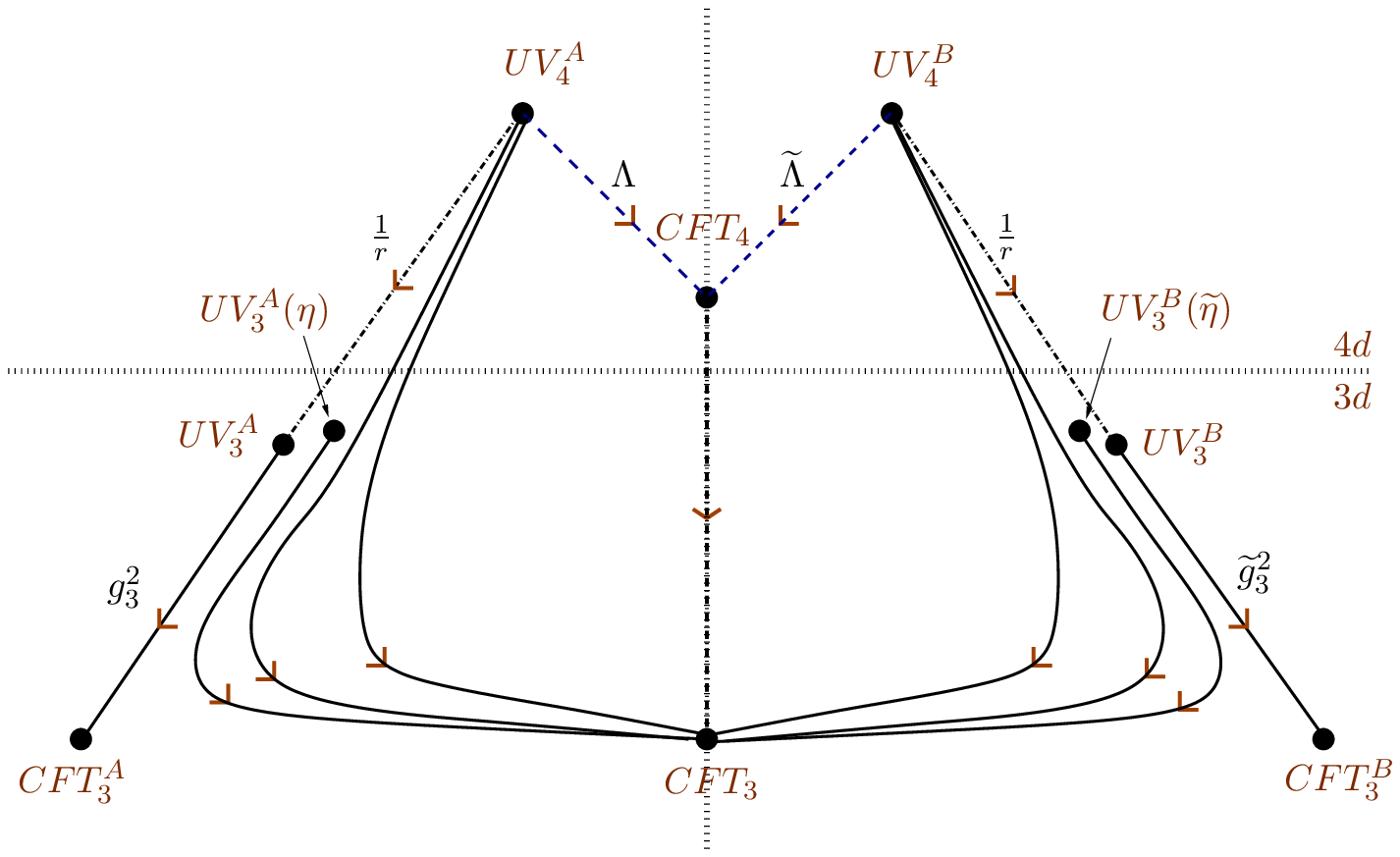}}
        \nobreak\bigskip
    {\raggedright\it \vbox{
{\bf Figure 1.}
{\it  Different ways to obtain $3d$ theories from $4d$ theories.
One starts with a duality in $4d$ between two UV theories $UV_4^{A,B}$, which both flow in $4d$ (below the scales $\Lambda$, ${\tilde \Lambda}$) to $CFT_4$ (dashed blue lines). One can dimensionally reduce the UV theories to $3d$ $UV_3^{A,B}$ Lagrangians with finite $3d$ gauge couplings, which then flow at low energies to $CFT_3^{A,B}$. Alternatively, one can keep the compactification radius and the scales finite and go to low energies. This gives an effective $3d$ description $CFT_3$ which differs from  $CFT_3^{A,B}$ by an $\eta$-superpotential. A $3d$ UV-completion of this theory is denoted by $UV_3^A(\eta)$ ($UV_3^B(\widetilde\eta)$) in the plot. As discussed in the text, $CFT_3$ may or may not be the same conformal field theory (deformed by a dangerously irrelevant operator) as $CFT_3^A$ or $CFT_3^B$.
}}}}}}
\bigskip\endinsert

So, at this point we understand why the naive dimensional reduction does not give a duality, but we do get a duality between some other $3d$ theories, with compact Coulomb branches and with extra terms in the effective superpotential\foot{In particular, in the example of the pure $SU(N_c)$ supersymmetric Yang-Mills (SYM) theory, the extra term in the superpotential stabilizes the runaway, and leads to $N_c$ supersymmetric vacua also in the effective $3d$ theory \SeibergNZ.}. The next question is whether we can turn  this into a bona fide $3d$ duality, {\it i.e.}\ a duality between $3d$ gauge theories that are well-defined at high energies. One obstruction to this is the compact Coulomb branch. In many theories, like $SU(N_c)$ SQCD, this is not really an obstruction, since the $W = \eta Y_{low}$ superpotential lifts the Coulomb branch completely (so it is not important at low energies). In other theories, like $U(N_c)$ SQCD, some of the Coulomb branch remains, but we can take a limit that focuses only on the low-energy theory near specific points on the Coulomb branch, and there again the compactness of the Coulomb branch becomes irrelevant. The second obstruction is the extra superpotential. We will show that in all cases, we can find monopole operators $Y_{high}$ that are well-defined in the $3d$ gauge theories, and that reduce to the effective operators $Y_{low}$ at low energies. Using these operators, we will be able to formulate a $3d$ duality, between $3d$ gauge theories deformed by a superpotential involving monopole operators, whose validity (at low energies) follows from the validity of the $4d$ IR dualities. By this mechanism, we can relate any $4d$ IR duality to an IR duality of $3d$ gauge theories, which are deformations of the naive $3d$ gauge theories (that arise by dimensional reduction) by monopole operators; theory A is deformed by
\eqn\highsuppot{W_A = \eta Y_{high},}
and theory B now has
\eqn\highsuppotb{W_B = M_{i}^{{\tilde i}} q^i {\tilde q}_{\tilde i} + {\tilde \eta} {\tilde Y}_{high},}
with $\eta {\tilde \eta} = (-1)^{N_f-N_c}$. The resulting $3d$ dualities can now be tested by a variety of methods; their $\S^3$ partition functions are guaranteed to agree from the discussion above, but we can compare their $3d$ superconformal indices (related to $\S^2\times \S^1$ partition functions), their chiral operators and moduli spaces, etc.  One can view these tests as new tests of the $4d$ IR dualities (since the $3d$ dualities follow from them), and the duality passes all of these tests. The relations between all the theories we discussed so far are shown in Figure 1.

Note that the $3d$ theories with the superpotentials \highsuppot, \highsuppotb\ are not really well-defined in the UV, since the term with $Y_{high}$ in the superpotential is not a relevant or a marginal deformation of the high-energy gauge theory (indeed, a naive computation of its UV dimension gives infinity). The claim of the duality is that any UV completion of the theory with these superpotentials will lead to a duality at low energies (this is similar to the $4d$ dualities with $\tr(X^n)$ superpotentials with $n>3$, discussed in \refs{\KutasovNP,\KutasovSS}; the same issue arises also in the $3d$ dualities of \AharonyGP). One UV completion is given by the $4d$ gauge theories on a circle, but since the operators $Y_{high}$ and ${\tilde Y}_{high}$ are well-defined in three dimensions, one can use them to construct $3d$ UV completions of these theories (for instance, by introducing a cutoff in the $3d$ theory).  We claim that $3d$ completions exist, so that there is a meaningful $3d$ duality\foot{Examples of $3d$ UV completions of theories with monopole operators in their superpotential appear in \newIS.}; the details of the UV completion will not be important for anything we discuss. At low energies, at the $3d$ fixed point without the extra superpotential, one can often determine the dimension of the deforming operator $Y$ by F-maximization \refs{\JafferisUN,\JafferisZI,\ClossetVG} (similar computations for $U(N_c)$ gauge theories were performed in \SafdiRE). In some cases this operator is relevant in theory A, and then one can flow from the fixed point without \highsuppot\ to the fixed point with this term.  In other cases the operator ${\tilde Y}$ may be relevant in theory B, and then one can flow from the fixed point without the second term in \highsuppotb\ to the theory with it. In all cases, even when the deformation is irrelevant, it is dangerously irrelevant, since it becomes relevant on the Coulomb branch, where $Y$ or ${\tilde Y}$ have expectation values. In particular, even when the deformation operator is irrelevant so that the low-energy CFTs with and without this term have the same correlation functions, their moduli spaces are different.

After analyzing the $3d$ dualities with the deformed superpotentials (which for $SU(N_c)$ are \highsuppot, \highsuppotb), we next show that we can flow from them by relevant deformations (that we can match between the two dual theories) to many other $3d$ dualities. In the example of $SU(N_c)$ SQCD, we will first show that there is a deformation that removes the superpotential \highsuppot. This leads to a new dual for the simplest $3d$ $SU(N_c)$ SQCD theory. The dual that we find is somewhat elaborate, explaining why it was not guessed in the 1990s. By adding additional mass terms, we can then flow to many other $3d$ dualities. In particular, turning on ``real masses'' for some of the fundamental or anti-fundamental chiral multiplets, we can flow to new dualities for chiral $SU(N_c)$ theories, and for $SU(N_c)$ theories with Chern-Simons (CS) terms.

There are many possible generalizations of this analysis. In this paper we discuss in detail what happens when we repeat this procedure for the $U(N_c)$ and $USp(2N_c)$ SQCD theories.  We reproduce the known duals of these theories, and clarify their relation to the $4d$ dualities. The case of the duality between $SO(N_c)$ gauge theories is somewhat more complicated (already in the known $4d$ \refs{\SeibergPQ,\IntriligatorID,\IntriligatorAU} and $3d$ \AharonyCI\ cases), since in that case the Coulomb branch is not lifted in the $4d$ theory on a circle, and various subtleties involving the precise definition of the $4d$ gauge theory \AST\ are important. We postpone the analysis of this case to a future paper \toappearthree; it leads to a $3d$ duality between an $SO(N_c)$ gauge theory and an $SO(N_f-N_c-2)$ gauge theory, which agrees with their index computations (generalizing the $O(N_c)$ dualities of \refs{\KapustinGH,\BeniniMF,\HwangHT,\AharonyCI}).

It is important to emphasize that our procedure of obtaining a $3d$ duality from a $4d$ duality is completely general, but the details are very different in different theories. In some theories superpotentials like \highsuppot, \highsuppotb\ arise in the low-energy effective action, while in other cases more complicated superpotentials arise, and there are also cases (like theories with extended supersymmetry) where the only effects of the circle compactification at low energies are in the K\"ahler potential. The details of how to remove the extra effects of the circle, and to obtain dualities for
$3d$ theories with no monopole operators in their superpotential, also depend on the precise theory one is analyzing, but in all examples we analyzed it is possible to do this. One example which is more complicated is the reduction of the $4d$ duality of \refs{\KutasovVE,\KutasovNP,\KutasovSS}.  Here the presence of the adjoint chiral superfields significantly modifies the form of the effective superpotential discussed above, both in the $3d$ theory and in the circle compactification of the $4d$ theory. It is possible also in this case to reduce the $4d$ duality to a $3d$ duality, reproducing the results of \refs{\NiarchosJB\NiarchosAA\MoritaCS\KapustinVZ-\KimCMA}, but we will not discuss this here.

We conjecture that all $3d$ dualities follow from $4d$ dynamics.  Our examples based on the reduction of a $4d$ dual pair and a subsequent flow to the IR demonstrate this for many cases.  Although we could not prove it for all known $3d$ dualities (for example, we do not discuss the non-Abelian ${\cal N}=4$ mirror symmetries), we suspect that such a relation always exists.  This situation is reminiscent of recent advances in ${\cal N}=2,\ 4$ dualities in $4d$, identifying their origin in the $6d$ ${\cal N}=(2,0)$ theory.  It would be extremely satisfying if all dualities, including the $4d$ ${\cal N}=1$ dualities and their $3d$ ${\cal N}=2$ descendants, follow from the same $6d$ ancestor.

The methods we discuss should also be useful for relating $4d$ and $3d$ dualities to $2d$ dualities \workinprogresstwo, including and generalizing the dualities discussed in~\refs{\HoriDK,\HoriPD,\GaddeDDA}. There are also other possible relations between $4d$ and $3d$ theories, when the latter arise on defects or boundaries of $4d$ theories, and it is interesting to see if these have any relation to our discussion.

We begin in section 2 by reviewing some of the background on $3d$ ${\cal N}=2$ gauge theories, their Coulomb branches and their monopole operators. In section 3 we discuss in detail the example of $SU(N_c)$ SQCD theories, showing how to go from the known $4d$ dualities to $3d$ dualities for these theories. In section 4 we discuss the same questions for $U(N_c)$ gauge theories, reproducing the known dualities and also some new ones. In section 5 we discuss how the supersymmetric partition functions realize the various dualities we find. In section 6 we describe the generalization to $USp(2N_c)$ gauge groups. Appendix A contains some technical clarifications and details about the computation of the $3d$ supersymmetric indices that we use to test our $3d$ dualities.
Appendix B contains a discussion on the reduction of the $4d$ index to the $3d$ partition function
for a chiral superfield.

After this paper was first posted, \ParkWTA\ appeared, which includes some overlap with the results of sections 3.4, 4.3 and appendix A.

\newsec{Background}
\seclab\background

In this section we provide some useful background about $3d$ ${\cal N}=2$ gauge theories, their low-energy effective descriptions, and their monopole operators. More details may be found in \refs{\AharonyBX,\newIS} and references therein.

\subsec{The classical Coulomb branch of $3d$ ${\cal N}=2$ supersymmetric gauge theories}
\subseclab\classcoul

The supersymmetry multiplets of $3d$ ${\cal N}=2$ theories may be viewed as dimensional reductions of $4d$ ${\cal N}=1$ multiplets. For the chiral multiplets the reduction is straightforward. The dimensional reduction of a vector field  gives a scalar $\sigma$ (coming from the $A_3$ component of the vector) and a $3d$ gauge field $A_{\mu}$. In a free $3d$ Abelian theory with gauge coupling $e_3^2$, the photon $A_{\mu}$ can be dualized into a scalar $a$ defined by
\eqn\dualscalar{\partial_{\mu} a = {\pi\over {e_3^2}} \epsilon_{\mu \nu \rho} F^{\nu \rho},}
where $F$ is the field strength of $A_{\mu}$, and the gauge field is normalized to have $1/4e_3^2$ in front of its kinetic term. The quantization of the magnetic flux implies that $a$ is periodic, and we normalize it so that $a \sim a + 2\pi$. The vacuum expectation values of $\sigma$ and $a$ label a two dimensional manifold known as the ``Coulomb branch'' of vacua. The supersymmetric version of this duality turns the vector multiplet (containing $\sigma$, $A_{\mu}$ and their fermionic partners) into a chiral multiplet, whose lowest component is given by
\eqn\defx{X = \exp\left({{2\pi \sigma}\over e_3^2} + i a\right),}
such that it is single-valued and its expectation value parameterizes the Coulomb branch. The relative coefficient between $\sigma$ and $a$ follows from supersymmetry and the requirement that $X$ should be annihilated by appropriate supercharges to be in a chiral multiplet.  There is also an anti-chiral multiplet whose lowest component is $X^*$.

In the presence of charged matter fields it is not known how to dualize the photon. However, in the supersymmetric theory (or in any other theory arising by dimensional reduction from $4d$) the scalar $\sigma$ couples to charged fields as a mass term, so in the low-energy theory below the scale of this mass it is still possible to dualize the gauge field and to describe the Coulomb branch in this way. The main difference is that once quantum corrections are taken into account, the effective kinetic term depends on $\sigma$, so $e_3^2$ that appears in the definition \dualscalar\ of $a$, and also the precise form of the dual chiral multiplet $X$, depend on $\sigma$ in a more complicated way (and in addition there is a non-trivial K\"ahler potential that arises for $X$). But at generic points on the Coulomb branch, where all charged fields are massive, it is still possible to dualize the vector multiplet, and to describe the low-energy physics by the chiral multiplet $X$.

In non-Abelian gauge theories the story is similar. The scalar $\sigma$ is in the adjoint representation of the gauge group $G$, and classically has no potential (just couplings to charged matter fields). A generic vacuum expectation value for $\sigma$ breaks the gauge group $G\to U(1)^{r_G}$ where $r_G$ is the rank of $G$, and for generic values of $\sigma$ where all matter fields and all off-diagonal vector fields are massive, one can dualize the $r_G$ massless vector multiplets into chiral multiplets $Y_i$ ($i=1,\cdots,r_G$). The low-energy theory at generic points on the classical Coulomb branch thus includes $r_G$ massless chiral multiplets $Y_i$.

Let us discuss two examples in detail. For $G=U(N_c)$, one can diagonalize the adjoint scalar $\sigma$ to $\sigma = {\rm diag}(\sigma_1,\cdots,\sigma_{N_c})$, and, for generic values of the $\sigma$'s, one can dualize the $N_c$ massless gauge fields into photons $a_i$, and parameterize the Coulomb branch by chiral operators\foot{Here and in the $SU(N_c)$ case discussed below we normalize the vector multiplet kinetic term to be ${1\over {2g_3^2}} {\rm tr}(F_{\mu \nu}^2 + ({D_{\mu} \sigma})^2 + {\rm fermions})$, with the coupling to fundamentals given by matrices $T^a$ obeying ${\rm tr}(T^a T^b) = \delta^{ab} / 2$, and we define ${\hat g_3}^2 \equiv g_3^2 / 4\pi$.}
\eqn\uncoords{X_i \sim \exp\left({\sigma_i \over {\hat g}_3^2} + i a_i\right).}
The dependence of $X_i$ written above on the dual photons $a_i$ is exact (since it follows from their periodicity), but the full dependence on $\sigma$ is more complicated in the quantum theory because of the non-trivial effective coupling $g_3^2(\sigma)$. The classical Coulomb branch may be parameterized either by the expectation values of $\{\sigma_i, a_i\}$ ($i=1,\cdots,N_c$), or by those of $X_i$, subject to the identification coming from Weyl transformations that permute the indices $i$. We will generally fix this Weyl freedom by choosing $\sigma_1 \geq \sigma_2 \geq \cdots \geq \sigma_{N_c}$. Note that in $3d$
Abelian gauge theories there is a global symmetry current $J=*F$, and a similar current can be formed from the $U(1)$ gauge field in $U(N_c)$; the corresponding ``topological'' global symmetry is usually denoted by $U(1)_J$, and it acts by shifting the scalar dual to the photon, such that the $X_i$ operators are charged under it.

For $G=SU(N_c)$, one can use the same coordinates $\sigma_i$ and $a_i$, subject to the constraint $\sum_i \sigma_i = \sum_i a_i=0$. So there are $(N_c-1)$ complex coordinates overall. One can choose the coordinates to be related to the simple roots of $SU(N_c)$, by defining Coulomb branch coordinates
\eqn\suncoords{Y_j \sim \exp\left({{\sigma_j-\sigma_{j+1}}\over {\hat g}_3^2} + i(a_j-a_{j+1})\right)}
for $j=1,\cdots,N_c-1$. These mix in a complicated way under Weyl transformations, and again it is convenient to fix this freedom by choosing $\sigma_1 \geq \sigma_2 \geq \cdots \geq \sigma_{N_c}$.

There are two different ways to give a mass to a chiral multiplet in $3d$ ${\cal N}=2$ theories. One way is to add a quadratic term in the superpotential, as in four dimensions; this is only possible for matter fields in non-chiral representations. Alternatively, even for chiral matter fields, one can couple a global symmetry (which is not an R-symmetry) to a background vector multiplet, and give a vacuum expectation value ${\hat m}$ to the scalar in that multiplet. This preserves supersymmetry, and gives to any chiral multiplet with charge ${\hat q}$ under the global symmetry a mass ${\hat q}\cdot {\hat m}$. The ``real mass'' parameters ${\hat m}$ are parity-odd.
Note that these parameters are real, and belong to background vector multiplets rather than background chiral multiplets, so they cannot appear in holomorphic objects like superpotentials or chiral ring relations. A field with charge $q$ under a $U(1)$ gauge symmetry also gets a mass $q\sigma$ from its coupling to the vector multiplet. In the absence of superpotential mass terms, the full (classical) mass $m_Q$ of some chiral multiplet $Q$ arises from the sum of all the contributions $q_i \sigma_i$ from the vector multiplets it couples to, and of all the contributions ${\hat q}_j {\hat m}_j$ from the background vector multiplets coupled to global symmetry currents.

Integrating out a chiral multiplet $Q$ that carries charges $q_1,q_2$ under two $U(1)$ gauge symmetries with gauge fields $A_{\mu}^{(1)}$, $A_{\mu}^{(2)}$ (which may or may not be the same) induces (at one-loop) a mixed Chern-Simons term proportional to $A^{(1)} \wedge dA^{(2)}$, with a coefficient $k = \frac{1}{2} q_1 q_2\, {\rm sign}(m_Q)$ \refs{\NiemiRQ,\RedlichDV} (note that our classical action can also contain Chern-Simons terms, which are consistent with the ${\cal N}=2$ supersymmetry \refs{\ZupnikRY,\IvanovFN}). When we integrate out a fermion charged under both a global symmetry and a gauge symmetry, the same mechanism induces a Fayet-Iliopoulos (FI) term for the gauge field, with a coefficient ${1\over 2} {\hat q} \,{\hat m}\, q\, {\rm sign}(m_Q)$, and it gives the dual photon a transformation under the global symmetry, such that $e^{ia}$ carries a global symmetry charge ${1\over 2} {\hat q} \, q\, {\rm sign}(m_Q)$. On the Coulomb branch, this mechanism gives the Coulomb branch coordinates $Y_i$ global symmetry charges, arising both from integrating out massive chiral multiplets and massive gauginos.

\subsec{The quantum Coulomb branch of $3d$ ${\cal N}=2$ gauge theories, and of $4d$ theories on $\S^1$}
\subseclab\quantum

In non-Abelian gauge theories, most of the classical Coulomb branch described above is lifted by quantum corrections. While the classical physics is invariant under shifts of the dual photons $a_i$, this is not true in the quantum theory due to non-perturbative effects~\PolyakovFU. In ${\cal N}=2$ supersymmetric gauge theories this was first analyzed in \AffleckAS\ for a pure $SU(2)$ SYM theory. On the Coulomb branch of this theory, the VEV of $Y_1$ classically breaks $SU(2)$ to $U(1)$. In $4d$ there is a classical static 't Hooft-Polyakov monopole solution associated to this breaking, and the same solution (without the time direction) may be viewed as an instanton solution of the $3d$ gauge theory. It was shown in \AffleckAS\ that this instanton generates an effective superpotential on the Coulomb branch of the form
\eqn\weffsutwo{W_{eff} = {1 \over {Y_1}}}
(up to an unimportant normalization). In this theory the Coulomb branch coordinate $Y_1$ has R-charge $(-2)$, which ensures that this superpotential is exact. The superpotential causes a repulsion of the two eigenvalues of $\sigma$, which drives $Y_1 \to \infty$, so that this theory has no supersymmetric vacua.

In pure $SU(N_c)$ or $U(N_c)$ gauge theories, the same effect arises whenever two eigenvalues of $\sigma$ approach each other leading to an unbroken $SU(2)$, and causes these eigenvalues to repel each other. So, in a pure $SU(N_c)$ gauge theory, there is a term $W_{eff}=1/Y_i$ in the effective superpotential as any $Y_i \to 0$, and these terms completely lift the Coulomb branch so that the theory has no supersymmetric vacuum. In the pure $SU(N_c)$ SYM theory the symmetries alone are not enough to fix the form of the effective superpotential. However, the R-symmetry, together with the requirement that the non-perturbative effects should not grow whenever any $Y_i\to \infty$, since this leads to a breaking of the gauge symmetry at high energies where the theory is weakly coupled, implies that the exact effective superpotential is
\eqn\weffsun{W_{eff} = \sum_{i=1}^{N_c-1} {1\over Y_i}\,.}

In the theory with flavors, the instantons described above sometimes have extra fermion zero modes which prevent them from generating a superpotential (a superpotential arises only when there are exactly two fermion zero modes, coming from the gauginos of the enhanced $SU(2)$).
For the instanton associated with $Y_j \simeq \exp((\sigma_j-\sigma_{j+1})/{\hat g}_3^2+i(a_j-a_{j+1}))$, this happens precisely if there is a matter field that becomes massless on the Coulomb branch at some $\sigma_j > \sigma > \sigma_{j+1}$. In an $SU(N_c)$ (or $U(N_c)$) theory with massless matter fields in the fundamental representation, whose $k$'th component is massless when $\sigma_k=0$, this means that any configuration on the Coulomb branch in which two eigenvalues are positive, or two eigenvalues are negative, is still lifted by the superpotential. All that remains of the Coulomb branch is the subspace on which
\eqn\unlifted{\sigma_1 > 0 = \sigma_2 = \cdots = \sigma_{N_c-1} > \sigma_{N_c}.}
In the $SU(N_c)$ theory we also have on this unlifted Coulomb branch $\sigma_1 = -\sigma_{N_c}$, and it can naturally be parameterized by the operator
\eqn\defy{Y \equiv \prod_{j=1}^{N_c-1} Y_j \simeq \exp\left({{\sigma_1-\sigma_{N_c}}\over {\hat g}_3^2}+i(a_1-a_{N_c})\right).}
In the $U(N_c)$ theory $\sigma_1$ and $\sigma_{N_c}$ are independent, so the quantum Coulomb branch is naturally parameterized by
\eqn\defx{X_+ \simeq \exp\left({\sigma_1\over {\hat g}_3^2}+ia_1\right),\qquad\qquad X_- \simeq \exp\left(-{\sigma_{N_c}\over {\hat g}_3^2}-ia_{N_c}\right).}
In addition to these instanton effects there are additional non-perturbative effects involving the flavor fields that lift the Coulomb branch in $SU(N_c)$ theories with $N_f < N_c-1$ flavors, and in $U(N_c)$ theories with $N_f < N_c$ flavors \AharonyBX.

As described above, the quantum numbers of the Coulomb branch coordinates are determined by those of the
matter fields, so they change when some matter fields go to infinite mass and decouple. In such cases we have a relation between the high-energy coordinates $Y_{high}$ and the low-energy coordinates $Y_{low}$, which can usually be uniquely determined by matching their quantum numbers. When we give a superpotential mass $m$ to some matter field and look at the low-energy theory that does not include this field, we obtain a relation of the schematic form $Y_{high} = m \cdot Y_{low}$. When we break an $SU(N_c)$ gauge group to $SU(N_c-1)$ by a VEV for fundamental and anti-fundamental fields $Q$ and ${\tilde Q}$, we obtain a relation of the form $Y_{high} \cdot \langle Q {\tilde Q} \rangle = Y_{low}$. Note that since the $Y$'s are chiral superfields, relations of this type can only depend on the VEVs of chiral superfields (and on superpotential couplings that can be thought of as background chiral multiplets), and not on real masses (which are background vector multiplets).

If we consider the $4d$ ${\cal N}=1$ gauge theory on a circle of radius $r$, at energies low compared to $1/r$, we have a similar low-energy effective action, with the scalars $\sigma$ coming from the holonomy on the circle $P \exp(i \oint A_3)$ (namely, $\sigma \simeq A_3$). The main difference is that since only the holonomy is gauge-invariant, the eigenvalues of $\sigma$ are periodic with a period $1/r$ (they can be shifted by $1/r$ by a large gauge transformation). This means that, if we order the eigenvalues cyclically on the circle, we can get an enhanced $SU(2)$ not only when $\sigma_i \to \sigma_{i+1}$ as above, but also when $\sigma_{N_c} + 1/r \to \sigma_1$. Locally on the Coulomb branch this enhanced $SU(2)$ is identical to the others, so the analysis of \AffleckAS\ implies that the non-perturbative effects associated with the corresponding instantons generate a superpotential \SeibergNZ
\eqn\circlesuppot{W \simeq {1\over {\exp\left({{\sigma_{N_c}+1/r-\sigma_1}\over {\hat g}_3^2}+i(a_{N_c}-a_1)\right)}} \simeq \eta Y,}
where $\eta = \exp(-4\pi / r g_3^2) = \Lambda^b$ is related to the strong coupling scale of the four dimensional gauge theory (here we set the $4d$ theta angle to zero, if it is present this relation is multiplied by $e^{i\theta}$). In the pure $4d$ $SU(N_c)$ SYM theory on a
circle, the exact effective superpotential is given by
\eqn\weffcircle{W = \sum_{i=1}^{N_c-1} {1\over Y_i} + \eta Y\,,}
leading to $N_c$ supersymmetric vacua.
The form of \weffcircle\ is fixed by the need to reproduce \weffsun\ in the $3d$ limit $\eta \to 0$, and the fact that the
theory on a circle has a global symmetry\foot{If the gauge group is $SU(N_c)/\Z_{N_c}$ then this is actually a large gauge transformation, and the number of vacua in the theory on a circle is different, as discussed in \AST.} which takes the eigenvalues $\{\sigma_1,\cdots,\sigma_{N_c}\} \to \{\sigma_{N_c}+1/r,\sigma_1,\cdots,\sigma_{N_c-1}\}$, implying that the superpotential must be invariant under cyclic permutations of the set $\{1/Y_i, \eta Y\}$.
The full set of identifications on the Coulomb branch is now rather complicated, since they involve the Weyl transformations permuting the eigenvalues as well as the shifts of $\sigma_i$; but in $SU(N_c)$ theories the superpotential \circlesuppot\ generally lifts the Coulomb branch, so we do not need to study them in detail.

Note that the term \circlesuppot\ in the effective superpotential arises even when there are massless flavors. In many theories with flavors, including the ones we will discuss in this paper, one can use symmetry arguments to argue that \circlesuppot\ is the exact superpotential on
the part of the Coulomb branch that was unlifted in $3d$ (namely, that it is the only non-singular term consistent with the symmetries of the $4d$ theory that can be written in terms of $Y$ and in terms of the matter fields).

In the $4d$ theory, there is another chiral operator $S = -{\rm tr}(W_{\alpha}^2)$, which couples
in the high-energy action to $8\pi^2/g_4^2$, so that it may be viewed as the derivative of the effective
superpotential with respect to $\log(\eta)$ (see, for instance, \IntriligatorAU). In the $4d$
 theory this field is massive so it does not have to be included in the discussion of
the low-energy dynamics, and its expectation value in different vacua can be computed using the
effective superpotential. In the effective theory on $\R^3\times \S^1$, the dependence on $\eta$
comes in through \circlesuppot, and often (in particular in the $SU(N_c)$ SYM and SQCD theories)
this is the only term depending on $\eta$ in the low-energy effective action. Thus, in these theories we have a relation
\eqn\SYrel{S = \eta Y~,}
meaning that the manifestation of the high-energy operator $S$ in the low-energy
effective action is through $Y$ (which is classically massless). For example, in the $4d$
pure SYM theory there is a chiral ring relation $S^{N_c} = \eta$, while in the effective
theory on $\R^3\times \S^1$, the superpotential \weffcircle\ leads to a relation
$Y^{N_c} = \eta^{1-N_c}$. If we take the standard $3d$ limit keeping the $3d$ gauge
coupling fixed, then $\eta$ goes to zero and \SYrel\ simply states $S=0$, demonstrating that the relevant object in $3d$ is $Y$. In the next subsection we will see that in $3d$ we have a rather different UV
definition for $Y$.

\subsec{Monopole operators}
\subseclab\monopolesec

The full low-energy effective action of $3d$ ${\cal N}=2$ theories is described using the gauge-invariant operators made from the matter chiral multiplets, and the $Y_i$ that come from dualizing the vector multiplets on the Coulomb branch. Our description of the $Y_i$ so far was in terms of the low-energy effective action, and it is natural to wonder whether we can also construct local operators in the high-energy gauge theory that would flow to the $Y_i$ that we described at low energies. We do not know how to do this for all the $Y_i$ that parameterize the classical Coulomb branch. However, there exist ``monopole operators'' in the high-energy theory, which flow to all the $Y_i$'s that are needed for describing the part of the Coulomb branch that is not lifted by the Affleck-Harvey-Witten superpotential.

The relation of the $Y_i$ to ``monopole operators'' follows from the fact that the insertion of $\exp(ia(x))$ into the path integral (where $a$ is a dual photon) creates one unit of magnetic flux on the $\S^2$ that surrounds the point $x$. Thus, instead of describing this operator in the effective theory by dualizing the photon, we can alternatively describe it as a disorder operator at high energies, generalizing the construction of a 't Hooft loop in four dimensions, by removing the point $x$ and requiring that we sum over gauge field configurations that have one unit of magnetic flux on the $\S^2$ around the point $x$ \refs{\BorokhovIB,\BorokhovCG,\KapustinPY}. This high-energy operator flows to $\exp(ia(x))$ in the low-energy effective action.

In the supersymmetric case we argued that it is natural to consider the low-energy chiral operators $X(x) \sim \exp(2\pi \sigma(x)/e_3^2+ia(x))$ (starting with the Abelian case for simplicity). As discussed in \refs{\BorokhovIB,\BorokhovCG,\KapustinPY,\newIS}, this operator may also be given a high-energy definition, by summing over field configurations in which the field $\sigma(y)$ has the singular behavior
\eqn\sigmabeh{\sigma(y) \simeq {1\over {2|x-y|}}}
as $y\to x$. One can show that this definition preserves half of the supersymmetry, so that it defines a chiral operator, and it flows to the Coulomb branch coordinate $X$ at low energies.

In a non-Abelian theory, as for 't Hooft lines in four dimensions, the definition of a monopole operator requires specifying the magnetic flux around the point $x$; this flux is in $U(1)^{r_G} \subset G$, and is labeled by a weight of the dual magnetic group $G^L$ modulo Weyl transformations. For the group $U(N_c)$, the simplest monopole operator $X_+$ breaks $U(N_c)\to U(1)\times U(N_c-1)$ and carries one unit of flux in the $U(1)$; its supersymmetric version also pushes one of the eigenvalues of $\sigma$ to $+\infty$. In the effective theory on the moduli space, since we fixed $\sigma_1$ to be larger than all the other eigenvalues, this operator flows to $X_+ \simeq
\exp(\sigma_1/{\hat g}_3^2+ia_1)$ of \defx. Similarly there is an operator $X_-$ carrying $(-1)$ units of flux in the $U(1)$ that pushes one of the eigenvalues of $\sigma$ to $-\infty$, and which on the moduli space flows to $X_- \simeq \exp(-\sigma_{N_c}/{\hat g}_3^2-ia_{N_c})$ of \defx. In the special case of $U(1)$ theories, these operators flow to the Coulomb branch coordinates $V_+ \simeq \exp(2\pi \sigma/e_3^2+ia)$ and $V_- \simeq \exp(-2\pi \sigma/e_3^2-ia)$ that were discussed in \AharonyBX; classically these obey $V_+ V_- = 1$, but quantum mechanically this is modified. In particular, whenever there are massless flavors, it is modified to $V_+ V_- = 0$.

For gauge group $SU(N_c)$, the dual group is $SU(N_c)/\ZZ_{N_c}$ so only magnetic charges corresponding to roots of the dual group are allowed. The monopole operator corresponding to a simple root breaks $SU(N_c)\to SU(N_c-2)\times U(1)\times U(1)$, and takes one eigenvalue of $\sigma$ to $+\infty$ and another one to $-\infty$; it flows on the moduli space precisely to the operator $Y \simeq \exp((\sigma_1-\sigma_{N_c})/{\hat g}_3^2+i(a_1-a_{N_c}))$ that we defined in \defy. One can also define more general monopole operators, that have more complicated fluxes and that take several eigenvalues of $\sigma$ to $+\infty$ (possibly at different rates) and several eigenvalues to $-\infty$. The global quantum numbers of the high-energy monopole operators may be computed in a similar way to those of the low-energy Coulomb branch coordinates, by summing the contributions from the fermions that become massive when the corresponding eigenvalues of $\sigma$ go to infinity.

In our discussion in the paper we will not always be careful to distinguish the high-energy monopole operators from the low-energy coordinates on the moduli space; whenever we write an operator in an action (rather than in an effective action) we will mean the high-energy monopole operator defined in this subsection.

Note that all the monopole operators discussed in
this section exist in the $3d$ theory, but they are not well-defined at high energies in the $4d$ theory on $\R^3\times \S^1$. In that theory the $\sigma$'s are compact so they cannot go to infinity. The $4d$ theory on $\S^1$ does not have BPS 't~Hooft line operators, and the chiral operators $Y_i$ only arise in the effective description at low energies. More generally, whenever there is a compact region in the moduli space, there is no high-energy operator that parameterizes it \newIS.

\newsec{Dualities for $SU(N_c)$ SQCD Theories}
\seclab\susection

\subsec{Reduction with $\eta$ for $SU(N_c)$ SQCD}
\subseclab\etaduality

As our first example of reducing $4d$ dualities to $3d$ dualities, we take
the original example of a $4d$ duality \SeibergPQ: the duality between the $4d$ $\cN=1$ SQCD theory with gauge group $SU(N_c)$ and $N_f > N_c+1$ flavors $Q$ and $\tilde Q$ in the fundamental and anti-fundamental representations (theory A), and the SQCD theory with gauge group $SU(N_f-N_c)$, $N_f$ flavors $q$ (fundamental) and $\tilde q$ (anti-fundamental), and $N_f^2$ singlets $M$, coupled to the quarks by a superpotential $W = M q {\tilde q}$ with obvious index contractions (theory B). The singlets of theory B are identified with the mesons $M \equiv Q {\tilde Q}$ of theory A. The quantum numbers of the different fields of the two theories in four dimensions under the non-anomalous global symmetry $SU(N_f)_L\times SU(N_f)_R\times U(1)_B\times U(1)_R$ are listed below, and they remain the same also for the $4d$ theory on a circle.

\eqn\Ahae{
\vbox{\offinterlineskip\tabskip=0pt
\halign{\strut\vrule#
%%%%%%%%%%%%%%%%%%
&~$#$~\hfil\vrule
&~$#$~\hfil\vrule
&~$#$~\hfil
&~$#$~\hfil%\vrule
%&~$#$\hfil
&~$#$\hfil
&~$#$\hfil
&\vrule#
\cr
%%%%%%%%%%%%%%%%%
\noalign{\hrule}
&  &  SU(N_c) & SU(N_f)_L &   SU(N_f)_R & U(1)_B & U(1)_R &\cr
\noalign{\hrule}
%%%%%%%%%%%%%%%%%%
&  Q         & \; {\bf  N_c}     & \; {\bf N_f}    &\; {\bf 1} & \quad 1  &    \quad  1 - {N_c\over N_f}   &\cr
& \tilde Q                & \; {\bf \bar N_c}     & \; {\bf 1}   & \; {\bf\bar N_f}   & \quad -1  &   \quad 1 - {N_c\over N_f}   &\cr
\noalign{\hrule}
%%%%%%%%%%%%%%%%%%
& M              & \; {\bf  1}     & \; {\bf N_f}   & \; {\bf \bar N_f}   & \quad 0  &    \quad  2 (1 - {N_c\over N_f})   &\cr
}\hrule}}

\eqn\Aham{
\vbox{\offinterlineskip\tabskip=0pt
\halign{\strut\vrule#
%%%%%%%%%%%%%%%%%%
&~$#$~\hfil\vrule
&~$#$~\hfil\vrule
&~$#$~\hfil
&~$#$~\hfil%\vrule
&~$#$\hfil
&~$#$\hfil
%&~$#$\hfil
&\vrule#
\cr
%%%%%%%%%%%%%%%%%%
\noalign{\hrule}
&  &  SU(N_f-N_c) & SU(N_f)_L&   SU(N_f)_R& \ \ U(1)_B & U(1)_R &\cr
\noalign{\hrule}
%%%%%%%%%%%%%%%%%%
&  q        & \; {\bf {N_f-N_c}}     & \; {\bf \bar N_f}    &\; {\bf  1} & \quad {N_c\over {N_f-N_c}}  &  \quad  {N_c\over N_f}   &\cr
& \tilde q                & \; {\bf \overline {N_f-N_c} }    & \; {\bf 1}   & \; {\bf N_f}   & \quad -{N_c\over {N_f-N_c}}  &   \quad {N_c\over N_f}   &\cr
%%%%%%%%%%%%%%%%%%
}\hrule}}

Let us first review the dynamics of the $3d$ $SU(N_c)$ SQCD theory (the undeformed dimensional reduction of theory A). Classically this theory has an $(N_c-1)$-dimensional Coulomb branch,
but almost all of this Coulomb branch is lifted by Affleck-Harvey-Witten-type superpotentials \AffleckAS. The remaining part of the Coulomb branch may be parameterized by the coordinate $Y$ \defy.
The useful gauge-invariant operators to describe the theory are the mesons $M \equiv Q {\tilde Q}$, $Y$, and the baryons (for $N_f \geq N_c$) $B \equiv Q^{N_c}$, ${\tilde B} \equiv {\tilde Q}^{N_c}$. Semi-classically along the Coulomb branch the gauge symmetry breaks to $SU(N_c-2)\times U(1)^2$, and the classical equations of motion imply that one can go on the Coulomb branch as long as ${\rm rank}(M) \leq N_c-2$ and $B={\tilde B}=0$. The full quantum picture is sometimes different \AharonyBX. For $N_f < N_c-1$ there is a runaway superpotential along the Coulomb branch, and the theory has no supersymmetric vacuum. For $N_f = N_c-1$ the classical moduli space is deformed, and the quantum moduli space is described by a constraint $Y \det(M) = 1$ (up to normalizations), which smoothly connects what remains of the Coulomb branch to the Higgs branch. For $N_f \geq N_c$ the quantum moduli space is the same as the semi-classical moduli space, and there is a superconformal fixed point at the origin $Y = M = B = {\tilde B} = 0$. For $N_f = N_c$ there is an effective description of the theory using the gauge-invariant operators and a superpotential
\eqn\suppotnfnc{W = Y (B {\tilde B} - \det(M))}
which implements the classical constraints on the moduli space. For $N_f > N_c$ there is no known effective description of this theory (we will find a dual description for it in the next subsection).

When we discuss instead the $4d$ theory on a circle, the Coulomb branch is compact, but in any case the extra superpotential that arises from instanton-monopoles \SeibergNZ, $W = \eta Y$, lifts the Coulomb branch for any $N_f$.
For $N_f = N_c-1$ there is no supersymmetric vacuum remaining; we can introduce a Lagrange multiplier $\lambda$ to implement the constraint, and the superpotential
\eqn\effnfncone{W = \lambda (Y \det(M) - 1) + \eta Y}
does not have solutions for its F-term equations. For $N_f = N_c$ we have an effective superpotential description of the form
\eqn\neffnfnc{W = Y (B {\tilde B} - \det(M) + \eta),}
which implies that the Coulomb branch is lifted while the classical Higgs branch is deformed. For $N_f > N_c$ the Coulomb branch is lifted but the Higgs branch remains.

The general arguments that we described in the introduction imply that the $3d$ theory A deformed (at high energies) by a superpotential $W_A = \eta Y$ should be the same at low energies as the $4d$ theory A on a circle, and that for $N_f > N_c+1$ this should be dual to the $3d$ theory B deformed by $W_B = {\tilde \eta} {\tilde Y}$.

We can first verify the consistency of our general arguments above for the cases with a small number of flavors. For $N_f = N_c-1$ there is no supersymmetric vacuum in $4d$, so our general arguments imply that the deformed $3d$ theory A should also have no such vacuum, and this is indeed what we find from \effnfncone. For $N_f=N_c$, the $4d$ effective description involves a quantum modification of the Higgs branch \SeibergBZ, which can be implemented by a Lagrange multiplier in the form
\eqn\fourdweff{W = \lambda (B {\tilde B} - \det(M) + \Lambda^{2 N_c}) = \lambda (B {\tilde B} - \det(M) + \eta).}
Our general arguments suggest that the reduction of this theory on a circle should be dual to the $3d$ SQCD theory deformed by $\eta Y$, and using \neffnfnc\ we see that this is indeed the case, if we identify the $4d$ Lagrange multiplier $\lambda$ with the $3d$ operator $Y$ (it is easy to check that the two fields have the same quantum numbers in this case).

For $N_f > N_c+1$ we can use the dual description, which is the dimensional reduction of theory B deformed by the extra term in the superpotential, such that
\eqn\wb{W_B = M q {\tilde q} + {\tilde \eta} {\tilde Y}.}
The four dimensional relation of couplings \IntriligatorAU\ implies that ${\tilde \eta} = (-1)^{N_f-N_c} / \eta$. In this theory again almost all of the Coulomb branch is lifted by instantons, and the second term in \wb\ lifts the rest (here ${\tilde Y}$, defined at high energies as in section \monopolesec, flows at low energies to the standard Coulomb branch coordinate \defy\ for the $SU(N_f-N_c)$ SQCD theory). Our claim is that this $3d$ theory
is dual to the deformed theory A.

It is easy to check that the two theories have the same chiral gauge-invariant operators (the identification of the mesons and baryons was used to set the quantum numbers of the dual quarks above). The extra superpotential that lifts the Coulomb branch implies that $Y$ and ${\tilde Y}$ are not good chiral operators in these theories. As mentioned in the introduction, it is clear from our construction that the two theories have the same partition function on $\S^3$, and this can also be directly verified.

Next we can compare the moduli spaces of vacua. In theory A we have the Higgs branch, which includes VEVs for $M$ of rank up to $N_c$, such that when the rank is equal to $N_c$, one must turn on also VEVs for $B$ and ${\tilde B}$, as implied by the classical relations of these chiral operators. There is also a baryonic branch, where we give VEVs only to $B$ or only to ${\tilde B}$. In theory B the naive Higgs branch is lifted by the $M q {\tilde q}$ superpotential. Suppose that we turn on a VEV for $M$ of rank $N_c$; we can write this VEV as an $(N_c\times N_c)$ matrix $M_{massive}$. At low energies we remain in theory B with an $SU(N_f-N_c)$ gauge theory with $(N_f-N_c)$ massless flavors, and with a low-energy superpotential for the remaining massless fields of the form
\eqn\wlow{W_B = M q {\tilde q} + {\tilde \eta} {\tilde Y}_{low}\, \det(M_{massive}).}
The last term incorporates the relation between the high-energy and low-energy Coulomb branch coordinates that we get when integrating out quarks with a mass matrix $M_{massive}$. The effective dynamics of the gauge theory that has the same number of massless flavors as colors, summarized in \suppotnfnc, means that we can replace $W_B$ by an effective description of the form (with $N \equiv q {\tilde q}$)
\eqn\wlown{W_B = {\tilde Y}_{low} (B {\tilde B} - \det(N) + {\tilde \eta}\, \det(M_{massive})) + M N,}
where $B$ and ${\tilde B}$ are the baryon operators of theory B (which are identified, up to constants, with those of theory A). The last term implies $N = 0$, and ${\tilde Y}_{low}$ then acts as a Lagrange multiplier that forces $B {\tilde B}$ to be proportional to $\det(M_{massive})$, as expected from the classical relations of theory A (to obtain the precise matching we need to carefully normalize the relation between the baryons on the two sides).
Similarly we can easily match the baryonic branches, and check that when $M$ has a rank larger than $N_c$ there is no supersymmetric vacuum in theory B, as expected.

As in $4d$, we can also compare the deformations of these theories. Adding a complex mass $m$ in theory A leads to a flow to an $SU(N_c)$ theory with $N_f-1$ flavors. The relation between the high-energy and low-energy monopole operators implies that we end up with an effective superpotential
\eqn\waeff{W_A = \eta m Y_{low}}
(where $Y_{low}$ is now the low-energy monopole operator). In the dual theory B, the mass term $W = m M_{N_f N_f}$ leads to a VEV for the quarks, $\langle q_{N_f} \tilde{q}_{N_f} \rangle = -m$, such that the low-energy theory is an $SU(N_f-N_c-1)$ theory with $N_f-1$ flavors, and with a superpotential
\eqn\wbeff{W_B = M q {\tilde q} - {\tilde  \eta} ({\tilde Y}_{low} / m).}
Thus we obtain the same deformed duality with $N_f-1$ flavors, with deformation parameters $\eta_{low} = \eta m$, ${\tilde \eta}_{low} = - {\tilde \eta} / m$, obeying $\eta_{low} {\tilde \eta}_{low} = (-1)^{N_f-1-N_c}$ as they should. Note that this change in $\eta$ is interpreted from the $4d$ point of view as the standard rescaling of $\Lambda$, which is usually written as $\Lambda_{high}^{3 N_c - N_f} = \Lambda_{low}^{3 N_c - (N_f - 1)} / m$. Deformations by real masses will be analyzed in the next subsection.

Note that in the deformed $3d$ theories the R-charges are completely fixed. Of course, in the IR fixed point the R-symmetry can differ from this by an accidental $U(1)$ symmetry. In particular, this must happen when $N_c < N_f < 4 N_c / 3$, since otherwise the dimension of the meson operators would violate the $3d$ unitarity bound. Presumably, as in the un-deformed $3d$ SQCD theory with $N_f = N_c \geq 3$, the mesons in this case are free fields in the low-energy theory, but the full theory does not seem to be free. Note that this is different from what happens in the un-deformed $3d$ SQCD theory, where the mesons are not free at low energies for any $N_f > N_c$ (but the monopole operator $Y$, which in this case is a chiral primary, does sometimes become free) \SafdiRE.

\subsec{A duality for $3d$ SQCD}
\subseclab\pureduality

In the previous subsection we found a dual description of the $3d$ SQCD theory deformed by $\eta Y$. It would be nice to obtain a dual for the $3d$ SQCD theory itself, but we cannot find it simply by taking $\eta \to 0$, since in the dual theory we need to take the deformation parameter $\tilde\eta$ to infinity, which is ill-defined.

It turns out that we can obtain such a duality by starting from the duality of the previous subsection, for $N_f+1$ flavors, and giving a vector-like real mass $\hat m$ to a single flavor (say, the last one) in theory A. This is just a background field for the diagonal $SU(N_f+1)\times U(1)_B$ flavor symmetry, and as such we can translate it easily to theory B (which is now an $SU(N_f+1-N_c)$ gauge theory). The mapping in \Ahae, \Aham\ of the $SU(N_f+1)\times U(1)_B$ quantum numbers implies that in theory B the first $N_f$ flavors of quarks get a real mass ${\hat m}_1$, and the $(N_f+1)$'th flavor gets a real mass ${\hat m}_2$, where
\eqn\realmasses{{\hat m}_1 = {{\hat m} \over {N_f - N_c + 1}},\qquad\qquad {\hat m}_2 = {{\hat m} (N_c -N_f) \over {N_f - N_c + 1}}.}

In theory A we want to focus on the configurations that remain (as ${\hat m}\to \infty$) at a finite distance on the Coulomb branch. In these vacua all the components of the last quark flavor are massive, and we can ignore them at low energies.  We obtain an effective $SU(N_c)$ theory with $N_f$ massless flavors. When we integrate out this final quark, the low-energy monopole operator $Y_{low}$ is related to the high-energy $Y$ by
$Y_{low} = Y_{high} / m$ (where $m$ is the complex mass of this quark). Note that this relation cannot depend on the real mass $\hat m$, since it is in a background vector (or linear) multiplet rather than in a chiral multiplet. So, the superpotential $W_A = \eta Y_{high}$ vanishes in the low-energy variables (since in our case $m=0$). This is related to the fact that the $SU(N_c)$ Coulomb branch ``pinches'' whenever some eigenvalue of $\sigma$ is equal to ${\hat m}$, as in a $U(1)$ theory with a single flavor \AharonyBX\ (see the top of Figure 2 below), and the superpotential $W = \eta Y_{high}$ lifts the part of the Coulomb branch that is above the ``pinch'' (this is the part that is parameterized by $Y_{high}$) but not below it.
As an alternative argument, the real mass deformation preserves a $U(N_f)^2$ global symmetry acting on the light fields, with the quantum numbers of the various fields and of $Y_{low}$ given in the table below, and there is no non-singular superpotential that is consistent with this symmetry.

Thus, in theory A we get in the limit ${\hat m} \to \infty$ precisely the $3d$ $SU(N_c)$ SQCD theory with no superpotential. The quantum numbers of the chiral operators in this theory are listed below (choosing a specific R-symmetry); note that now we do not have an anomaly, nor an $\eta$-term, so we have an extra global $U(1)$ symmetry compared to the theories discussed in the previous subsection.

\eqn\Ahaethree{
\vbox{\offinterlineskip\tabskip=0pt
\halign{\strut\vrule#
%%%%%%%%%%%%%%%%%%
&~$#$~\hfil\vrule
&~$#$~\hfil\vrule
&~$#$~\hfil
&~$#$~\hfil%\vrule
&~$#$\hfil
&~$#$\hfil
&~$#$\hfil
&\vrule#
\cr
%%%%%%%%%%%%%%%%%
\noalign{\hrule}
&  &  SU(N_c) & SU(N_f)_L &   SU(N_f)_R & U(1)_B & U(1)_A & U(1)_R &\cr
\noalign{\hrule}
%%%%%%%%%%%%%%%%%%
&  Q         & \; {\bf  N_c}     & \; {\bf N_f}    &\; {\bf 1} & \quad 1  &    \quad  1 &  \quad  0   &\cr
& \tilde Q                & \; {\bf \bar N_c}     & \; {\bf 1}   & \; {\bf\bar N_f}   & \quad -1  &    \quad  1&   \quad 0   &\cr
\noalign{\hrule}
%%%%%%%%%%%%%%%%%%
& M              & \; {\bf  1}     & \; {\bf N_f}   & \; {\bf \bar N_f}   & \quad 0  &    \quad  2 &  \quad  0   &\cr
& Y              & \; {\bf  1}     & \; {\bf 1}   & \; {\bf  1}   & \quad 0  &    \quad -2N_f &  \quad  2(N_f-N_c+1)   &\cr}
\hrule}}

What do we get in the $SU(N_f-N_c+1)$ gauge theory B? This theory no longer has any vacuum at the origin of its Coulomb branch, $\tilde\sigma=0$, since this is lifted by instantons (none of the quarks are massless there). The vacuum with the most massless fields that is not lifted by instantons involves taking the $\tilde\sigma$ matrix to have $(N_f-N_c)$ eigenvalues equal to $(-{\hat m}_1)$, and one eigenvalue equal to $(-{\hat m}_2)$. Note that this choice is traceless, as it should be. With this choice the gauge symmetry is broken to $SU(N_f-N_c)\times U(1)$; we can view this as $U(N_f-N_c)$, and then the components of the first $N_f$ flavors of quarks $q$ that are massless in this vacuum transform in the fundamental $({\bf {N_f-N_c}})_1$ representation of this $U(N_f-N_c)$ (with $\tilde{q}$ in the conjugate representation), while the massless component of the final flavor (which we will denote by $b$) transforms as ${\bf 1}_{-(N_f-N_c)}$ (with ${\tilde b}$ in the conjugate representation). The
off-diagonal singlet operators $M$ obtain real masses and become massive; we are left with an $N_f\times N_f$ matrix of singlets that may be identified with the remaining mesons $M$ of theory A, and one extra massless singlet coming from the meson $M_{N_f+1}^{N_f+1}$ of the final flavor, which with some foresight we denote by $Y$.

The superpotential of theory B includes the terms $W = M q {\tilde q} + Y b {\tilde b}$ coming from the first term in \wb, and also two extra terms. The extra terms can naturally be written in terms of the
Coulomb branch coordinates ${\tilde X}_+$ and ${\tilde X}_-$ of $U(N_f-N_c)$ theories, defined in \defx.
Note that here we need to perform a change of variables from the $SU(N_f-N_c+1)$ variables to the $U(N_f-N_c)$ variables; the relative $\hat\sigma$'s of the $U(N_f-N_c)$ theory take values in the unbroken $SU(N_f-N_c)$ group, but the trace of $\hat\sigma$ of $U(N_f-N_c)$ involves shifting together the first $N_f-N_c$ eigenvalues of the original $\tilde\sigma$ of $SU(N_f-N_c+1)$, and shifting the final eigenvalue in the opposite direction. The $W = {\tilde \eta} {\tilde Y}$ superpotential that we had in the high-energy $SU(N_f-N_c+1)$ theory then translates (assuming $\hat m > 0$) in the effective $U(N_f-N_c)$ theory precisely into $W = {\tilde \eta} {\tilde X}_-$. In addition, there is an Affleck-Harvey-Witten type monopole-instanton related to the breaking of $SU(N_f-N_c+1)$ to $SU(N_f-N_c)\times U(1)$, which leads as in \AffleckAS\ to a superpotential term proportional to $W = {\tilde X}_+$. The full effective superpotential of theory B is thus
\eqn\bsuppot{W_B = M q {\tilde q} + Y b {\tilde b} + {\tilde \eta} {\tilde X}_- + {\tilde X}_+.}
The parameter $\tilde\eta$ here does not play any role and can be absorbed in a rescaling of ${\tilde X}_-$ -- so we will set it to one from here on.

The global symmetries of the effective low-energy theory B, expanded around this vacuum, obviously include $SU(N_f)_L\times SU(N_f)_R\times U(1)_R$ factors. In addition, there is the usual topological $U(1)_J$ symmetry associated with the $U(1)$ factor in our gauge group, and six $U(1)$ flavor symmetries acting on our six matter fields ($M$, $q$, ${\tilde q}$, $Y$, $b$ and $\tilde b$), but one combination of these flavor symmetries is gauged, and the four terms in the superpotential break four combinations of the other symmetries. Thus, we have just two $U(1)$ symmetries, as in theory A. In order to determine the precise combination of $U(1)$ symmetries that matches with theory A, we first match the mesons $M$, determining most of the quantum numbers of $M$, $q$ and $\tilde q$ (except for their $U(1)_B$ charge). The baryons $B = Q^{N_c}$ that involve $N_c$ of the first $N_f$ flavors of quarks mapped in the $SU(N_f-N_c+1)$ theory to quarks made from $(N_f-N_c)$ of the first $N_f$ flavors and also the final flavor, which means that now we need to identify them with $q^{N_f-N_c} b$. This is indeed a singlet of the $U(N_f-N_c)$ gauge group, and similarly ${\tilde B} = {\tilde Q}^{N_c} = {\tilde q}^{N_f-N_c} {\tilde b}$. This identification determines the global charges of all the matter fields in theory B, up to a possible mixing of the $U(1)_B$ with the $U(1)$ gauge symmetry that we fix in an arbitrary way. We can then compute the global symmetry quantum numbers of ${\tilde X}_+$ and ${\tilde X}_-$ in the usual way (as monopole operators in theory B). We find the following quantum numbers in theory B:

\eqn\Ahamthree{
\vbox{\offinterlineskip\tabskip=0pt
\halign{\strut\vrule#
%%%%%%%%%%%%%%%%%%
&~$#$~\hfil\vrule
&~$#$~\hfil\vrule
&~$#$~\hfil
&~$#$~\hfil%\vrule
&~$#$\hfil
&~$#$\hfil
&~$#$\hfil
&\vrule#
\cr
%%%%%%%%%%%%%%%%%%
\noalign{\hrule}
&  &  U(N_f-N_c) & SU(N_f)_L&   SU(N_f)_R & U(1)_B & U(1)_A & U(1)_R &\cr
\noalign{\hrule}
%%%%%%%%%%%%%%%%%%
&  q        & \; ({\bf {N_f-N_c}})_1     & \; {\bf \bar N_f}    &\; {\bf  1} & \quad 0  &    \quad  -1 &  \quad  1   &\cr
& \tilde q                & \; ({\bf \overline {N_f-N_c} })_{-1}    & \; {\bf 1}   & \; {\bf N_f}   & \quad 0  &    \quad  -1 &   \quad 1   &\cr
&  b        & \; {\bf {1}}_{N_c-N_f}     & \; {\bf 1}    &\; {\bf  1} & \quad N_c  &    \quad  N_f &  \quad  N_c-N_f &\cr
& \tilde b                & \; {\bf {1} }_{N_f-N_c}    & \; {\bf 1}   & \; {\bf 1}   & \quad -N_c  &    \quad  N_f &   \quad N_c-N_f   &\cr
%\noalign{\hrule}
%%%%%%%%%%%%%%%%%%
& M             & \; {\bf  1}     & \; {\bf N_f}   & \; {\bf \bar N_f}   & \quad 0  &    \quad  2 &  \quad  0   &\cr
& Y             & \; {\bf  1}     & \; {\bf 1}   & \; {\bf 1}   & \quad 0  &    \quad  -2 N_f &  \quad  2(N_f-N_c+1)   &\cr
\noalign{\hrule}
& {\tilde X}_\pm              & \; {\bf  1}     & \; {\bf 1}   & \; {\bf 1}   & \quad 0  &    \quad 0 &  \quad 2   &\cr}
\hrule}}

Note that even though we determined the quantum numbers by the matching of mesons and baryons, using the first two terms in \bsuppot, the quantum numbers of ${\tilde X}_\pm$ (determined from those of $q$, $\tilde q$, $b$ and $\tilde b$) came out to be consistent with the last two terms in \bsuppot, which is a nice consistency check confirming that the unbroken symmetries of theory B exactly match with those of theory A. In addition, the quantum numbers of the singlet $Y$ come out to be exactly the same as those of the monopole operator $Y$ in theory A, so it is natural to identify them. After this identification the chiral operators of the two theories match exactly, noting that the last two terms in \bsuppot\ lift the Coulomb branch of theory B and prevent ${\tilde X}_\pm$ from being good chiral operators. Note that without the superpotential the Coulomb branch of this theory would be quite complicated, since the quarks are massless when some ${\hat \sigma}_i$ vanishes, while $b$ and $\tilde b$ are massless when $\sum_i {\hat \sigma}_i = 0$.

\midinsert\bigskip{\vbox{{\epsfxsize=3.2in
        \nobreak
    \centerline{\epsfbox{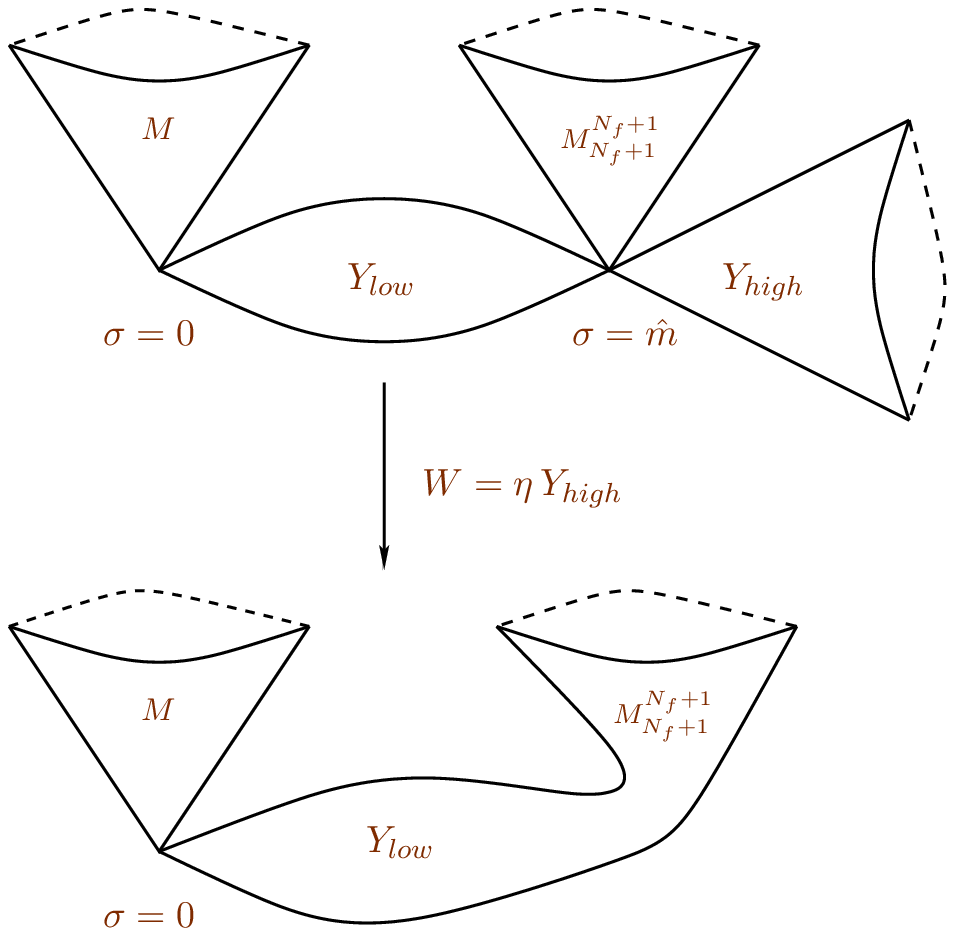}}
        \nobreak\bigskip
    {\raggedright\it \vbox{
{\bf Figure 2.}
{\it The moduli space of the $SU(N_c)$ SQCD theory with a real mass, first without
the $W_A = \eta Y$ superpotential and then with it.
}}}}}}
\bigskip\endinsert

In the discussion above, the monopole operator $Y$ originated in theory B from the meson $M_{N_f+1}^{N_f+1}$.  This fact can be understood in theory A as follows.
Consider first theory A with the real mass, but without the $W_A = \eta Y$ superpotential. The Higgs branch splits into a branch where the mesons of the first $N_f$ flavors acquire VEVs, and another branch where $M_{N_f+1}^{N_f+1}$ acquires a VEV. The former intersects the Coulomb branch at $\sigma=0$, and the latter at $\sigma={\hat m}$. The Coulomb branch splits into the component with $\sigma > {\hat m}$, which can be parameterized by $Y_{high}$, and the component with $0 < \sigma < {\hat m}$, which can be parameterized by $Y_{low}$; the two fields $Y_{low}$ and $Y_{high}$ are independent in the low-energy effective action, and label different branches (see the top part of Figure 2). The field $Y_{high}$ comes from the high-energy monopole operator $Y$, but there does not seem to be a high-energy operator that flows to $Y_{low}$. (The part of the classical moduli space parameterized by $Y_{low}$ is compact, and hence there cannot be a corresponding monopole operator \newIS.)  The low-energy theory near the intersection point at $\sigma={\hat m}$ is a $U(1)$ theory with $N_f=1$, which may be alternatively described \AharonyBX\ by $W_{eff} = - V_+ V_- M_{N_f+1}^{N_f+1}$. Here $V_+ \simeq Y_{high}$, $V_- \simeq 1 / Y_{low}$. Now we can reinstate $W_A = \eta Y$, which in this low-energy effective theory looks like $\eta V_+$. The equation of motion now gives $V_- M_{N_f+1}^{N_f+1} = \eta$, leading to the identification of $Y_{low}$ with $M_{N_f+1}^{N_f+1}$ as we found above. The component of the Coulomb branch parameterized by $Y_{high}$ is lifted, and the other component merges with one of the Higgs branches, such that the only non-trivial low-energy theory lives at $\sigma=0$, see the bottom part of Figure 2.

Until now we discussed a duality between low-energy field theories, but
we can now lift the duality we found between effective field theories to a duality between $3d$ gauge theories. We claim that the $3d$ $SU(N_c)$ SQCD theory A is dual to the $3d$ $U(N_f-N_c)$ gauge theory we found, with the superpotential \bsuppot.\foot{As we mentioned in the introduction, strictly speaking the presence of monopole operators in the $3d$ Lagrangian makes the theory ill-defined in the UV.  However, such a theory can be embedded in any $3d$ UV completion without affecting our conclusions.} One can verify that the $\S^3$ partition functions of these two theories are equal; this is not obvious since in order to get the new theories from those of the previous subsection we had to take some real masses to infinity, and to scale the Coulomb branch coordinates $\sigma$ and $\tilde\sigma$ in a very particular
 way with the real mass. However, when one takes the limit in this way one obtains a precise match of the partition functions (see section 5).

As a test of the duality, we can deform both theories by $W = \eta Y$. Theory A clearly goes over to the theory discussed in the previous subsection. In theory B this superpotential induces a VEV $\langle b \tilde{b} \rangle = -\eta$, which breaks the $U(N_f-N_c)$ symmetry to $SU(N_f-N_c)$. The monopole operators of $U(N_f-N_c)$ become the monopole operator $\tilde Y$ of $SU(N_f-N_c)$; matching the quantum numbers implies that (up to constants) ${\tilde X}_+ = {\tilde X}_- = {\tilde Y} / b {\tilde b}$, so that the last two terms in \bsuppot\ become precisely the ${\tilde \eta} {\tilde Y}$ superpotential that we had in theory B in the previous subsection (together with the $M q {\tilde q}$ term). The behavior of the theory under mass deformations is similar to that discussed in the previous subsection.

 Our derivation above is valid for $N_f > N_c+1$, but let us see what we get for lower values of $N_f-N_c$. For $N_f=N_c$ the duality still formally works, but in theory B we have no gauge group and just a $W = Y b {\tilde b}$ superpotential. In this case the $b$'s are simply the baryons of theory A, so this is almost what we expect for this case, but we
actually expect to get another term as in \neffnfnc, $W = - Y \det(M)$. Can we understand how this term is generated when we flow down from higher values of $N_f$, where the duality should give the complete picture? For $N_f=N_c+1$, theory B is a $U(1)$ gauge theory with $N_f+1$ flavors and the superpotential \bsuppot\ (where now the ${\tilde X}_{\pm}$ operators are really ${\tilde V}_{\pm}$ monopoles of the $U(1)$ theory). We want to show that upon deforming by $m_{N_f} M_{N_f}^{N_f}$ we generate the terms we need in the low-energy theory. Suppose we turn on an expectation value of rank $N_c=N_f-1$ for $M$ (involving its components $M_{others}$ with indices $1,\cdots,N_f-1$), and also to $Y$. Then we remain in theory B with a single massless quark $q_{N_f}$, ${\tilde q}^{N_f}$, and by matching the monopole operators we get an effective superpotential of the form
\eqn\effpotmass{W = M_{N_f}^{ N_f} q_{N_f} {\tilde q}^{N_f} + \sqrt{Y \det(M_{others})} ({\tilde V}_- + {\tilde V}_+) + m_{N_f} M_{N_f}^{ N_f}.}
The $U(1)$ theory with one flavor has an effective description \AharonyBX\ using $W = -{\tilde V}_+ {\tilde V}_- N_{N_f}^{ N_f}$, where $N_{N_f}^{ N_f} \equiv q_{N_f} {\tilde q}^{N_f}$. So, the full effective description is given by the sum of these two superpotentials. All the fields are now massive and can be integrated out, and we find at low energies $W = - Y \det(M_{others}) / m_{N_f}$. Since $Y / m_{N_f}$ is exactly the monopole operator of the low-energy $SU(N_c)$ theory A with $(N_f-1)$ flavors, this is precisely the missing term we need.

Next, we can test the duality by matching the moduli space of both theories.
As mentioned above, the moduli space of theory A is just the classical Higgs branch parameterized by $M$, $B$ and $\tilde B$ with their classical relations, joined to a Coulomb branch parameterized by $Y$ whenever $B={\tilde B}=0$ and ${\rm rank}(M) \leq N_c-2$. In theory B the naive Higgs branch labeled by $q {\tilde q}$ (or by $b {\tilde b}$) is lifted, but we can naively turn on any VEV we want for $M$ and $Y$.
In general it is rather complicated to analyze the moduli space in this theory, as it involves the strong coupling dynamics of $U(N_f-N_c)$, but for the case of $N_f=N_c+1$ it is relatively easy to see that we obtain the same constraints as in theory A.

First, let us turn on a VEV for $M$ of rank $N_c$. At low energies we remain in theory B with a $U(1)$ theory with two massless flavors and with
\eqn\utwosup{W = M q {\tilde q} + Y b {\tilde b} + \sqrt{\det(M_{others})} ({\tilde V}_+ + {\tilde V}_-).} The theory including the first two terms in $W$ is related by mirror symmetry \refs{\IntriligatorEX,\deBoerMP,\deBoerKA,\AharonyBX} to another $U(1)$ theory with two massless flavors and no superpotential, such that $M$ and $Y$ map to the diagonal meson operators of this mirror theory, while ${\tilde V}_+$ and ${\tilde V}_-$ map to its off-diagonal meson operators. The last two terms in \utwosup\ thus become mass terms for the quarks of the mirror theory, making it clear that this theory does have a supersymmetric vacuum, but that one cannot turn on additional expectation values for $M$ and/or $Y$ (since the resulting theory of massive quarks has no Higgs branch). This agrees with theory A. The resulting low-energy mirror theory does have a Coulomb branch, along which the product of its monopole operators is equal to the determinant of the quark mass matrix (which in our case is $\det(M_{others})$). The mirror map \refs{\IntriligatorEX,\deBoerMP,\deBoerKA,\AharonyBX} maps these monopoles to the baryons $q b$ and ${\tilde q} {\tilde b}$ in our original theory, so we find precisely the correct constraint $B {\tilde B} = \det(M_{others})$ in this case.

Now, suppose that we try to turn on $Y$ and also give a VEV to $M$ of rank $(N_c-1)$. In this case we are left again in theory B with a $U(1)$ theory with 2 massless flavors $q^i$ and $\tilde{q}_{\tilde i}$, but now the effective superpotential is different, taking the form
\eqn\suppotvevym{W = M_{i}^{{\tilde i}} q^i {\tilde q}_{\tilde i} + \sqrt{Y \det(M_{others})} ({\tilde V}_+ + {\tilde V}_-).}
We can now use the duality of \AharonyGP\ for the $U(1)$ theory (we will rederive this duality from four dimensions in the next section) to map it to another $U(1)$ theory with two massless flavors, with monopole operators $V_+$ and $V_-$, and with
\eqn\suppotvevymn{W = {\tilde V}_+ V_- + {\tilde V}_- V_+ + \sqrt{Y \det(M_{others})} ({\tilde V}_+ + {\tilde V}_-).}
The F-term equations now imply that both $V_+$ and $V_-$ should be non-zero, which is impossible (since whenever there are massless quarks in a $U(1)$ gauge theory, $V_+ V_- = 0$). Thus, there is no supersymmetric vacuum with non-zero $Y$ and $M$ of rank $(N_c-1)$, as in theory A.

We can also compare the $3d$ indices of theories A and B, and we will describe this
in appendix A.

\subsec{Flows to chiral theories}
\subseclab\chiralsec

We can flow from the $SU(N_c)$ SQCD theory to many other theories by turning on real masses. In theory A we are allowed to turn on arbitrary real masses for the various $Q$'s and ${\tilde Q}$'s, which are background fields for the $SU(N_f)_L\times SU(N_f)_R\times U(1)_B\times U(1)_A$ global symmetry. By turning on various real masses, we can flow to theories that have Chern-Simons terms at low energies, and also that have different numbers of fundamental and anti-fundamental chiral multiplets (since there is no anomaly that forbids this in three dimensions). We will analyze here just a few examples of these flows, though general flows can be analyzed by similar methods.

Suppose that we want to flow in theory A to a low-energy theory with only chiral multiplets in the fundamental representation. We can do this by giving real masses to all the ${\tilde Q}$'s, and taking them to infinity while staying at the origin of the Coulomb branch. If we give a positive real mass to $n$ ${\tilde Q}$'s and a negative real mass to $(N_f-n)$ ${\tilde Q}$'s, we induce at low energies a Chern-Simons level $k_A = (2n - N_f) / 2$ for the $SU(N_c)$ gauge group. As long as $n \neq 0$ and $n\neq N_f$ (namely, for $|k_A| < N_f/2$), we can always take the sum of the real masses to zero, so that they involve a background field that is purely in $SU(N_f)_R$. In theory B we then find that all the ${\tilde q}$'s obtain a real mass, as well as all the mesons $M$. We want to remain in theory B near the origin of the moduli space, where the $q$'s are light (since we need to keep the baryonic operators $B = Q^{N_c} = q^{N_f-N_c} b$). In theory B we induce at low energies a Chern-Simons term at level $k_B = (N_f - 2n) / 2$ for the $U(N_f-N_c)$ gauge group, as we did in theory A. However, in this theory we also induce a FI-term for the $U(1)$ gauge group, proportional to the difference between the sum of the positive real masses and the sum of the negative ones.

The sign of the FI-term is such that the D-term equations can be solved by turning on an expectation value for $b$ (but not for $q$ or ${\tilde b}$). For $k_B=0$ this is the only way to solve the D-term equations; when $k_B \neq 0$ one can also solve them by moving on the Coulomb branch of the $U(1)$ theory by a distance proportional to the real masses, but this makes the quarks massive and leads to other vacua which do not match to the specific vacuum of theory A that we are interested in (at the origin of the Coulomb branch).

The VEV for $b$ breaks the $U(N_f-N_c)$ gauge group to $SU(N_f-N_c)$, and $Y$ and ${\tilde b}$ become massive through the superpotential \bsuppot. We thus remain at low energies with an $SU(N_f-N_c)$ gauge theory with $N_f$ chiral flavors $q$, with no superpotential, and with $k_B = -k_A$. So, we claim that the dual of an $SU(N_c)$ theory with $N_f$ chiral multiplets $Q$ in the fundamental representation and Chern-Simons level $k_A$ ($|k_A| < N_f/2$) is an $SU(N_f-N_c)$ theory with $N_f$ chiral multiplets $q$ and level $k_B = -k_A$. This is very similar to the $U(N)$ dualities of \BeniniMF. The global symmetry on both sides is $U(N_f)$, and the baryons in the two theories clearly match. For $k_A \neq 0$ these are the only chiral operators.

For $k_A=0$ (which we can only get for even values of $N_f$, as required by the parity anomaly) there are sometimes additional chiral operators, labeling Coulomb branches in these theories. The original Coulomb branch of theory A, labeled by $Y$ of \defy, is always lifted, since on this branch we break the $SU(N_c)$ symmetry to $SU(N_c-2)\times U(1)\times U(1)$, and a Chern-Simons term is induced for the $U(1)$ vector multiplet whose scalar parameterizes this Coulomb branch. On the other hand, for even values of $N_c$, there is another component of the Coulomb branch that is not lifted, where we turn on $N_c/2$ eigenvalues of $\sigma$ to be equal to $\sigma_0 > 0$, and $N_c/2$ eigenvalues equal to $(-\sigma_0)$. On this branch we break the gauge group to $SU(N_c/2)\times SU(N_c/2)\times U(1)$, and we induce Chern-Simons terms for the $SU(N_c/2)$ groups (with level $k=\pm N_f/2$) but not for the $U(1)$ group. This means that this branch remains in the moduli space. Moreover, in our original non-chiral theory this branch was lifted by Affleck-Harvey-Witten instantons, but now these are not present due to the Chern-Simons terms in the low-energy $SU(N_c/2)$ theories. Thus, this branch remains and is labeled by
\eqn\defyprime{Y' \simeq \exp\left({{\sigma_1+...+\sigma_{\frac{N_c}{2}}-\sigma_{\frac{N_c}2+1}-...-\sigma_{N_c}}\over {{\hat g}_3^2}}+i(a_1+...+a_{\frac{N_c}{2}}-a_{\frac{N_c}{2}+1}-...-a_{N_c})\right).}
When $N_c$ is even, $N_f-N_c$ is also even, and we have a similar Coulomb branch labeled by some ${\tilde Y}'$ in theory B, along which the $SU(N_f-N_c)$ gauge group is spontaneously broken to $SU((N_f-N_c)/2)\times SU((N_f-N_c)/2)\times U(1)$. We can compute the quantum numbers of $Y'$ and ${\tilde Y}'$ and see that they are equal, so that these two extra chiral operators can be identified between theories A and B. Note that here, unlike in our previous dualities, we directly match the Coulomb branches of theories A and B. One can also match the effective theories on the moduli space; the number of supersymmetric vacua in the $SU(N_c/2)$ Chern-Simons theory with level $N_f/2$ is the same as that of the $SU((N_f-N_c)/2)$ theory at the same level. For odd values of $N_c$ there is no
Coulomb branch (and no extra chiral operators) in these theories even for $k_A=0$.

The flow above only gave us theories with $|k| < N_f/2$. We can get more general theories by continuing to flow, by giving real masses to some of the $Q$'s. Naively, such a flow reduces $N_f$ on both sides, changes $k$ by some amount, but does not change the dual $SU(N_f-N_c)$ gauge group. If this was true we would get a contradiction, since the baryon operators would no longer match. What actually happens, as in \Shamirthesis, is that after such a flow there are several different supersymmetric vacua on both sides, and the vacuum in which $SU(N_c)$ is unbroken no longer maps to the vacuum where $SU(N_f-N_c)$ is unbroken. The analysis is rather
complicated, and we will not perform it in detail here. If we end up with $|k| < N_f/2$ then we reproduce the duality described above, while in other cases we find more complicated duals that involve also extra
$U(1)$ factors (on one side or the other).

\subsec{Flows to Chern-Simons-matter theories}
\subseclab\csmdualsec

As another example, consider the flow to a non-chiral $SU(N_c)$ gauge theory with a non-zero Chern-Simons level $k > 0$. In order to obtain this, we need to start from theory A which is an $SU(N_c)$ theory with $N_f+k$ flavors, and give positive real masses to $k$ $Q$'s and also to $k$ ${\tilde Q}$'s. For simplicity, let's assume that these real masses are all equal so that this involves background vector fields in $SU(N_f+k)_L$, $SU(N_f+k)_R$ and $U(1)_A$, but not in $U(1)_B$. In theory A, integrating out the massive quarks induces a Chern-Simons level $k_A=k$ for the low-energy theory with $N_f$ massless flavors. In theory B, which is now a $U(N_f+k-N_c)$ theory, we induce real masses (with opposite signs compared to theory A) to $k$ $q$'s and $k$ ${\tilde q}$'s, to the mesons that are not in the first $N_f\times N_f$ block, and to $Y$, $b$ and $\tilde b$. This theory has a vacuum at the origin of its Coulomb branch, where there is an induced Chern-Simons term of level $k_B=-k$ for the $SU(N_f+k-N_c)$ gauge fields, and of level $k_{U(1)}=-k+(N_f+k-N_c)=N_f-N_c$ for the $U(1)$ gauge field (this gets extra contributions from $b$ and ${\tilde b}$). There is no induced FI-term in this case, and the only term remaining in the superpotential is $W_B = M q {\tilde q}$.

Naively the Chern-Simons term means that the topological $U(1)_J$ symmetry of theory B, whose current is $J=*F_{U(1)}$, is not a good global symmetry, since the monopoles (that carry the global $U(1)_J$ charge) carry $k_{U(1)}$ units of electric $U(1)$ charge, so they are not gauge-invariant. However, if we consider the $U(1)_{\tilde B}$ flavor symmetry acting (with opposite charges) on $q$ and ${\tilde q}$, then the precise statement is that a combination of $U(1)_{\tilde B}$ and $U(1)_J$ is gauged, but the other combination remains a good global symmetry, and we can identify it with the $U(1)_B$ global symmetry of theory A. In particular, the product of a monopole operator ${\tilde X}_+$ of $U(N_f+k-N_c)$ (defined as in \defx) with $(N_f-N_c)$ quarks $q$ is neutral under the $U(1)$, and carries this global symmetry charge. Because of the different Chern-Simons levels for $SU(N_f+k-N_c)$ and $U(1)$, this operator is not neutral under $SU(N_f+k-N_c)$. However, we can make it neutral by adding to it $(k-1)$ gaugino superfields $W_{\alpha}$ (if we take the monopole ${\tilde X}_+$ of \defx, and multiply it by $(N_f-N_c)$ quarks, whose color indices are the bottom $(N_f-N_c)$ indices in $U(N_f+k-N_c)$, then we need to add $(k-1)$ gauginos with indices $(1,i)$ for $i=2,\cdots,k$).
The resulting operator ${\tilde X}_+ q^{N_f-N_c} W_{\alpha}^{k-1}$ then has exactly the same quantum numbers as $B = Q^{N_c}$ (in order to see that it has the correct spin one needs to carefully follow the spin of monopoles in the Chern-Simons theory), and similarly for ${\tilde B}$. The quantum numbers of all operators mentioned above are listed in the following tables (with a specific combination of $U(1)_{\tilde B}$ and $U(1)_J$ under which the monopoles carry no charge chosen to be $U(1)_B$ in theory B):

\eqn\Ahaethreek{
\vbox{\offinterlineskip\tabskip=0pt
\halign{\strut\vrule#
%%%%%%%%%%%%%%%%%%
&~$#$~\hfil\vrule
&~$#$~\hfil\vrule
&~$#$~\hfil
&~$#$~\hfil%\vrule
&~$#$\hfil
&~$#$\hfil
&~$#$\hfil
&\vrule#
\cr
%%%%%%%%%%%%%%%%%
\noalign{\hrule}
&  &  SU(N_c)_k & SU(N_f)_L &   SU(N_f)_R & U(1)_B & U(1)_A & U(1)_R &\cr
\noalign{\hrule}
%%%%%%%%%%%%%%%%%%
&  Q         & \; {\bf  N_c}     & \; {\bf N_f}    &\; {\bf 1} & \quad 1  &    \quad  1 &  \quad  0   &\cr
& \tilde Q                & \; {\bf \bar N_c}     & \; {\bf 1}   & \; {\bf\bar N_f}   & \quad -1  &    \quad  1&   \quad 0   &\cr
\noalign{\hrule}
%%%%%%%%%%%%%%%%%%
& M              & \; {\bf  1}     & \; {\bf N_f}   & \; {\bf \bar N_f}   & \quad 0  &    \quad  2 &  \quad  0   &\cr
& B              & \; {\bf  1}     & \; {\bf N_f}^{N_c}   & \; {\bf  1}   & \quad N_c  &    \quad N_c &  \quad  0   &\cr
& {\tilde B}              & \; {\bf  1}     & \; {\bf 1}   & \; {(\bf \bar N_f)^{N_c}}   & \quad -N_c  &    \quad N_c &  \quad  0   &\cr
}
\hrule}}

\eqn\Ahamthreek{
\vbox{\offinterlineskip\tabskip=0pt
\halign{\strut\vrule#
%%%%%%%%%%%%%%%%%%
&~$#$~\hfil\vrule
&~$#$~\hfil\vrule
&~$#$~\hfil
&~$#$~\hfil%\vrule
&~$#$\hfil
&~$#$\hfil
&~$#$\hfil
&\vrule#
\cr
%%%%%%%%%%%%%%%%%
\noalign{\hrule}
&  &  U(N_f+k-N_c)_{-k,N_f-N_c} & SU(N_f)_L &   SU(N_f)_R & U(1)_B & U(1)_A & U(1)_R &\cr
\noalign{\hrule}
%%%%%%%%%%%%%%%%%%
&  q         & \; ({\bf  N_f+k-N_c})_1     & \; {\bf\bar N_f}    &\; {\bf 1} & \quad {N_c\over {N_f-N_c}}  &    \quad  -1 &  \quad  1   &\cr
& \tilde q                & \; ({\bf \overline {N_f+k-N_c}})_{-1}     & \; {\bf 1}   & \; {\bf N_f}   & \quad -{N_c\over {N_f-N_c}}  &    \quad  -1&   \quad 1   &\cr
&  W_{\alpha}         & \; ({\bf  (N_f+k-N_c)^2})_0     & \; {\bf 1}    &\; {\bf 1} & \quad 0  &    \quad  0 &  \quad  1   &\cr
%\noalign{\hrule}
%%%%%%%%%%%%%%%%%%
%& {\tilde X}_\pm              & \;      & \; {\bf 1}   & \; {\bf 1}   & \quad \pm {k N_c \over {N_f+k-N_c}} \  &    \quad  N_f &  \quad  -(N_f+k-N_c-1)   &\cr
}
\hrule}}

All other operators, including monopole operators, are not chiral in these theories.
Thus, we find that all the chiral operators match, and we claim that we can lift this duality to high energies and that the $SU(N_c)$ theory with $N_f$ non-chiral flavors and level $k>0$ is dual to the $U(N_f+k-N_c)$ theory at levels $(-k)$ (for $SU(N_f+k-N_c)$) and $(N_f-N_c)$ (for the $U(1)$ factor, when it is normalized as in $U(N_f+k-N_c)$), with $N_f$ non-chiral flavors, $N_f^2$ singlets and a $W_B = M q {\tilde q}$ superpotential:
\eqn\sunkd{SU(N_c)_{k} \ {\rm /w}\ N_f \quad \longleftrightarrow \quad U(N_f+k-N_c)_{-k,N_f-N_c} \ {\rm /w}\ N_f, \quad W_B=M q {\tilde q}.}
This is quite similar to the $U(N_c)$ duality of \GiveonZN, and we will discuss the precise relation between them in the next section.

The flat directions of theory A are identical to the ones of the four dimensional SQCD theory. There is a mesonic flat direction, that is easy to match to theory B. There is also a baryonic flat direction, where we turn on (say) a single baryon operator $B=Q^{N_c}$. Can we describe this in theory B? This seems hard to do using the description of the baryon operators in theory B that we gave above, which includes gauginos. However, if we just look for flat directions in theory B, we find that despite the presence of the Chern-Simons term, which naively lifts the whole Coulomb branch, there is a flat direction with $k$ equal non-zero eigenvalues of $\tilde\sigma$, breaking the gauge group to $SU(k)\times SU(N_f-N_c)\times U(1)\times U(1)$. On this flat direction, labeled by a monopole
\eqn\kmonopole{{\tilde X}_{+k} \simeq \exp\left({{{\tilde\sigma}_1+\cdots+{\tilde\sigma}_k}\over \hat{\tilde g}_3^2}+i({\tilde a}_1+\cdots+{\tilde a}_k)\right)}
(where $\hat{\tilde g}_3\equiv \tilde g_3^2/4\pi$ is the rescaled gauge coupling in theory B), the contributions of the $SU(k+N_f-N_c)$ and $U(1)$ factors to the Chern-Simons term of the $U(1)$ that labels the flat direction exactly cancel, so we can turn on its $\hat\sigma$ and still solve its D-term equation. On the other hand, there is an induced mixed Chern-Simons term between the two $U(1)$'s, which forces us when we move by an amount $\hat\sigma$ along this flat direction, to also turn on expectation values for $(N_f-N_c)$ $q$'s (proportional to $\sqrt{\hat\sigma}$), completely breaking $SU(N_f-N_c)$ and the second $U(1)$. The gauge-invariant operator labeling this branch of the moduli space is ${\tilde X}_{+k} q^{k(N_f-N_c)}$, and it has the correct global symmetry charges to match with $B^k$ in theory A. Thus, we identify this branch of the moduli space (which is visible already classically) with the baryonic flat direction of theory A.

In order to get a precise match we need to show that on this branch there are $k$ distinct supersymmetric vacua, with different VEVs for $B$ (that lead to the same value of $B^k$). This
comes from the $SU(k)_{-k}$ factor, and we can analyze it far on the moduli space where we can ignore the quarks. The $SU(k)_{-k}$ theory on its own is trivial and has a unique supersymmetric vacuum, as can be seen from its Witten index \WittenDS. However, what we get on the moduli space is not really an $SU(k)$ theory, but rather a $U(k)$ theory, with level $(-k)$ for $SU(k)$ and level zero for the $U(1)$ factor. This means that this theory has
a chiral operator ${\hat b} = {\tilde X}_+ W_{\alpha}^{k-1}$ (with ${\tilde X}_+$ as in \defx) which is gauge-invariant. This operator cannot be written as a product of operators in the $U(1)$ part and the $SU(k)_{-k}$ part, since its $U(1)$ part by itself carries a fractional magnetic charge in the $U(1)$ theory (which is not allowed). However, if we look at ${\hat b}^k$, then this can be decomposed as a legal operator in the $U(1)$ theory, which is simply the operator ${\tilde X}_{+k}$ \kmonopole, times an operator in the $SU(k)_{-k}$ theory. Since the latter theory is trivial, we identify the $SU(k)_{-k}$ component with one, and we find that ${\hat b}^k = {\tilde X}_{+k}$. This relation is consistent with having $k$ different vacua in the $U(k)_{-k,0}$ theory, such that the phase of ${\hat b}$ takes $k$ different values.
In this way we can obtain a precise matching of the moduli spaces of the two theories. We can also show that if we turn on a vacuum expectation value of rank $N_c$ for the mesons, we get also in theory B the relation $B {\tilde B} \propto \det_{N_c\times N_c}(M)$ (this follows from the relation $V_+ V_- = \det(m)$ in a $U(1)$ theory with a quark mass matrix $m$).

In the special case of $N_f=0$, \sunkd\ becomes a duality for the pure supersymmetric Chern-Simons theories,
which arise at low energies from the pure Yang-Mills-Chern-Simons theory.
This is similar to the known level-rank dualities of $U(N_c)$ Chern-Simons theories \refs{\Naculich,\Camperi,\Mlawer}. In our case it maps
\eqn\levelrank{SU(N_c)_k \qquad \longleftrightarrow \qquad U(|k|-N_c)_{-k,-N_c\, {\rm sign}(k)}.}
At low energies we can integrate out the massive gauginos and obtain a duality for pure bosonic Chern-Simons theories.
Taking into account the shift of the $SU(N_c)$ level by $k\to k - N_c\, {\rm sign}(k)$, this implies for the pure bosonic Chern-Simons theories the duality $SU(N_c)_{|\tilde k|} \longleftrightarrow U(|\tilde k|)_{-N_c}$. This is the standard example of level-rank duality \refs{\NakanishiHJ,\NaculichNC},
\eqn\levelrankb{SU(n)_m \qquad\longleftrightarrow \qquad U(m)_{-n}}
with $m,n>0$, which can be proven by studying $mn$ free complex fermions in two dimensions.

\newsec{Dualities for $U(N_c)$ theories}
\seclab\undualities

In four dimensions, $U(1)$ gauge theories are always IR-free, so they do not lead to
any interesting low-energy dynamics. This is not true in three dimensions, where these theories do have non-trivial dynamics (such as confinement). Can we use our procedure, described above, to learn also about the IR dynamics of $3d$ $\cN=2$ theories with $U(1)$ or $U(N_c)$ gauge groups? There are two ways we can try to do this, which lead to equivalent answers.  We will discuss them in the next two subsections.

\subsec{Gauging $U(1)_B$}

One way to get a $3d$ $U(N_c)$ theory is to start from a $3d$ $SU(N_c)$ theory and to gauge the $U(1)_B$ global symmetry group. More precisely, we assign to the quarks charge $1/N_c$ under this $U(1)_B$, and view the gauge group as $U(N_c) \simeq (SU(N_c)\times U(1))/\Z_{N_c}$. The fact that the gauge group is $U(N_c)$ rather than $SU(N_c)\times U(1)$ is crucial in understanding the monopole operators.  Here, with the $U(N_c)$ gauge theory, the allowed monopole operators (or Coulomb branch coordinates) carrying the minimal charge are $X_+$ and $X_-$ defined in \defx.

From every duality discussed in the previous section, we can construct a duality for $U(N_c)$ by gauging the $U(1)_B$ global symmetry, assigning the minimal allowed charge to the baryon operator. As a first example, consider the duality with $\eta$ discussed in section \etaduality\ (in the next subsection we will also relate this to $4d$ $U(N_c)$ theories on a circle). The normalization of $U(1)_B$ is such that when we gauge it we obtain precisely a duality between a $U(N_c)$ theory and a $U(N_f-N_c)$ theory, with the field content of tables \Ahae\ and \Aham. The monopole operator $Y$ \defy\ of the $SU(N_c)$ theory is still a legal operator also in the $U(N_c)$ theory, but in this theory it decomposes into a product of two gauge-invariant monopole operators \defx, $Y = X_+ X_-$. This corresponds to the fact that (in the absence of any superpotential), as we reviewed in section \quantum, the $U(N_c)$ theory with flavors has a two dimensional Coulomb branch, corresponding to making one eigenvalue of $\sigma$ positive and one negative with no relation between them. The $U(N_c)$ theory has an extra global $U(1)_J$ symmetry, whose current is the Hodge dual of the $U(1)$ field strength; the monopole operators $X_\pm$ carry charge $\pm 1$ under this symmetry. We find that the duality maps the $U(N_c)$ theory with $N_f$ flavors and
\eqn\wuna{W_A = \eta X_+ X_-}
to a $U(N_f-N_c)$ theory with $N_f$ flavors, singlet mesons $M$, and
\eqn\wunb{W_B = {\tilde \eta} {\tilde X}_+ {\tilde X}_- + M q {\tilde q}.}
The dual theory also has a $U(1)_J$ symmetry, identified with the one of theory A, under which ${\tilde X}_\pm$ carry charge $\pm 1$. In both theories the superpotential lifts generic points on the Coulomb branch for all $N_f > N_c$, but a one dimensional branch (turning on either $X_+$ or $X_-$, and either ${\tilde X}_+$ or ${\tilde X}_-$) remains. This comes from the fact that the superpotential affects only the $SU(N_c)$ factor, not the $U(1)$ factor, so it does not lift the Coulomb branch of the $U(1)$ theory.
We can identify the Coulomb branches between the two theories, with $\eta X_+$ identified with ${\tilde X}_+$ and $\eta X_-$ with ${\tilde X}_-$ (one can check that these identifications are consistent with all the quantum numbers). We can also compare the other chiral operators and other components of the moduli space as in section \etaduality, just without the baryons and the baryonic branches, and find an exact match.

Next, we can do the same thing for the duality of the standard $SU(N_c)$ SQCD theory (with no $\eta$ term), discussed in section \pureduality. In theory A, gauging $U(1)_B$ gives the standard $U(N_c)$ SQCD theory, with no superpotential. In theory B it is convenient to use the choice of the $U(1)_B$ symmetry exhibited in \Ahamthree\ (rather than other choices that are linear combinations of this with the $U(1)$ gauge symmetry). Using this choice the only fields charged under the new $U(1)$ are $b$ and ${\tilde b}$, whose charge we normalize to one. Thus, the new sector of the theory is simply a $U(1)$ gauge theory with a single (non-chiral) flavor. This theory was analyzed in \AharonyBX, and it was found that if we denote its monopole operators by $V_+$ and $V_-$, and its gauge-invariant ``meson'' $b {\tilde b}$ by $N$, it is equivalent at low energies to a theory of three chiral superfields with $W = -V_+ V_- N$, such that under the new global $U(1)_J$ symmetry $V_\pm$ carry charges $\pm 1$. Note that the global charges of the $b$'s in table \Ahamthree\ imply that $V_\pm$ carry a $U(1)_A$ charge of $(-N_f)$, and an R-charge of $(N_f-N_c+1)$. The superpotential \bsuppot\ contains an extra term $W = Y N$, such that the equation of motion of $N$ sets the singlet field $Y$ in theory B to be equal to $V_+ V_-$. Next, we integrate out all the massive fields. As part of integrating out $b$ and $\tilde b$ (by replacing them by their bound state $N$) we need to correct the quantum numbers of the monopoles ${\tilde X}_\pm$.  Denoting the monopoles of the low-energy $U(N_f-N_c)$ by ${\hat X}_\pm$, they are related to the monopoles of the high-energy $U(N_f-N_c)\times U(1)_B$ theory through
${\tilde X}_\pm \simeq {\hat X}_\pm V_\mp$ (it is easy to see that this is consistent with all global symmetry charges, if we assign to ${\hat X}_\pm$ $U(1)_J$ charges of $\pm 1$). Gathering everything together, we find that the superpotential in theory B (after absorbing $\tilde\eta$ of \bsuppot\ by a rescaling), with gauge group $U(N_f-N_c)$, is given by
\eqn\weffb{W_B = M q {\tilde q} + {\hat X}_- V_+ + {\hat X}_+ V_-.}
This is exactly the duality originally suggested in \AharonyGP, if we identify the singlets $V_\pm$ in theory B with $X_\pm$ in theory A. We have thus derived this duality from the $4d$ $SU(N_c)$ duality.

The dualities of the previous two paragraphs are related in several ways. First, it is obvious that if we take the duality of the standard $U(N_c)$ theory, discussed in the previous paragraph, and deform it by $W = \eta X_+ X_-$, we obtain (after integrating out $V_\pm=X_\pm$ in theory B) precisely the duality discussed earlier, of the $U(N_c)$ theory with this superpotential. Conversely, as in the previous section, we can start from the duality with $\eta$ and add real masses to flow to the duality without $\eta$. This works
similarly to our discussion in section \pureduality, with the vacuum at the origin of theory A mapping to a vacuum away from the origin in theory B.
Note that in this discussion we assumed that $N_c > 1$ and $N_f-N_c > 1$, but we can flow to the theories
with $N_c=1$ and $N_f-N_c=1$ by adding masses and/or Higgsing, obtaining the same dualities for these theories as well.

Next, we can start from the chiral duality of section \chiralsec\ and gauge the $U(1)_B$ global symmetry. We immediately get a duality between a $U(N_c)$ theory with $N_f$ quarks in the fundamental representation and level $k_A$
(for the $SU(N_c)$ factor), and a $U(N_f-N_c)$ theory with $N_f$ quarks in the fundamental and level $k_B=-k_A$ (for the $SU(N_f-N_c)$ factor), with $|k_A| < N_f/2$. To figure out the Chern-Simons levels for the $U(1)$ parts of the gauge theories we get, we need to be careful. Since the theories we discuss here are not parity-invariant, there can be contact terms for the $U(1)_B$ symmetry \refs{\ClossetVG,\ClossetVP} that become Chern-Simons terms when we gauge this symmetry, and that need to be properly taken into account. The flow from the non-chiral duality that we discussed in section \chiralsec\ leads to contact terms which give a $U(1)$ level such that the theories mentioned above become precisely $U(N_c)_{k_A}$ and $U(N_f-N_c)_{k_B}$ (where this is defined such that the Chern-Simons term is a single trace in the fundamental representation of $U(N)$). Of course, when we gauge the $U(1)$ we can always add some extra
Chern-Simons term for it, as long as we do it consistently on both sides.
The result we find precisely reproduces the dualities of \BeniniMF\ (note that these dualities take a different form in the chiral theories with $|k| > N_f/2$ and with $|k| < N_f/2$).

We can also take the $SU(N_c)$ Chern-Simons-matter dualities of section \csmdualsec\ and gauge the $U(1)_B$ symmetry there (with some choice of its Chern-Simons level). We can choose the $U(1)_B$ in theory B to act in various ways, by mixing it with the $U(1)$ gauge symmetry. In particular, if we choose the specific combination of $U(1)_B$ and the gauge symmetry such that $U(1)_B$ acts in theory B only on the monopole operators, then gauging the $U(1)_B$ is the same as adding a mixed Chern-Simons term between the original $U(1)$ (in $U(N_f+k-N_c)$) and the new $U(1)$ that we are gauging. In order to match the Chern-Simons levels of $U(1)_B$ in theories A and B we need to take into account the ``background Chern-Simons terms'' that this symmetry had as a global symmetry \refs{\ClossetVG,\ClossetVP}. We end up with dualities between $U(N_c)$ and $U(N_f+k-N_c)\times U(1)$ theories, which resemble the ones of \GiveonZN, but are not exactly the same; presumably they are related to the dualities of \GiveonZN\ by a mirror symmetry transformation (their equivalence can be tested by all the methods we discussed).

\subsec{$4d$ $U(N_c)$ theories on a circle}

An alternative way to obtain $3d$ dualities for $U(N_c)$ theories is to start from $4d$ dualities for such theories and to carefully reduce them on a circle, as in section \etaduality. Analyzing $4d$ dualities for $U(N_c)$ theories is a bit subtle, because the $U(1)$ factor is not asymptotically free, and in the presence of massless charged matter fields its coupling flows to zero in the IR. Since we will eventually be interested in $3d$ theories with a fixed $3d$ $U(1)$ coupling (such that we work at energy scales well below this gauge coupling), we will assume that we have some fixed and small $U(1)$ coupling in the $4d$ physics (say, at the scale $1/r$), which then becomes strong at low energies once we compactify the theory on a circle.

Since the $U(1)$ theory is free at low energies, the $4d$ effective description includes the $U(1)$ vector multiplet. For instance, if we take the $U(N_c)$ theory in $4d$ with $N_f=N_c$, we can describe its IR dynamics by going to the low-energy dynamics of the $SU(N_c)$ factor (which gives a constrained moduli space satisfying $\det(M)-B {\tilde B}=\eta$ \SeibergBZ), and then weakly coupling the $U(1)$ gauge field to the baryons. The D-term of this $U(1)$ forces $|B|=|\tilde B|$, and the $U(1)$ gauge field is massless when $B={\tilde B}=0$. Similarly, we can treat the $SU(N_c)$ dualities of \SeibergPQ\ as $U(N_c)$ dualities, with a weakly coupled $U(1)$ gauge field coupling to the baryon number at low energies.

Since the low-energy dynamics contains massless photons we need to discuss how to reduce these on a circle. We can start by discussing $4d$ supersymmetric $U(1)$ gauge theories and their compactification on a circle.
Let's begin with a free $4d$ photon. Compactifying on a circle we get a $3d$ photon and the holonomy $\sigma = A_3$, which is periodic due to large gauge transformations around the circle, $\sigma \sim \sigma + 1/r$ (such that the holonomy $\exp(i \oint A_3) = \exp(i 2\pi r \sigma)$ is invariant). As in section \classcoul, we can also dualize the $3d$ photon to a periodic scalar field $a$. In the supersymmetric theory, the duality turns the vector multiplet into a chiral multiplet. The $3d$ theory we obtain is just a sigma model on a torus (which is the Coulomb branch). The size of the torus goes as $1/r$, while the complex structure of the torus is given by the $4d$ complex coupling $\tau \equiv \theta / 2\pi + 4\pi i / e_4^2$ (the theta dependence arises when carefully dualizing the field strength). This theory is free so there is no interesting dynamics in this case. The low-energy theory has a global $U(1)_J$ symmetry corresponding to shifting the dual photon.
 Before we take the circle size to zero there is also another global symmetry, whose charge is the momentum in the $x^3$ direction.

Next we couple this photon to charged matter fields (chiral multiplets). For theories that we get by reduction from $4d$ we have the same number of positively charged and negatively charged matter multiplets (say, $N_f$ of each). In four dimensions there is now a Higgs branch of the moduli space, parameterized by the meson matrix $M_i^{\tilde i} = Q_i {\tilde Q}^{\tilde i}$ of rank one, with the photon massless at the origin of the moduli space. This remains true also after compactification on a circle, but now we have an extra Coulomb branch. The reduction of the $4d$ gauge theory implies that the matter fields are massless at $\sigma=0$ (and its images), so this is where the Coulomb branch intersects the Higgs branch. The metric on the Coulomb branch torus is corrected such that the circle of the dual photon shrinks to zero size there, so the Coulomb branch is a pinched torus (topologically a sphere with two singular points that are identified), see Figure 3. Locally near $\sigma=0$ the low-energy theory is just the $3d$ $U(1)$ theory with $N_f$ flavors. This theory has \AharonyBX\ two components to its Coulomb branch, one parameterized by $V_+$ containing the region of positive $\sigma$, and one parameterized by $V_-$ containing the region of negative $\sigma$ (with $V_\pm$ defined as in section \background).

\midinsert\bigskip{\vbox{{\epsfxsize=3.2in
        \nobreak
    \centerline{\epsfbox{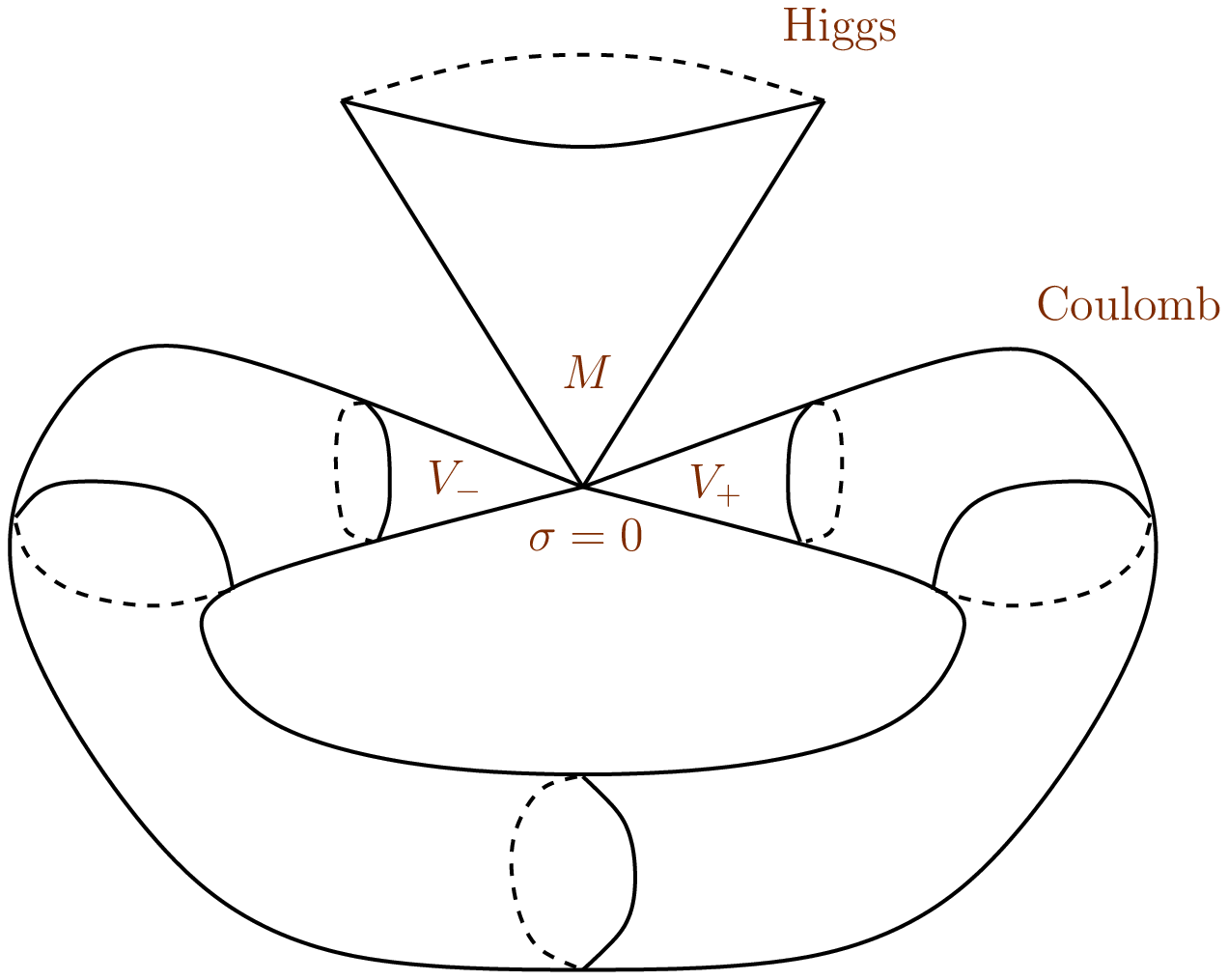}}
        \nobreak\bigskip
    {\raggedright\it \vbox{
{\bf Figure 3.}
{\it  The moduli space of the $U(1)$ theory with flavors on a circle.
 The Higgs branch pinches the Coulomb branch at $\sigma=0$.
}}}}}}
\bigskip\endinsert

In the $4d$ theory on a circle, the two components are joined together.
Obtaining a Lagrangian description of the resulting moduli space is somewhat complicated.  One way to do this is to start from the space of all values of $\sigma$, containing an infinite number of copies of the Higgs branch and of the Coulomb branch, and then to impose the identification coming from the large gauge transformations. However, we do not need this global description for our purposes here.

The reduction from four dimensions does not help us with analyzing the IR dynamics of the $U(1)$ theory
itself, near the origin of the moduli space. The dynamics for $N_f=1$ was analyzed in \AharonyBX, while for $N_f > 1$ there is a conjectured dual description of the theory near the origin \AharonyGP, that we derived from $4d$ in the previous subsection.
In any case, if we take the $4d$ $U(1)$ SQED theory on a circle and take the $3d$ limit, the compact Coulomb branch of Figure 3 becomes infinitely large and we obtain the $3d$ SQED theory.

We can now move on to discussing $4d$ $U(N_c)$ gauge theories on a circle. The reduction on the circle now leads to several different effects. First, we classically obtain a compact $N_c$-complex-dimensional Coulomb branch. The Coulomb branch is compact due to similar identifications to those described above.
As discussed in section \background, most of the Coulomb branch is lifted by monopole-instantons (including the extra instanton related to the $4d$ instanton on the circle).
 A one dimensional Coulomb branch survives (turning on either $\sigma_1$ or $\sigma_{N_c}$, but not both), and requires similar identifications as we had in the $U(1)$ theories on a circle (since in the theory on a circle, turning on a positive $\sigma_1$ is related by a large gauge transformation to turning on a negative $\sigma_{N_c}$). If we focus on the $3d$ low-energy conformal field theory at the origin of the Coulomb branch,
 then the compactness of the Coulomb branch is not important; it only affects irrelevant operators in the K\"ahler potential. On the other hand, we still need to keep the contribution from
the extra monopole-instanton, giving us an extra term $W = \eta X_+ X_-$ (which is just the $W = \eta Y$ superpotential that we had in the $SU(N_c)$ case, rewritten in the language of the $U(N_c)$ Coulomb branch coordinates; note that this term is not present for $N_c=1$).
This term, while it is sometimes irrelevant in the conformal field theory, is always relevant on the Coulomb branch near the origin, so it has to be kept.
If we now start from the $4d$ $SU(N_c)$ duality of \SeibergPQ, interpret it as a $U(N_c)$ duality, compactify it on a circle and take the low-energy limit focusing on the point at the origin of the Coulomb branch, we regain the $U(N_c)$ duality with the $\eta$ term, described at the beginning of the previous subsection. From there we can flow, as mentioned in the previous subsection, to the duality without $\eta$.

Starting with $SU(N_c)$ dualities we can obtain $U(N_c)$ dualities only with
$N_c > 1$, but by going on the Higgs branch we can flow from there also to the duality for $N_c=1$. In this sense the dynamics of $3d$ $U(1)$ theories also follows from $4d$, except for the case of $N_c=N_f=1$ which we had to put in by hand in our analysis of the previous section (we will discuss how to derive also this duality from $4d$ below).

We can also discuss the case of $N_f=N_c$, where in the $SU(N_c)$ case we saw that the instanton dynamics was already included in the $4d$ effective description. In $4d$ for $SU(N_c)$ we had a constraint $B {\tilde B} - \det(M) + \eta = 0$ \SeibergBZ. When we gauge the $U(1)_B$ symmetry and compactify on a circle, the dynamics of the $U(1)$ theory with one flavor (coming from the baryons) leads to a superpotential $W = -X_+ X_- N$, with $N \equiv B {\tilde B}$ (usually we would write this superpotential with the monopoles $V_\pm$ of $U(1)$, but in the $U(N_c)$ theory we have to dress these with $SU(N_c)$ monopoles to get the $X_\pm$ monopoles, since the $V$'s by themselves, which would be legal monopole operators in $SU(N_c)\times U(1)$, are not legal monopole operators in the $U(N_c)$ theory). We can thus write the effective action as $W = -X_+ X_- N + \lambda (N - \det(M) + \eta)$, and integrating out $N$ we can rewrite this as $W = X_+ X_- (\eta -\det(M))$. The first term is again the instanton term that we expect to get in the theory on a circle, and in the $3d$ limit of $\eta \to 0$ this reproduces the expected IR dynamics for $U(N_c)$ with $N_f=N_c$ flavors \AharonyGP. As before, we can see from this superpotential that in the theory on a circle part of the Coulomb branch is not lifted -- we can turn on either $X_+$ or $X_-$, but not both. To give a full description of the theory on a circle we would also need to take into account the periodic identifications on the Coulomb branch, that are similar to the $U(1)$ case.

\subsec{Ungauging $U(1)$ symmetries}
\subseclab\gaugetopo

As discussed above, starting from any duality for $SU(N_c)$ we can derive a duality for $U(N_c)$ by gauging the $U(1)_B$ global symmetry. We can also try to go in the other direction. $3d$ $U(N_c)$ gauge theories have a $U(1)_J$ global symmetry, whose current is the Hodge dual of the $U(1)$ field strength. We can gauge this $U(1)_J$ symmetry, which is equivalent to adding a mixed Chern-Simons term of level one between the new $U(1)_{new}$ and the original $U(1)$ in $U(N_c)$. We then have a new $U(1)_{J'}$ global symmetry whose current is the Hodge dual of the $U(1)_{new}$ gauge field. The mixed Chern-Simons term implies that monopoles of $U(1)_{new}$ are charged under the original $U(1)$ symmetry, so they can be used to make new gauge-invariant operators out of states (like baryons) that were not invariant under the original $U(1) \subset U(N_c)$. The two $U(1)$ gauge fields are massive due to the off-diagonal Chern-Simons coupling. In the absence of any extra Chern-Simons terms involving the $U(1)$'s, at low energies we obtain simply a $SU(N_c)$ gauge theory, in which the $U(1)_{J'}$ symmetry can be interpreted as the $U(1)_B$ baryon number global symmetry. Thus, this gives us a general procedure to take a $U(N_c)$ theory and turn it into an $SU(N_c)$ theory \KapustinSim.

We can now try to take the known dual of $U(N_c)$ \AharonyGP, gauge $U(1)_J$ and obtain a new dual for the $SU(N_c)$ theory. Recall that this dual theory B is a $U(N_f-N_c)$ SQCD theory, with singlets $M$, $X_+$ and $X_-$, and with a superpotential $W_B = M q {\tilde q} + X_+ {\tilde X}_- + X_- {\tilde X}_+$. Naively one may think that also in theory B, when we gauge its $U(1)_J$ symmetry we would find an $SU(N_f-N_c)$ theory, but this is not correct, since there are two extra singlets $X_\pm$ that carry the $U(1)_J$ charge. Thus, we cannot just forget about the $U(1)$ dynamics. However, we still find a dual description for $SU(N_c)$ in terms of the $U(N_f-N_c)\times U(1)_{new}$ theory, with the superpotential and Chern-Simons terms mentioned above. And we can still identify the $U(1)_{J'}$ global symmetry with the $U(1)_B$ symmetry of the original $SU(N_c)$ theory A.

This description is similar, but not identical to, the $U(N_f-N_c)$ dual description that we found for the $SU(N_c)$ theory in section \pureduality. In our previous duality we had elementary fields $b$ and ${\tilde b}$ in theory B that were used to form the baryons of theory A, while in the new duality the same role is played by the monopoles of $U(1)_{new}$. And, in the previous duality we had a singlet $Y$ appearing in theory B that mapped to the Coulomb branch coordinate of theory A, while in the new duality this is identified with $X_+ X_-$ (while $X_+$ and $X_-$ separately are not gauge-invariant).

This procedure of ``ungauging'' $U(1)$ symmetries does not work when we have a Chern-Simons term for the original $U(1) \subset U(N_c)$, since in that case the product of the two $U(1)$'s that we get after gauging gives a non-trivial topological theory that cannot be ignored.

\subsec{$U(1)$ dualities from $4d$}

We saw above how starting with the $4d$ dualities we can derive $3d$ dualities for $SU(N_c)$ and $U(N_c)$ SQCD theories. However, there are two dualities that we did not derive above. The first is the duality \AharonyBX\ of the $U(1)$ theory with $N_f=1$ to the theory of three chiral superfields with $W = -V_+ V_- M$. This duality follows from the general $U(N_c)$ dualities that we described above, but we actually used it to derive these dualities above, so it would be nice to have an independent derivation of this duality from $4d$. Another duality is $3d$ mirror symmetry \refs{\IntriligatorEX,\deBoerMP,\deBoerKA,\AharonyBX}, relating (in the simplest example) the $3d$ $U(1)$ SQCD theory with $N_f$ flavors $Q_i$, ${\tilde Q}^i$ to a $U(1)^{N_f-1}$ theory with $N_f$ pairs of chiral multiplets $q^i$, ${\tilde q}_i$, $N_f$ singlets $S_i$ and a superpotential $W = \sum_{i=1}^{N_f} S_i q^i {\tilde q}_i$. In the second theory, we take $N_f$ pairs $q^i$ and ${\tilde q}_i$ ($i=1,\cdots,N_f$), and we gauge the $U(1)^{N_f-1}$ under which $U(1)_i$ ($i=1,\cdots,N_f-1$) assigns charge $(+1)$ to $q^i$ and $\tilde q_{i+1}$ and charge $(-1)$ to $\tilde q_i$ and $ q^{i+1}$. For $N_f=1$ this is identical to the dual description mentioned above. We used above a specific example of this mirror symmetry, for $N_f=2$, to check the consistency of our $SU(N_c)$ dualities, and it would be nice to show that this duality (which exchanges mesons with monopoles) also follows from the $4d$ dualities.

The duality of $U(1)$ with $N_f=1$ follows from the low-energy description of the $SU(2)$ SQCD theory with $N_f=2$, as described in \newIS. We derived above, using a reduction from $4d$, the fact that the $3d$ $SU(2)$ theory with $N_f=2$ is described by the superpotential\foot{For the special case of $SU(2)$ we can also write this as $(- Y {\rm Pf}({\hat M}))$, manifesting the full $SU(2N_f)$ global symmetry of the $SU(2)$ gauge theory.}
\eqn\weffsutwo{W = Y (B {\tilde B} - \det(M))~.}
When we give a complex mass to one of the flavors, this implies that the low-energy theory of $SU(2)$ with $N_f=1$ is described by a constraint $Y_{low} M_{1}^{1} = 1$. Now, suppose that instead we turn on a real mass $\hat m$ for the second flavor. We then have three branches meeting at $\sigma = {\hat m}$. One is the Higgs branch labeled by $M_{2}^{2}$. For $\sigma > {\hat m}$ we have the original Coulomb branch labeled by $Y$. For $\sigma < {\hat m}$ we can integrate out the massive flavor, and obtain a new theory in which the original $Y$ vanishes but we have $Y_{low} M_{1}^{1} = 1$. This means that we can parameterize this part of the Coulomb branch by $M_{1}^{1} = 1 / Y_{low}$. Since the $SU(2)$ gauge symmetry is broken to $U(1)$ at $\sigma = {\hat m}$, we can also rewrite near this point the $SU(2)$ Coulomb branch coordinates as $U(1)$ Coulomb branch coordinates, and one finds $V_+ = Y$, $V_- = 1 / Y_{low}$. To compute the low-energy superpotential we now note that in our effective description above using $Y$, $B$, ${\tilde B}$ and $M$, the only fields that did not acquire real masses are $Y$, $M_{1}^{1}$ and $M_{2}^{2}$, so the superpotential \weffsutwo\ becomes
\eqn\weffsutwon{W = - Y M_{1}^{1} M_{2}^{2} = - V_+ V_- M,}
where $M \equiv M_{2}^{2}$ is the low-energy meson of the effective $U(1)$ theory near $\sigma={\hat m}$ (the other flavor is massive at this point). Thus, we can derive from $4d$ the effective description of \AharonyBX\ for this case, that we used to derive the other dualities above.

We can now take $N_f$ copies of this, so that we have a $U(1)^{N_f}$ gauge theory with $N_f$ flavors $q^i$, ${\tilde q}_i$ (one for each $U(1)$). At low energies this is equivalent to a theory with no gauge symmetry and with $W = -\sum_{i=1}^{N_f} V_+^i V_-^i N_{i}^{i}$, where $N_{i}^{i} \equiv q^i {\tilde q}_i$. We can now gauge the diagonal $U(1)_J$ symmetry, whose current is the sum of all $U(1)^{N_f}$ field strengths. From the gauge theory point of view this is the same as ``ungauging'' the diagonal $U(1)$, so we get the $U(1)^{N_f-1}$ gauge theory with the same matter content as in the mirror symmetry reviewed above. In the low-energy effective description we have $N_f$ flavors $V_{\pm}^i$ charged under this $U(1)_J$, with the superpotential above. We can now add to this $N_f$ singlet fields $S_i$ and deform the theory by $W = \sum_{i=1}^{N_f} S_i q^i {\tilde q}_i$. In the low-energy description this becomes $\sum_{i=1}^{N_f} S_i N_{i}^{i}$, which enables us to eliminate the original superpotential we had (and to identify $S_i = V_+^i V_-^i$). We then reproduce precisely the usual $3d$ mirror symmetry. It would be interesting to derive also other examples of $3d$ mirror symmetry from four dimensional dynamics by similar methods.

\newsec{Partition functions on $\S^3$ and dualities}
\seclab\partitionfuncs

An extremely useful test of conjectured dualities is given by the supersymmetric index\foot{When a theory is conformal it coincides with the superconformal index \KinneyEJ.}.  In $4d$ this is related to the partition function on $\S^3\times \S^1$, and in $3d$ it is the partition function on $\S^2\times \S^1$.  In addition, in $3d$ we can also study the supersymmetric partition function on (a generally squashed) $\S^3$. The feature that makes these partition
functions useful for checking dualities is that they are independent of many details of the theories, allowing one to infer information easily about the IR physics from
the UV description \refs{\RomelsbergerEG, \KapustinKZ, \FestucciaWS}.
These partition functions can be used efficiently to give strong evidence for putative dualities, and/or to suggest new ones.  Here we will apply this technique to several of the theories discussed in the previous sections.

The validity of a certain $4d$ duality implies equality of the  supersymmetric partition functions on  $\S^3\times \S^1$ computed using the two dual descriptions.
When we take the limit in which the size of the $\S^1$ goes to zero, these reduce to $3d$ partition functions on $\S^3$. As we discussed in the preceding sections, the $3d$ theories obtained by this reduction also have
in general non-trivial superpotential terms, and with these superpotentials the $3d$ theories are dual to each other. The equality of the $3d$ partition functions is then a simple test of this duality.

We have argued that to obtain a dual of the $3d$ SQCD theory without any superpotential, certain real mass parameters have to be taken large in the dualities obtained by the reduction.
 The $\S^3$ partition function is a function of real mass parameters of global symmetries.
 It is given by a matrix integral over the scalar components of the vector multiplets of the gauged symmetries.  Thus, we can take the limit of large real masses directly at the level of the partition function, and observe what kind of a theory is obtained in the IR.  In this respect the $\S^3$ partition function is more useful than the $3d$ supersymmetric index, which is the partition function on $\S^2\times \S^1$.  The supersymmetric index does not depend on the real masses, and is given as an integral over the zero modes of a component of the gauge field. Thus, it is not possible to study the large real
mass limit directly at the level of the index. However, if one conjectures some $3d$ duality then the $3d$ index is a very useful tool in figuring out the precise map between the protected states in the two dual descriptions, as discussed in appendix~A.

In this section we will first review basic facts about the $4d$ index, the $3d$ partition function on $\S^3$ and the relation between them.
Next we will discuss  in detail the procedure of taking large real masses at the level of the partition function. Then we will consider the $U(N)$ and $SU(N)$ dualities discussed above, and see how simple manipulations of the partition functions echo the physical considerations that led us to consider them.

\subsec{$3d$ partition functions from $4d$ indices}

Let us begin by reviewing the basic properties of the $4d$ index (related to the
supersymmetric partition function on $\S^3\times \S^1$), the
partition function on $\S^3$, and the relation between them.
We will refer to these two objects as the index and the partition function, respectively.

\bigskip
\noindent{\it The $4d$ index}

\nobreak
We consider $4d$ theories with four supercharges on $\S^3\times \S^1$, with radii $r_3$ and $r_1$, such that supersymmetry is preserved. The $4d$ index is defined by the following sum over the states of the theory on $\S^3$
\refs{\RomelsbergerEG,\KinneyEJ}:
\eqn\indexDef{
\II(p,\,q;\,\{u_a\})=\Tr \left[(-1)^F\,e^{-\beta\,\delta}\,p^{j_1+j_2-\frac{R}{2}}
\,q^{j_1-j_2-\frac{R}{2}}\,\prod_{a}u_a^{e_a}\right]\,.
} Here $j_1$ and $j_2$ are the Cartan generators of the $SU(2)_1\times SU(2)_2$ isometry of the sphere,
$R$ is the $U(1)_R$ charge, and
the charges  $e_a$ correspond to $U(1)$ global symmetries (which could be in the Cartan subalgebra of non-Abelian global symmetries). Note that the theory on $\S^3$ depends on the choice of a specific $U(1)_R$ symmetry.
The chemical potential $\beta$ couples to
\eqn\deltaIn{
\delta\equiv \{{\cal Q},\,{\cal Q}^\dagger\}=E-2j_1+\frac{3}2\,R\,,
} where $E$ is the energy times $r_3$ (for conformal theories this is related by the state/operator map to the conformal dimension),
and where we choose ${\cal Q}$ to be the supersymmetry generator with $(j_1,\,j_2)=(-\frac{1}2,\,0)$
and $R=-1$. The index is actually independent of $\beta$, since
$j_1\pm j_2-\frac{R}{2}$ and $e_a$ commute with ${\cal Q}$.
  The fugacities in \indexDef\
are usually taken to satisfy the following reality conditions:
\eqn\realityF{
{\rm Im}(p\,q)=0\,,\qquad |p/q|=|u_a|=1\,.
} However, one can analytically continue the fugacities to arbitrary complex values, and we will do
so in what follows. For convergence of the trace formula \indexDef\ one has to assume that $|p\,q|<1$.

When $p=q$, \indexDef\ is equal to the partition function of the $4d$ theory on $\S^3\times \S^1$, up to an overall constant. From this point of view $\beta = 2 \pi r_1 / r_3$. For general values of the parameters, the index is related instead~\ImamuraWG\ to a partition function of the $4d$ theory on $\S^3_b\times \widetilde {\S}^1$, where $\S^3_b$ is a squashed 3-sphere with radius ${r}_3$ and squashing parameter $b$, and $\widetilde{\S}^1$ is a circle of radius ${\tilde r}_1$ (see appendix B for more details). The relation between the parameters is~\ImamuraWG
\eqn\squashparams{
b^2=\frac{\log\, (p)}{\log\, (q)}\,, \qquad
\tilde r_1 = \frac{2}{b+b^{-1}}\,r_1\,.
}

The index of a single chiral superfield is given by~\DolanQI
\eqn\indexChi{
{\cal I}=\Gamma((p\,q)^{\frac{R}{2}}\,\prod_a u_a^{e_a};p,\,q)\,.
} Here $R$ is the R-charge of the field, $u_a$ are fugacities for $U(1)$ symmetries
labeled by $a$, and $e_a$ are the charges of the chiral field under these symmetries.
The function on the right-hand side is the elliptic Gamma function
\eqn\ellG{
\Gamma(z;p,\,q)\equiv \prod_{i,j=0}^\infty\frac{1-p^{i+1}q^{j+1}\,z^{-1}}{1-p^iq^j\,z}\,.
}
In \indexChi\ one removes a quadratically divergent term, which can be absorbed in background local FI
counter-terms (and in the Einstein Hilbert term, which in this context is like an FI term for the background $U(1)_R$ in the gravitational multiplet). The finite part is fixed such that the vacuum contributes $(+1)$ to \indexDef.

To compute the index of a gauge theory one includes also the contribution to the index of the vector multiplet, and projects on gauge-invariant states. The combined contribution to the index of the
vectors and the projection is
\eqn\projectF{
\eqalign{
\II
=\frac{(p;p)^{r_G}(q;q)^{r_G}}{|W|}\oint_{{T}^{r_G}} \prod_{\ell=1}^{r_G}\frac{dz_\ell}{2\pi i \,z_\ell}\,
\prod_{\alpha\in R_+}\theta(e^{\alpha(\epsilon)};p)\;\theta(e^{-\alpha(\epsilon)};q)\,\cdots \, .
}
} Here $r_G$ is the rank of the gauge group $G$, $T^{r_G}$ is the maximal torus of the group,
 $W$ is the Weyl group of $G$, and $R_+$ is the set of positive roots of $G$.
The positive roots are linear combinations of the basis vectors $\e_i$, and we defined $z_l=\exp(\e_l)$, so that $e^{\alpha(\epsilon)}$ is a monomial in the $z_l$.
 We have used in the equation above the theta-function and the q-Pochhammer symbol,
\eqn\thetaDef{
\theta(z;p)\equiv \prod_{\ell=0}^\infty(1-z\,p^\ell)(1-z^{-1}p^{\ell+1})\,,\qquad
(a;b)\equiv \prod_{\ell=0}^\infty(1-a\,b^\ell)\,.
}
Given the matter content of the theory, the gauge interactions, and the exact global symmetries,
the index is completely determined.  In particular, the only effect of superpotential terms is to
reduce the number of global symmetries, and thus the possible fugacities.

There is another ingredient that may be present if the gauge group contains a $U(1)$ factor, namely, a Fayet-Iliopoulos (FI) term.  On Euclidean $\S^3 \times \R$, with a round $\S^3$ of radius $r_3$, this enters the action as a term \RomelsbergerEG:
\eqn\fourdfiterm{
\int d^4 x\,\sqrt{g} \, \xi^{(4)}\; (D - \frac{2 i A_4}{r_3})\,,
}
where $D$ and $A_4$ (the component of the gauge field along the $\R$ direction) are components of the $U(1)$ vector multiplet, and we denote the FI parameter $\xi^{(4)}$ with a superscript to distinguish it from the $3d$ FI parameter, discussed below.  When we compactify the $\R$ direction on a circle of radius $r_1$, $A_4$ gets compactified, with $A_4 \sim A_4 + 1/r_1$.
To ensure invariance of the path integral under this shift, we should impose\foot{ If the gauge group is $\R$ rather than $U(1)$, there is no such compactification of $A_4$ and $\xi^{(4)}$ is not quantized.  However, in that case the index vanishes for generic values of $\xi^{(4)}$.  This situation is similar to the quantization of the FI-term in \refs{\SeibergQD,\BanksZN}.}
\eqn\fiquant{
\int  d^4 x\,\sqrt{g}\, \xi^{(4)}\; \frac{2 i}{r_1\, r_3} = 8 \, \pi^3\,i\, \xi^{(4)}\, {r_3}^2 = 2\, \pi\, i\, n\,, {\rm\ with\ } n\, \in \ZZ\, .
}
The contribution of such a term to the partition function is then given by inserting a factor of $z^n$ into the integral \projectF\ over the parameter $z$ corresponding to this $U(1)$ gauge multiplet.

Let us discuss a couple of simple examples. Given
two chiral superfields $\Phi_{1,2}$, with R-charges $R$ and $(2-R)$ and with opposite charges, $\pm e$,
for a $U(1)_u$ global symmetry, the index is given by
\eqn\massChi{
\II=\Gamma((p\,q)^{\frac{R}{2}}\, u^{e};p,\,q)\,
\Gamma((p\,q)^{\frac{2-R}{2}}\, u^{-e};p,\,q)=1\,.
} The above choice of charges is consistent with a superpotential mass term $W = m\, \Phi_1\,\Phi_2$.
The massive fields disappear from the IR physics and thus should not contribute to the index:
the identity \massChi\ is the index  manifestation of this physical fact.

As an example of an index of a gauge theory, let us write down the index of $SU(N)$ SQCD with $N_f$ flavors whose $R$-charge is $R$,
\eqn\SQCDex{
\eqalign{
\II=&\frac{(p;p)^{N-1}(q;q)^{N-1}}{N!}\oint_{{T}^{N-1}} \prod_{\ell=1}^{N-1}\frac{dz_\ell}{2\pi i \,z_\ell}\,
\prod_{i>j=1}^N \theta(z_i/z_j;p)\;\theta(z_j/z_i;q)\times\cr
&\qquad\qquad\prod_{m=1}^{N_f}\prod_{i=1}^N\Gamma((p\,q)^{\frac{R}{2}}\,c\,a_m\,z_i;p,\,q)
\Gamma((p\,q)^{\frac{R}{2}}\,c^{-1}\,b_m\,z^{-1}_i;p,\,q)\,.
}
} Here $a_m$ (with $\prod_m a_m = 1$), $b_m$ (with $\prod_m b_m = 1$) and $c$ parameterize the Cartan subalgebra of the $SU(N_f)_L\times SU(N_f)_R\times U(1)_B$ flavor symmetry.
The fugacities $z_i$ ($i=1,\cdots,N$) with $\prod_{i=1}^N z_i=1$ parameterize the Cartan subalgebra of  the $SU(N)$ gauge group.
The R-charges used here should be anomaly free, namely $R=1-\frac{N}{N_f}$.
The existence of $4d$ IR dualities implies a huge variety of identities that are satisfied by contour integrals of elliptic
Gamma functions \DolanQI\ (see also \refs{\SpiridonovZA,\SpiridonovHF}).

\bigskip
\noindent{\it The $3d$ partition function}

\nobreak
The $3d$ partition function on a squashed sphere $\S^3_b$ is very similar in structure to the $4d$ index.  Rather than the fugacities $p,q,$ and $u_a$, it is a function of real mass parameters $m_a$ for the global symmetries of the theory, and of the squashing parameter $b$.  The partition function of a $3d$ chiral superfield is given by~\HamaEA
\eqn\partChi{
{\cal Z}=\Gamma_h(\omega\,R+\sum_a m_a\,e_a;\omega_1,\,\omega_2)\,.
} Here $R$ is the R-charge, $m_a$ are real mass parameters related to $U(1)$ global symmetries
labeled by $a$, and $e_a$ are the charges of the chiral superfield under
those symmetries.  The $m_a$ can be thought of as constant values for the scalars $\sigma$ in background vector multiplets coupled to the global symmetries. For the squashed sphere we define $\omega_1=i\,b \,{ r}_3^{-1}$ and $\omega_2=i\,b^{-1} \, {r}_3^{-1}$, where $b$ is the squashing parameter ($b$ is either real or a phase~\refs{\ImamuraWG,\ClossetRU}). The parameter $\omega$ is defined as $\omega\equiv \frac{\omega_1+\omega_2}2$.  Finally, the hyperbolic Gamma function is defined by the following infinite product (assuming ${\rm Im} (\frac{\omega_2}{\omega_1})>0$)
\eqn\hypG{
\Gamma_h(z;\omega_1,\,\omega_2)=
e^{\frac{\pi i}{2\omega_1\omega_2}\left((z-\omega)^2-\frac{\omega_1^2+\omega_2^2}{12}\right)}\,
\prod_{\ell=0}^\infty
\frac{1-e^{\frac{2\pi i}{\omega_1}(\omega_2-z)}\,e^{\frac{2\pi i\omega_2\,\ell}{\omega_1}}}
{1-e^{-\frac{2\pi i}{\omega_2}\,z}\,e^{-\frac{2\pi i\omega_1\,\ell}{\omega_2}}}\,.
}

The partition function of a gauge theory is given as a matrix integral over the scalar component $\sigma$ of the vector multiplet, with eigenvalues $\sigma_\ell$,
\eqn\partInt{
\frac{1}{|W|}\int\prod_{\ell=1}^{r_G}\frac{d\sigma_\ell}{\sqrt{-\omega_1\omega_2}}
e^{\frac{2\pi i \xi_i^{(3)} \, \Tr_i\, (\sigma)}{\omega_1\omega_2}}
e^{\frac{\pi i\,  \Tr\, (\sigma^2)}{\omega_1\omega_2}}
\prod_{\alpha\in R_+}
\frac{1}{\Gamma_h(\alpha(\sigma);\omega_1,\omega_2)\Gamma_h(-\alpha(\sigma);\omega_1,\omega_2)}\,\dots
}
As in $4d$, the integrand includes a contribution from the vector multiplet, and the dots denote contributions from the chiral multiplets.  Here we have included FI parameters $\xi_i^{(3)}$ for $U(1)$ factors in the gauge group (picked out by the traces $\Tr_i$), which can be thought of as real mass parameters for the corresponding $U(1)_J$ global symmetries.\foot{For later convenience, our normalization of $\xi^{(3)}$ here differs by a factor of $(-2\pi)$ from the standard one, namely we have in the Lagrangian $(-{1\over {2\pi}}\xi^{(3)} D)$.}  A new ingredient that appears in $3d$ is that one may have a Chern-Simons term for the gauge group, which contributes a factor $e^{\frac{\pi i\,  \Tr (\sigma^2)}{\omega_1\omega_2}}$.\foot{Here $\Tr$ is an invariant scalar product of the Lie algebra,
which we take to include also the Chern-Simons levels $k$ of the different gauge groups. For example, for
a $U(N)$ theory with level $k$, we will take $\Tr$ to be the usual trace in the fundamental representation
times the level $k$.  The normalization in the general case is discussed in \KapustinKZ.}  As in the index in $4d$, the only effect
of the superpotential is to restrict the allowed symmetries, {\it i.e.} the allowed real mass parameters.

As two examples let us again consider a massive field and a simple gauge theory. The partition function
of two chiral superfields that can be joined by a superpotential mass term is given by:
\eqn\massChiT{
{\cal Z}=\Gamma_h(\omega\,R+m\,e;\omega_1,\,\omega_2)\,
\Gamma_h(\omega\,(2-R)-m\,e;\omega_1,\,\omega_2)=1\,,
}  and is again trivial, as expected.  The partition function of $3d$ $U(N)$ SQCD with $N_f$ flavors of R-charge $R$, and Chern-Simons level $k$, is given by
\eqn\threeDQCD{
\eqalign{
&{\cal Z}=\frac{1}{N!}\int\prod_{\ell=1}^{N}\frac{d\sigma_\ell}{\sqrt{-\omega_1\omega_2}}
e^{\frac{2\pi i \xi\, \sum_\ell \sigma_\ell}{\omega_1\omega_2}}
e^{\frac{\pi i k\, \sum_\ell \sigma_\ell^2}{\omega_1\omega_2}}
\prod_{i\neq j}
\frac{1}{\Gamma_h(\sigma_i-\sigma_j;\omega_1,\omega_2)}\times\cr
&\qquad\qquad \prod_{i=1}^N\prod_{a=1}^{N_f} \Gamma_h(\omega R+\sigma_i+m_a+m_A;\omega_1,\omega_2)\,
\Gamma_h(\omega R-\sigma_i+\tilde{m}_a+m_A;\omega_1,\omega_2)\,.
}
} The real masses $m_a$ (with $\sum_a  m_a=0$), $\tilde{m}_a$ (with $\sum_a \tilde{m}_a=0$), and $m_A$ correspond to the $SU(N_f)_L\times SU(N_f)_R\times U(1)_A$ global symmetries.
There is no analogue here to the $4d$ restrictions on charges coming from anomalies.  In fact, the parameter $R$ can be redefined by shifting some of the real mass parameters by an amount proportional to $\omega$, which has the physical interpretation of mixing the R-symmetry with the corresponding flavor symmetry.\foot{A similar property holds in the $4d$ index, but in that case anomalies and discrete symmetries forbid any such mixing in all the theories we consider in this paper.}  As we will see below, $3d$ dualities imply numerous identities relating integrals of hyperbolic Gamma
functions. For a very useful resource for such identities one can consult~\debult.

\bigskip
\noindent {\it From the $4d$ index to the $3d$ partition function}

\nobreak
Given the $4d$ index one can interpret it as a partition function on $\S^3_b\times \widetilde{\S}^1$, and take the radius of $\widetilde{\S}^1$ to
zero to obtain the partition function  on $\S^3_b$ of a $3d$ theory with the same matter
content and gauge interactions as the $4d$ theory we start with.
This procedure was discussed in \refs{\DolanRP,\GaddeIA,\ImamuraUW,\NiarchosAH,\ImamuraWG}, and we review it in appendix B.
We parameterize the $4d$ fugacities as follows:
\eqn\fugmass{
u_a=e^{2 \pi i{\tilde r}_1\,m_a},\qquad p=e^{2 \pi i{\tilde r}_1\,\omega_1},\qquad q=e^{2 \pi i{\tilde r}_1 \,\omega_2}.
}
%Since we have set $r_3=1$, $r$ is the (dimensionless) ratio of the radii of $\S^1$ and $\S^3$.
The $3d$ reduction is obtained by taking  the limit ${\tilde r}_1\to 0$, keeping $m_a$ and $\omega_{1,2}$ fixed.\foot{
The reality condition of the fugacities~\realityF\ is consistent with  $\omega_1=ib {r}_3^{-1}$
and $\omega_2=ib^{-1} {r}_3^{-1}$ with $|b|=1$.}
Under the reduction, the $4d$ index of a $4d$ chiral superfield becomes, up to a divergent exponential, the partition function of a $3d$ chiral superfield:
\eqn\threeDlim{\eqalign{
\lim_{{\tilde r}_1\to 0}\;
&\left[\Gamma(e^{2 \pi i {\tilde r}_1\,(\omega\,R+\sum_a m_a\,e_a)};e^{ 2 \pi i {\tilde r}_1 \,\omega_1},\,e^{ 2 \pi i {\tilde r}_1\,\omega_2})\,e^{\frac{\pi i}{6\,\omega_1\,\omega_2\;{\tilde r}_1}\left(\sum_a m_a\,e_a-\omega(1-R)\right)}\right]=\cr
&\qquad\qquad\qquad\qquad\qquad\qquad\qquad
\Gamma_h(\omega\,R+\sum_a m_a\,e_a;\omega_1,\,\omega_2)\,.
}
} A way to derive this relation is by using the $SL(3;\ZZ)$ properties of the elliptic Gamma function~\slthreeZ\ (see appendix B for a physical explanation of this relation).
The integration in the $4d$ equation \projectF\ is performed over the zero mode of the $\widetilde{\S}^1$ component of the gauge field, with a holonomy that we denoted by $z_\ell$.
 In $3d$ this compact integral becomes a non-compact integral over the scalar component of the gauge field $\sigma_\ell$, with $z_l = \exp(2\pi i {\tilde r}_1 \sigma_\ell)$.
We can think of this integration as integrating over the real mass corresponding to the symmetry that is gauged. In more detail, the
contribution of the gauging in the index~\projectF, taking the $3d$ limit, becomes
\eqn\gaugeThree{
\eqalign{
&\lim_{{\tilde r}_1\to 0} \, \frac{(p;p)^{r_G}(q;q)^{r_G} e^{\frac{\pi i\,|G|\,\omega}{6{\tilde r}_1\,\omega_1\omega_2}}}{|W|}\oint_{{T}^{r_G}} \prod_{\ell=1}^{r_G}\frac{dz_\ell}{2\pi i \,z_\ell}\,
\prod_{\alpha\in R_+}\theta(e^{\alpha(\epsilon)};p)\;\theta(e^{-\alpha(\epsilon)};q)\,\dots\sim\cr
&\qquad
\frac{1}{|W|}\int\prod_{\ell=1}^{r_G}\frac{d\sigma_\ell}{\sqrt{-\omega_1\omega_2}}\prod_{\alpha\in R_+}
\frac{1}{\Gamma_h(\alpha(\sigma);\omega_1,\omega_2)\Gamma_h(-\alpha(\sigma);\omega_1,\omega_2)}\,\dots
}
} Here $|G|$ is the number of generators of the group. To derive this relation one uses the standard $SL(2;\ZZ)$ transformations of the theta-function and the q-Pochhammer symbol. Here we have interchanged the order of taking the limit and performing the integration.\foot{ In some cases this prescription gives a divergent result, beyond the phase factor we explicitly pulled out above: an example is ${\cal N}=4$ SYM.  We
will comment on the physics of such divergences  elsewhere~\toappearthree.}

Let us now comment on the extra divergent exponential factors we obtain in the reduction.
The divergent prefactor of the full partition function is
\eqn\phases{
\eqalign{
\exp\left(
\frac{\pi i\,\,}{6\,{\tilde r}_1\,\omega_1\omega_2}
\left\{\omega\left(\sum_\alpha(1-R_\alpha)-|G|\right)
-\sum_{a}m_a\,\sum_{\alpha}{e^{(\alpha)} }_a
\right\}
\right)\,.
}
}
Here the sum over $\alpha$ is over chiral matter fields. The exponent is proportional to the $U(1)_R$-gravity-gravity and flavor-gravity-gravity $4d$ anomalies.\foot{This kind of relation of $4d$ anomalies to the $SL(3;\ZZ)$ transformations of elliptic Gamma functions was discussed in~\SpiridonovWW\ (see also~\RazamatUV).}
 It is also proportional to finite
induced background FI terms (see the comment after \ellG). These terms diverge as ${\tilde r}_1\to 0$. We will
simply subtract them. This can be interpreted as changing the induced $3d$ background FI terms.\foot{See~\refs{\ImamuraUW,\CveticXN}\ for related discussions.} Any two dual theories in $4d$ will give rise to equal factors \phases, due to 't Hooft's anomaly matching argument, and therefore removing the divergent factor is consistent with the duality.

We can also see that the $4d$ FI parameter reduces in the expected way to the $3d$ FI parameter. In our normalizations we have
\eqn\fourdthreedfi{
\xi^{(3)} = - 4 \pi^2 \, {\tilde r}_1 \, \xi^{(4)} = n\, {\tilde r}_1\, \omega_1 \, \omega_2 \,. }
Here $n$ is an integer \fiquant.  Although this parameter is quantized in the $4d$ index, we see that as ${\tilde r}_1$ is sent to zero, the range of $\xi^{(3)}$ effectively becomes continuous, and the insertion of $z^n$ in the $4d$ index reduces precisely to the contribution of a $3d$ FI-term, as in \partInt.

As discussed above, the partition functions of the reduced theories that we obtain in this way are not defined for all choices of real mass parameters, but only for those consistent with the non-anomalous $4d$ symmetries.  Such a restriction on the allowed real mass parameters is exactly what one finds when a superpotential is present, and thus the partition functions are consistent with such superpotentials. In cases that the partition functions of a putative dual pair agree only if such a restriction is imposed,
one may take this as an indication of the presence of a superpotential breaking the symmetries that are anomalous in $4d$, which is precisely what we have seen in the preceding sections.

\subsec{Partition functions and real masses}

The real mass parameters of a $3d$ theory are labeled by an element $\mu$ in the Cartan subalgebra of the flavor group $H$, with the real mass of a chiral superfield that transforms with weight $\tau$ under $H$ given by $\tau(\mu)$.
The partition function of a theory with gauge group $G$, flavor group $H$, and a collection of chiral multiplets is then:\foot{
Here we suppress the dependence on $\omega_i$ and write $\Gamma_h(z) \equiv \Gamma_h(z,\omega_1,\omega_2)$ for the hyperbolic Gamma function. We also do not include bare FI-terms or Chern-Simons terms; it is straightforward to modify the following argument to include them.}
\eqn\genPart{
{\cal Z}=\frac{1}{|W|} \int \prod_{\ell=1}^{r_G}\frac{d\sigma_\ell}{\sqrt{-\omega_1\omega_2}}  {
\prod_{a} \Gamma_h(\omega R_a + \rho_a(\sigma) + \tau_a(\mu)) \over \prod_{\alpha \in R_+} \Gamma_h(\alpha(\sigma))\,\Gamma_h(-\alpha(\sigma))}\,,
}
where $r_G$ is the rank of the gauge group, $W$ is the Weyl group, $(\rho_a,\tau_a)$ is the weight under $G \times H$ of the $a$'th chiral multiplet, and $R_a$ is its R-charge.\foot{
For example, if $G=SU(N_c)$, $H=SU(N_f)$, and the chiral field is in the fundamental representation of both, then $a$ has $N_c\times N_f$ values, $a=(a_G,a_H)$, with $\rho_{(a_G,a_H)}(\sigma)=\sigma_{a_G}$
and $\tau_{(a_G,a_H)}(\mu)=\mu_{a_H}$.
}
We would like to consider the limit of the partition function as some of these real masses are taken large~ (see~\refs{\DolanRP,\WillettGP,\BeniniMF,\NiarchosAH} for previous discussions).
We pick a direction $\mu_o$ in the Cartan subalgebra of $H$, shift the real masses by
\eqn\massShift{
\mu \rightarrow \mu + s\;  \mu_o \,,
}
and consider the limit where $s$ becomes large and positive.

First let us discuss what happens to the partition function of a chiral field in this limit.
If the chiral multiplet is charged under the $U(1)$ subgroup corresponding to $\mu_o$,
it becomes heavy and decouples in the IR.  For large $s$, the partition function becomes (using the
asymptotics of the hyperbolic Gamma function, {\it e.g.} see~\debult)
\eqn\gammaAsymp{\eqalign{
&\log\;\left(  \Gamma_h \left( \omega R + \rho(\sigma) +\tau(\mu+s\,\mu_o) \right) \right) =\cr
&\qquad
{\rm sign}(\tau(\mu_o))\, \frac{\pi i}{2 \omega_1\omega_2} \bigg( \left[\omega (R-1) + \rho(\sigma)+\tau(\mu+s\,\mu_o) \right]^2 - \frac{{\omega_1}^2 + {\omega_2}^2}{12} \bigg) + {\cal O} (e^{-\alpha s}) \,,
}
} where  $\alpha$ is some positive constant.
This exponential contains terms linear and quadratic in the real mass parameters.  Those which are $s$-independent correspond to FI and CS terms,
respectively, which are generated when we integrate out this heavy chiral multiplet.
These, in general, will occur both for gauge symmetries and for flavor symmetries.
In addition we have contributions scaling linearly and quadratically with the large parameter $s$.
Contributions of this kind are interpreted as (mixed) CS contact terms \refs{\ClossetVG,\ClossetVP}
involving the symmetry associated with the real mass that acts only on heavy fields.

Now consider this limit of the partition function for a general gauge theory.  The partition function is given by an integral over the Cartan subalgebra of the gauge group, parameterized by the eigenvalues of the gauge scalar $\sigma$, as in~\genPart.  The integrand is a product of contributions from the chirals, which are functions of the real mass parameters and the eigenvalues of $\sigma$.
In the large $s$ limit, we can approximate this integral by finding the region(s) in $\sigma$-space that give the dominant contribution to the integral for large $s$, {\it i.e.}, the saddle points of the integral.

To determine these saddle points, suppose that as we take $s$ large, we shift $\sigma$ in a coordinated way
\eqn\sigmaShift{
\mu \rightarrow \mu + s\, \mu_o, \qquad\qquad \sigma  \rightarrow \sigma + s\, \sigma_o\,, }
where $\sigma_o$ is some direction in the space of $\sigma$'s\foot{We take $\mu_o,\sigma_o$ to be dimensionless so that $s$ has dimensions of mass.}.
 For finite $s$ this is just a change of variables, but when we take the large $s$ limit of the integrand for a given choice of $\sigma_o$, we are effectively focusing only on a specific region in $\sigma$ space, and ignoring the contributions from regions at a distance of order $s$ away.

The different choices of $\sigma_o$ correspond to focusing on different points in the Coulomb branch of the theory, and it is natural to expect that the dominant saddle points occur at one of the singularities in the Coulomb branch, where there is an interacting fixed point. The chiral multiplets and vector multiplets whose contributions remain finite (respectively, diverge with $s$)  correspond to the ones that remain light (respectively, heavy) in this vacuum.  The contributions from the heavy multiplets can be analyzed using \gammaAsymp. They contribute FI and CS terms, which arise from integrating them out. The light multiplets
comprise the low energy theory at this point in the Coulomb branch.  When a theory that is deformed by a real mass has several non-trivial vacua, the one that contributes the dominant saddle in the partition function computation depends on the choice of the parameters, {\it e.g.} R-charges.\foot{The properties of the partition function are known to depend on the choice of the parameters. In particular, even without taking the limit of large real masses, the partition function converges only for certain ranges of the parameters \refs{\WillettGP,\SafdiRE}.}

If we start from some duality and take the large real mass limit, the sets of vacua in the two sides of the duality are in general mapped in a
nontrivial way into each other~\Shamirthesis.
Since the  partition functions of the two dual theories are equal for any value of $s$,   the leading saddles in the large $s$ limit should also agree.  While this saddle may occur at the trivial vacuum $\sigma_o=0$ for one theory, this could map to a non-trivial vacuum for the dual, as discussed in \NiarchosAH.  If there is a range of parameters for which a given vacuum
gives the leading saddle in the large $s$ limit, one can directly deduce an
equality of the partition functions of the low-energy theories in that vacuum and its dual, which can be viewed as strong evidence for a putative duality between these low-energy theories.\foot{In fact, if this range includes an open set in the space of parameters, one can argue by analytic continuation that the partition functions of the low energy theories must be equal, as analytic functions, in the entire range of parameters.}

\subsec{An example of a large real mass limit}
\subseclab\partexample

To illustrate the general discussion above, let us consider the concrete  example of a $U(N_c)$ theory with $N_f+1$ fundamental flavors,
which has the following partition function:
\eqn\QCDthree{
 {\cal Z}^{U(N_c)}_{N_f+1}(m_a,\mu_a,\xi) = \frac{1}{N_c!} \int
 \prod_{\ell=1}^{N_c} \frac{d\sigma_\ell}{\sqrt{-\omega_1\omega_2}} \frac{ \prod_{j=1}^{N_c} e^{ \frac{2\pi i}{\omega_1\omega_2}  \xi \sigma_j}
  \prod_{a=1}^{N_f+1} \Gamma_h( \pm (\sigma_j + m_a) + {\hat \mu}_a) }{\prod_{i <j} \Gamma_h(\pm(\sigma_i - \sigma_j))}\,.
}
Here and in the rest of the paper we use a shorthand notation $\Gamma_h(x \pm y ) \equiv \Gamma_h(x+y) \Gamma_h(x-y)$.
The parameters $m_a$ and $\mu_a$ are the vector and axial real masses, respectively, which are related to the real masses ${\hat m}_a,\tilde{{\hat m}}_a$ of the chiral fields $Q_a,\tilde{Q}_a$ by:
\eqn\vecaxialmassdef{
 m_a = \frac{1}{2}({\hat m}_a - \tilde{{\hat m}}_a), \;\;\;\; \mu_a = \frac{1}{2}({\hat m}_a + \tilde{{\hat m}}_a).
}
In \QCDthree\ we defined for brevity a complexified axial real mass including the R-charges,
\eqn\complexAxial{
{\hat \mu}_a \equiv \mu_a+\omega\,R.
}
This will allow us to write formulas that are independent of the arbitrary choice of the R-charge of the fundamental fields.
The parameter $\xi$ is the FI-term.

Now consider giving a large positive vector-like real mass to the $(N_f+1)$'th flavor, {\it i.e.}, set $m_{N_f+1}=s$ and let $s\to +\infty$.
Using \gammaAsymp\ we find, for large $s$:
\eqn\UNlimita{
\eqalign{
&{{\cal Z}^{U(N_c)}_{N_f+1}}^{(1)} \rightarrow
{\exp \bigg(\frac{2\pi i}{\omega_1\omega_2} s N_c (-\omega+{\hat \mu}_{N_f+1})  \bigg)} \,\times  \cr
 &\qquad \frac{1}{N_c!}\;\int
 \prod_{\ell=1}^{N_c} \frac{d\sigma_\ell}{\sqrt{-\omega_1\omega_2}} \frac{ \prod_{j=1}^{N_c} e^{ \frac{2\pi i}{\omega_1\omega_2}
 (\xi-\omega +{\hat \mu}_{N_f+1}) \sigma_j}  \prod_{a=1}^{N_f} \Gamma_h( \pm (\sigma_j + m_a) + {\hat \mu}_a) }{\prod_{i <j} \Gamma_h(\pm(\sigma_i - \sigma_j))} \,,
}}
where the superscript denotes that this is the contribution from the first vacuum, namely, the one at the origin of moduli space.  Up to an overall $s$-dependent factor, this is the partition function of the $U(N_c)$
theory with $N_f$ flavors, which is the low energy theory at the origin of moduli space obtained by adding this
large real mass and integrating out the heavy fields.

Let us take again the limit inside the integral, but now at the same time we also shift $\sigma$ in order to explore the vacuum with  $\sigma_{N_c}\approx -s$,
\eqn\newshift{ \sigma_{N_c} \equiv \hat{\sigma} - s \,.}
This causes the $N_c$'th component of the first $N_f$ flavors to become massive, but it cancels the mass of the $N_c$'th
 component of the $(N_f+1)$'th flavor, such that it remains light.  In addition, it effectively gives a mass to the gauge multiplet components with
 an index $i$ or $j$ equal to $N_c$, corresponding to Higgsing the gauge group down to $U(N_c-1) \times U(1)$.  The contribution of these vectors becomes an exponential factor, as for the massive chiral multiplets in~\gammaAsymp.
 Using the asymptotic formula \gammaAsymp, we find in the limit of large $s$ and in the vacuum~\newshift\foot{Here the replacement $N_c! \rightarrow (N_c-1)!$ in the prefactor arises from summing
 over the Weyl-equivalent choices of $\sigma_o$.}
\eqn\UNlimitb{
\eqalign{
 &{{\cal Z}^{U(N_c)}_{N_f+1}}^{(2)} \rightarrow \exp \bigg( \frac{2 \pi i}{\omega_1 \omega_2} \bigg( s (-N_f \omega  + \sum_{a=1}^{N_f} {\hat \mu}_a + (N_c-1) {\hat \mu}_{N_f+1}) + \sum_{a=1}^{N_f} m_a
 ( \omega - {\hat \mu}_a) \bigg) \bigg) \times  \cr
& \frac{1}{(N_c-1)!}\int \prod_{j=1}^{N_c-1} \frac{d\sigma_j}{\sqrt{-\omega_1\omega_2}} \frac{d\hat \sigma}{\sqrt{-\omega_1\omega_2}}
  \frac{\prod_{j=1}^{N_c-1} e^{\frac{2\pi i}{\omega_1\omega_2}  (\xi + {\hat \mu}_{N_f+1}) \sigma_j} \prod_{a=1}^{N_f} \Gamma_h(\pm (\sigma_j + m_a) + {\hat \mu}_a) }{\prod_{i<j}^{N_c-1}
 \Gamma_h(\pm (\sigma_i - \sigma_j))}  \times  \cr
 &\qquad\qquad e^{2 \pi i (\xi + \omega(N_f+1-N_c) - \sum_{a=1}^{N_f} {\hat \mu}_a) \hat{\sigma}} \Gamma_h(\pm \hat{\sigma} +{\hat \mu}_{N_f+1})\,.
}}
Here the superscript denotes this is the contribution from the second vacuum.  This partition function corresponds to a $U(N_c-1) \times U(1)$ theory, with $N_f$ flavors in the fundamental of $U(N_c-1)$ and
one additional flavor with charge one under the $U(1)$ factor. This is the low-energy theory in the vicinity of the vacuum specified by the VEV $\vev{\sigma}={\rm diag}(0,\cdots,0,-m_{N_f+1})$ in the Coulomb branch.

Note that there are FI-terms for the $U(N_c-1)$ and $U(1)$ factors of the gauge group.  Since there was only one $U(1)_J$ symmetry in the UV, these IR FI parameters are related in a simple way to the FI parameter $\xi$ of the $U(N_c)$ gauge group in the UV theory
\eqn\fitermsflow{
\xi_{U(N_c-1)} =\xi+ {\hat \mu}_{N_f+1}  , \;\;\; \xi_{U(1)} = \xi + \omega(N_f+1-N_c) - \sum_{a=1}^{N_f} {\hat \mu}_a\,.
}
In general, such a relation among the real mass parameters of the theory (in this case the FI parameters $\xi_{U(N_c-1)}$ and $\xi_{U(1)}$) reflects a superpotential term in the action.   In this case, this term is generated by the monopole-instantons in the broken part of the $U(N_c)$ gauge group:
\eqn\neweqw{W = X_- \, V_+\,,}
where $X_\pm$ and $V_\pm$ are the monopoles for the $U(N_c-1)$ and $U(1)$ factors, respectively.

We have thus computed in \UNlimita\ and \UNlimitb\ the partition functions in two different vacua after turning on a large real mass for one of the flavors.
These two expressions have different scalings in the large real mass limit.
One can check, at least in some examples,
that the two different vacua give the leading contribution in the large $s$ limit
to the original
partition function~\QCDthree\ for different choices of the R-charges.\foot{
For example with $N_c=2$ and $N_f=4$, parameterizing the R-charges of the first three flavors by $R$ and the last one by $2-3R$ (motivated by having $\eta$ superpotentials in what follows),
the vacuum at the origin is dominant for $R<1/2$ and the vacuum with the Higgsed gauge group is dominant for $R>1/2$.
}
We will see in the next subsection that under the $U(N)$ duality discussed in section \undualities, the two partition functions \UNlimita\ and \UNlimitb\ will map into each other, correctly reproducing the behavior of this duality under real mass deformations \refs{\Shamirthesis,\BeniniMF,\newIS}.

\subsec{$U(N)$ duality}

Let us consider the duality of $U(N)$ theories obtained in section \undualities\ by reduction from $4d$.
  Here theory A is the $U(N_c)$ theory with $N_f$ fundamental flavors, with an additional superpotential:
\eqn\etaW{
W_A = \eta\, X_+ X_- \,,
}
where $X_\pm$ are the monopole operators parameterizing the Coulomb branch of the theory.
This is nearly the same as the $U(N)$ theory considered in the previous subsection,
 the only difference being the superpotential \etaW.
In the partition function this difference  manifests itself as an additional constraint on the real mass and R-charge parameters.
Namely, we must impose
\eqn\massConstr{
\sum_{a=1}^{N_f} {\hat \mu}_a = \omega(N_f -N_c).
}
The partition function is then given by
\eqn\UNelec{
{\cal Z}_A = \frac{1}{N_c!} \int
 \prod_{j=1}^{N_c} \frac{d\sigma_j}{\sqrt{-\omega_1\omega_2}}
 \frac{\prod_{j=1}^{N_c} e^{\frac{2\pi i}{\omega_1\omega_2}  \xi \sigma_j} \prod_{a=1}^{N_f} \Gamma_h(\pm (\sigma_j + m_a) + {\hat \mu}_a)}{\prod_{i<j}
 \Gamma_h( \pm (\sigma_i - \sigma_j))}\,. }
The dual of this theory is the
 $U(N_f-N_c)$ theory with $N_f$ flavors and ${N_f}^2$ uncharged chiral  multiplets ${M_a}^b$.  The superpotential is
\eqn\dualWU{ W_B = \sum_{a,b} \tilde{q}_a\, {M^a}_b\, q^b + \tilde{\eta} \tilde{X}_+ \,\tilde{X}_- \,.}
With the real mass parameters appropriately mapped from theory A, the partition function is given by:
\eqn\UNmag{
\eqalign{
{\cal Z}_B = &\prod_{a,b=1}^{N_f} \Gamma_h( m_a + {\hat \mu}_a - m_b + {\hat \mu}_b) \cdot\cr
 &\qquad \frac{1}{(N_f-N_c)!} \int \prod_{j=1}^{N_f-N_c} \frac{d\ts_j}{\sqrt{-\omega_1\omega_2}} \frac{\prod_{j=1}^{N_f-N_c}
e^{\frac{2\pi i}{\omega_1\omega_2}  \xi \ts_j} \prod_{a=1}^{N_f} \Gamma_h(\omega \pm (\ts_j - m_a) - {\hat \mu}_a)}{\prod_{i<j}\Gamma_h(\pm (\ts_i - \ts_j))}
\,.
}
}
The first factor is the contribution of the uncharged chiral multiplets ${M_a}^b$.
As a consequence of the duality in $4d$, the partition functions ${\cal Z}_A$ and ${\cal Z}_B$ are equal
as functions of the mass parameters, subject to the constraint \massConstr.
Mathematically, the equality of \UNelec\ and \UNmag\ follows from the equality of the supersymmetric indices of the corresponding $4d$ theories.\foot{Specifically, starting from the identity for the $SU(N)$ duality \DolanQI, one can gauge the $U(1)_B$ symmetry at the level of the index by integrating over the corresponding fugacity, $z$, on both sides, which gives the $U(N)$ identity.  Including also a factor of $z^n$, one further derives the equality as a function of the $4d$ FI-term, which upon taking the $3d$ limit gives the identity here.}

Next, as in sections \pureduality\ and \undualities, we add real masses and take them to be large. Let us start with the duality with $N_f+1$ flavors and
take $m_{N_f+1}=s \rightarrow \infty$ on both sides.
We can study two different vacua after performing the real mass deformation, so we will list the partition functions in both.
In the expressions below we do not impose the constraint~\massConstr, but we emphasize that the duality only holds when it is satisfied.
Here ${\cal Z}_A^{(1)}$ corresponds to the vacuum with $\sigma_o=0$, and ${\cal Z}_A^{(2)}$ corresponds to the vacuum with $\vev{\sigma}={\rm diag}(0,\cdots,0,-m_{N_f+1})$:
\eqn\UNlimitEa{
\eqalign{
{\cal Z}_A^{(1)} \rightarrow & \exp \bigg(\frac{2\pi i}{\omega_1\omega_2} s N_c (-\omega + {\hat \mu}_{N_f+1})  \bigg) \times\cr
&\qquad\qquad\qquad \frac{1}{N_c!} \int
 \prod_{j=1}^{N_c} \frac{d\sigma_j}{\sqrt{-\omega_1\omega_2}} \frac{ \prod_{j=1}^{N_c} e^{ \frac{2\pi i}{\omega_1\omega_2}  \xi' \sigma_j}  \prod_{a=1}^{N_f} \Gamma_h( \pm
(\sigma_j + m_a) + {\hat \mu}_a) }{\prod_{i <j}^{N_c} \Gamma_h(\pm(\sigma_i - \sigma_j))}\,,\cr
{\cal Z}_A^{(2)} \rightarrow & \exp \bigg( \frac{2 \pi i}{\omega_1 \omega_2} \bigg( s (-N_f \omega + (N_c-1) {\hat \mu}_{N_f+1} + \sum_{a=1}^{N_f} {\hat \mu}_a) +
\sum_{a=1}^{N_f} m_a ( \omega - {\hat \mu}_a - \xi'')  \bigg) \bigg) \times  \cr
&\times \frac{1}{(N_c-1)!}\int  \prod_{j=1}^{N_c} \frac{d\sigma_j}{\sqrt{-\omega_1\omega_2}} \frac{d\hat{\sigma}}{\sqrt{-\omega_1\omega_2}}
\frac{ \prod_{j=1}^{N_c-1} e^{\frac{2\pi i}{\omega_1\omega_2}  \xi''
  \sigma_j } \prod_{a=1}^{N_f} \Gamma_h(\pm (\sigma_j +m_a) + {\hat \mu}_a) }{\prod_{i<j}^{N_c-1} \Gamma_h(\pm (\sigma_i - \sigma_j))}  \times  \cr
& \times e^{\frac{2\pi i}{\omega_1\omega_2} (\xi'' + \omega(N_f+1-N_c) - \sum_{a=1}^{N_f+1} {\hat \mu}_a) \hat{\sigma}} \Gamma_h(\pm \hat{\sigma} +{\hat \mu}_{N_f+1})\,,
}
}
where we have defined $\xi' =\xi''-\omega =  \xi -\omega + {\hat \mu}_{N_f+1}$.
For theory B, we have the additional contribution of the uncharged chiral fields ${M_a}^b$.  Their asymptotic behavior as $s\to \infty$ is independent of the VEV, and is given by:
\eqn\anothereq{\exp \bigg( \frac{2\pi i}{\omega_1 \omega_2} \sum_{a=1}^{N_f} (s - m_a) (-\omega + {\hat \mu}_{N_f+1} + {\hat \mu}_a) \bigg)  \bigg( \prod_{a,b}^{N_f} \Gamma_h( m_a + {\hat \mu}_a - m_b +
{\hat \mu}_b) \bigg) \Gamma_h(2 {\hat \mu}_{N_f+1} )\,.}
We have thus the following partition functions in the two different vacua:
\eqn\UNlimitMa{
\eqalign{
&{\cal Z}_B^{(1)} \rightarrow \exp \bigg(\frac{2\pi i}{\omega_1\omega_2} \bigg( s ( -N_f \omega + (N_c-1) {\hat \mu}_{N_f+1} + \sum_{a=1}^{N_f} {\hat \mu}_a )
+ \sum_{a=1}^{N_f} m_a (\omega -{\hat \mu}_a - \xi'') \bigg) \bigg) \times  \cr
& \qquad \qquad \times \bigg( \prod_{a,b}^{N_f} \Gamma_h( m_a + {\hat \mu}_a - m_b + {\hat \mu}_b) \bigg) \Gamma_h(2 {\hat \mu}_{N_f+1} )\,\frac{1}{(N_f+1-N_c)!} \times  \cr
& \qquad \qquad \int \prod_{\ell=1}^{N_f-N_c+1} \frac{d\ts_\ell}{\sqrt{-\omega_1\omega_2}} \frac{ \prod_{j=1}^{N_f+1-N_c} e^{ \frac{2\pi i}{\omega_1\omega_2}
 \xi'' \ts_j}  \prod_{a=1}^{N_f} \Gamma_h( \omega\pm (\ts_j - m_a) - {\hat \mu}_a) }{\prod_{i <j} \Gamma_h(\pm(\ts_i - \ts_j))}\,,\cr
&{\cal Z}_B^{(2)} \rightarrow \exp \bigg(\frac{2\pi i}{\omega_1\omega_2}
\bigg( s (-N_c \omega + N_c {\hat \mu}_{N_f+1})  - \sum_{a=1}^{N_f} \xi' m_a \bigg)\bigg( \prod_{a,b}^{N_f} \Gamma_h( m_a + {\hat \mu}_a - m_b + {\hat \mu}_b) \bigg)
\times  \cr
& \frac{\Gamma_h(2 {\hat \mu}_{N_f+1} )}{(N_f-N_c)!}\int \prod_{\ell=1}^{N_f-N_c} \frac{d\ts_\ell}{\sqrt{-\omega_1\omega_2}} \frac{d\hat{\ts}}{\sqrt{-\omega_1\omega_2}}  \frac{\prod_{j=1}^{N_f-N_c} e^{\frac{2\pi i}{\omega_1\omega_2}  \xi' \ts_j}
 \prod_{a=1}^{N_f} \Gamma_h(\omega \pm (\ts_j - m_a )- {\hat \mu}_a) }{\prod_{i<j}^{N_f-N_c} \Gamma_h(\pm (\ts_i - \ts_j))}  \times  \cr
 &\qquad e^{\frac{2\pi i}{\omega_1\omega_2} ( \xi' + \omega(N_f+1-N_c)- \sum_{a=1}^{N_f+1} {\hat \mu}_a ) \hat{\ts}} \Gamma_h(\omega \pm \hat{\ts} -{\hat \mu}_{N_f+1}) \,.
}}
Now the key point is to notice that the large $s$ scaling of the first vacuum of theory A, ${\cal Z}_A^{(1)}$
in \UNlimitEa, and the second vacuum of theory B, ${\cal Z}_B^{(2)}$ in \UNlimitMa, are the same, as well as for the other pair.
We can then conclude that  the two vacua map into each other under the duality, strip
off the divergent factors (the background CS terms involving symmetries
acting on the heavy field), and take the strict $s \rightarrow \infty$ limit, while imposing \massConstr. The matching of ${\cal Z}_A^{(1)}$ and ${\cal Z}_B^{(2)}$ gives
\eqn\UNdualitynoeta{
\eqalign{
&\frac{1}{N_c!} \int \prod_{\ell=1}^{N_c} \frac{d\sigma_\ell}{\sqrt{-\omega_1\omega_2}} \frac{ \prod_{j=1}^{N_c} e^{ 2 \pi i\xi \sigma_j}
  \prod_{a=1}^{N_f} \Gamma_h( \pm (\sigma_j + m_a) + {\hat \mu}_a) }{\prod_{i <j} \Gamma_h(\pm(\sigma_i - \sigma_j))} = \cr
& \qquad\bigg(\prod_{a,b} \Gamma_h( m_a + {\hat \mu}_a - m_b + {\hat \mu}_b) \bigg) \Gamma_h(2 (N_f+1-N_c) \omega  - 2 \sum_{a=1}^{N_f} {\hat \mu}_a) \times \cr
&\qquad \frac{1}{(N_f-N_c)!}\int \prod_{\ell=1}^{N_f-N_c} \frac{d\ts_\ell}{\sqrt{-\omega_1\omega_2}}
  \frac{\prod_{j=1}^{N_f-N_c} e^{2 \pi i \xi \ts_j} \prod_{a=1}^{N_f} \Gamma_h(\omega \pm (\ts_j - m_a) - {\hat \mu}_a) }
{\prod_{i<j}^{N_f-N_c} \Gamma_h(\pm (\ts_i - \ts_j))} \times \cr
 &\qquad \int\frac{d\hat{\ts}}{\sqrt{-\omega_1\omega_2}}\, e^{2 \pi i \xi \hat{\ts}} \Gamma_h(- (N_f-N_c) \omega  \pm \hat{\ts} + \sum_{a=1}^{N_f} {\hat \mu}_a ) \,.
}}
This follows from the statement that the theories at the corresponding points in the moduli space are dual.
In theory A we have a partition function of the $U(N_c)$ gauge theory with $N_f$ flavors without any superpotential.
In theory B we have a partition function of a $U(N_f-N_c)\times  U(1)$ gauge theory. The matter content is $N_f$
 flavors in the fundamental representation of $U(N_f-N_c)$ and uncharged under the $U(1)$; a single flavor charged under the $U(1)$ and uncharged under $U(N_f-N_c)$; and $N_f^2+1$
singlet fields. The FI-terms of the two gauge group factors are the same, which is
consistent with the presence of the superpotential \neweqw\ involving the monopole operators of the two factors.

Note that there is no longer any constraint \massConstr\ on the real mass parameters of the theory.  This is because the original constraint involved the parameter ${\hat \mu}_{N_f+1}$, which does not appear in this theory, and so the remaining parameters are unconstrained. This implies that one does not have the $\eta$-dependent superpotential in the IR.

One could also focus on the second vacuum of theory A, which will be dual to the first vacuum of theory B. However, the duality obtained in this way will be essentially identical to the above.

So far we have found in theory B a $U(N_f-N_c)\times U(1)$ gauge theory. In the previous sections we saw how we can turn this into a $U(N_f-N_c)$ theory, by taking into account the dynamics of the $U(1)$ theory with a single flavor. Let us see how to do this for the partition function.
Consider the following piece of the partition function of theory B in~\UNdualitynoeta
\eqn\XYZpart{
 {\cal Z}_{U(1)}=\Gamma_h(2 (N_f+1-N_c) \omega  - 2 \sum_{a=1}^{N_f} {\hat \mu}_a) \int\frac{d\hat{\ts}}{\sqrt{-\omega_1\omega_2}}\,
 e^{2 \pi i \xi \hat{\ts}} \Gamma_h(- (N_f-N_c) \omega  \pm \hat{\ts} + \sum_{a=1}^{N_f} {\hat \mu}_a )\,.
}
In addition to an uncharged chiral superfield, this represents the partition function of a  $U(1)$ gauge theory with one flavor, which is mirror dual to the theory of three chiral multiplets with $W=XYZ$ \AharonyBX.
 Mirror symmetry implies the following identity of partition functions
\eqn\XYZident{
 \int \frac{d\lambda}{\sqrt{-\omega_1\omega_2}} e^{\frac{2 \pi i}{\omega_1 \omega_2} \xi \lambda} \Gamma_h( \pm \lambda + {\hat \mu})  = \Gamma_h(2 \mu) \Gamma_h(\omega \pm \xi - {\hat \mu})\,. }
Applying this here, we can rewrite \XYZpart\ as:
\eqn\mirrorresult{
\eqalign{
&{\cal Z}_{U(1)}=
\Gamma_h(2 (N_f+1-N_c) \omega  - 2 \sum_{a=1}^{N_f} {\hat \mu}_a)\times\cr
&\qquad\qquad  \Gamma_h(-2 (N_f-N_c) \omega  + 2 \sum_{a=1}^{N_f} {\hat \mu}_a )\; \Gamma_h( (N_f+1-N_c) \omega \pm \xi -\sum_{a=1}^{N_f} {\hat \mu}_a )\,.
} }
The first two terms cancel due to~\massChiT.
Substituting the third and fourth factors of \mirrorresult\ into \UNdualitynoeta, we arrive at
\eqn\UNdualitynoetaaha{
\eqalign{
&\frac{1}{N_c!} \int \prod_{j=1}^{N_c} \frac{d\sigma_j}{\sqrt{-\omega_1\omega_2}} \frac{ \prod_{j=1}^{N_c} e^{ 2 \pi i\xi \sigma_j}
  \prod_{a=1}^{N_f} \Gamma_h( \pm (\sigma_j +m_a) + {\hat \mu}_a) }{\prod_{i <j} \Gamma_h(\pm(\sigma_i - \sigma_j))} = \cr
& \qquad\qquad\bigg(\prod_{a,b} \Gamma_h( m_a + {\hat \mu}_a - m_b + {\hat \mu}_b) \bigg) \Gamma_h( (N_f+1-N_c) \omega \pm \xi -\sum_{a=1}^{N_f} {\hat \mu}_a )  \times \cr
&\frac{1}{(N_f-N_c)!}\int \prod_{j=1}^{N_f-N_c} \frac{d\ts_j}{\sqrt{-\omega_1\omega_2}}
  \frac{\prod_{j=1}^{N_f-N_c} e^{2 \pi i \xi \ts_j} \prod_{a=1}^{N_f} \Gamma_h(\omega \pm (\ts_j -m_a) - {\hat \mu}_a) }
{\prod_{i<j}^{N_f-N_c} \Gamma_h(\pm (\ts_i - \ts_j))}\,.
}}
This is precisely the duality of \AharonyGP, where theory B contains additional singlet fields $V_\pm$, which are charged under the $U(1)_J$ symmetry of the theory: their charge under $U(1)_J$ is given by the coefficient of $\xi$ and is given by $\pm1$, and their R-charge is the coefficient of $\omega$ and is given by $N_f-N_c+1$, as expected.  Turning on an $\eta$ superpotential and thus requiring~\massConstr, the fields $V_\pm$ become massive and disappear from the theory, and we are back to the starting point~\UNmag, as discussed in section \undualities.  Note that the identities~\UNdualitynoetaaha\ and \XYZident, which were proven in~\debult, imply that~\UNdualitynoeta\ is correct.

\subsec{$SU(N)$ duality}
\subseclab\indexsun

Let us now study the duality involving $SU(N)$ SQCD.
Here theory A, obtained by reducing the $4d$ SQCD on a circle, is $SU(N_c)$ with $N_f$
fundamental flavors, with the superpotential $W_A = \eta Y$.  The partition function is given by:
\eqn\SUpartE{ {\cal Z}_A = \frac{1}{N_c!} \int \prod_{j=1}^{N_c} \frac{d\sigma_j}{\sqrt{-\omega_1\omega_2}} \delta \left(\sum_{j=1}^{N_c} \sigma_j\right)
 \frac{ \prod_{j=1}^{N_c} \prod_{a=1}^{N_f}
\Gamma_h(  {\hat \mu}_a \pm (\sigma_j  +m_a+ \frac{\beta}{N_c}))}{\prod_{i<j}^{N_c} \Gamma_h(\pm(\sigma_i - \sigma_j)) }\,.
}
Here $\beta$ is the fugacity for the $U(1)_B$ baryonic symmetry, normalized so that the baryon has charge one.
The superpotential imposes a constraint on the real masses,
\eqn\massConstrSU{
\sum_{a=1}^{N_f}  {\hat \mu}_a = \omega (N_f-N_c)\,.
}
Theory B is an $SU(N_f-N_c)$ gauge theory with $N_f$ flavors and ${N_f}^2$ chiral multiplets ${M_a}^b$, with
 superpotential $W_B=\tilde{\eta}\, \tilde{Y} + \sum_{a,b} \tilde{q}_a\, {M^a}_b\, q^b$.
 The partition function, with real mass parameters mapped appropriately from theory A, is
\eqn\SUpartM{
\eqalign{
 {\cal Z}_B =& \prod_{a,b} \Gamma_h( m_a + {\hat \mu}_a - m_b + {\hat \mu}_b)
\;\frac{1}{(N_f-N_c)!}\times \cr
&
\int \prod_{\ell=1}^{N_f-N_c} \frac{d\ts_\ell}{\sqrt{-\omega_1\omega_2}} \delta\left(\sum_{j=1}^{N_f-N_c} \ts_j\right)
 \frac{ \prod_{j=1}^{N_f-N_c}\prod_{a=1}^{N_f} \Gamma_h( \omega
- {\hat \mu}_a \mp (m_a-\ts_j- \frac{\beta}{N_f-N_c}))}{\prod_{i<j}^{N_f-N_c} \Gamma_h(\pm(\ts_i - \ts_j)) }\,.
}
}
The partition functions ${\cal Z}_A$ and ${\cal Z}_B$ are equal as a function of the real mass parameters subject to the constraint \massConstrSU.
This equality again follows from the equality of the corresponding supersymmetric indices in four dimensions.

We now wish to  remove the $\eta$ superpotentials by turning on real masses, as discussed in section \pureduality. The analysis is very similar to the $U(N)$ case, with vacua that are exchanged by the duality, although some of the details
are more complicated because of the constraint $\sum_j \sigma_j=0$.  In both cases we will start with the theory with $N_f+1$ flavors and take the following limit of the mass parameters (as in section \pureduality):
\eqn\limitSU{ m_a \rightarrow m_a - \frac{1}{N_f+1}\, s, \qquad
 m_{N_f+1} \rightarrow  \frac{N_f}{N_f+1}\, s,\qquad
 \beta \rightarrow \beta + \frac{N_c}{N_f+1}\, s\,.}
Here $a$ runs over the first $N_f$ indices.
As discussed in section \pureduality, unlike in the $U(N)$ case, in theory A there is now only a single vacuum near $\vev{\sigma} = 0$, and in theory B there is one at $\vev{\ts} = -s \cdot {\rm diag}(\frac{1}{N_f+1-N_c},\cdots,\frac{1}{N_f+1-N_c},\frac{N_c-N_f}{N_f+1-N_c})$. The corresponding partition functions
${\cal Z}_A$ and ${\cal Z}_B$ are given by\foot{Here in ${\cal Z}_B$
 we have also made a convenient  finite
 shift of $\ts_j \rightarrow \ts_j + \frac{\beta}{(N_f-N_c)(N_f+1-N_c)}$, $j=1,\cdots,N_f-N_c$,
 and $\ts_{N_f+1-N_c} \rightarrow \ts_{N_f+1-N_c} - \frac{\beta}{N_f+1-N_c}$, which has the
 effect of properly normalizing the $\beta$ term in the partition function.}
\eqn\SUpartnoeta{
\eqalign{
&{\cal Z}_A\rightarrow \exp \bigg( \frac{2 \pi i}{\omega_1 \omega_2} (-\omega + {\hat \mu}_{N_f+1})(s + \frac{\beta}{N_c}) \bigg)\times \cr
 & \qquad \frac{1}{N_c!} \int \prod_{j=1}^{N_c} \frac{d\sigma_j}{\sqrt{-\omega_1\omega_2}}\,\delta \left(\sum_{j=1}^{N_c} \sigma_j\right)\,
 \frac{\prod_{j=1}^{N_c} \prod_{a=1}^{N_f} \Gamma_h(
 {\hat \mu}_a \pm (\sigma_j + m_a+\frac{\beta}{N_c}))}{\prod_{i<j} \Gamma_h(\pm (\sigma_i - \sigma_j))}\, }}
\eqn\SUpartnoetb{
\eqalign{&{\cal Z}_B \rightarrow
\exp \bigg( \frac{2\pi i}{\omega_1 \omega_2}  (-\omega + {\hat \mu}_{N_f+1})(s + \frac{\beta}{N_c}) \bigg) \bigg( \prod_{a,b}^{N_f} \Gamma_h( m_a + {\hat \mu}_a -m_b + {\hat \mu}_b) \bigg)
 \,\frac{1}{(N_f-N_c)!}\times  \cr
& \qquad \int \prod_{j=1}^{N_f-N_c} \frac{d\ts_j}{\sqrt{-\omega_1\omega_2}}\,\prod_{j=1}^{N_f-N_c} e^{\frac{2\pi i}{\omega_1 \omega_2} (\sum_{a=1}^{N_f+1} {\hat \mu}_a - (N_f+1-N_c) \omega) \ts_j}\,\times\cr
&\qquad \Gamma_h(\omega -{\hat \mu}_{N_f+1} \mp \sum_j \ts_j)\,\Gamma_h(2 {\hat \mu}_{N_f+1})\frac{
 \prod_{j=1}^{N_f-N_c}\prod_{a=1}^{N_f} \Gamma_h(\omega  - {\hat \mu}_a\pm (\ts_j - m_a + \frac{\beta}{N_f-N_c}))}{\prod_{i<j} \Gamma_h(\pm (\ts_i - \ts_j))}\,.
}
}
In theory A we find that the first $N_f$ flavors remain light and the theory is $SU(N_c)$ SQCD with $N_f$ flavors. In theory B
the choice of $\ts_o$ breaks the gauge group from $SU(N_f+1-N_c)$ to $U(N_f-N_c)$.
 It also leaves $N_f$ light fundamental flavors of $U(N_f-N_c)$, and one light flavor charged under the overall
 $U(1)$ of the gauge group.  Finally, theory B also contains mesons ${M_a}^b$, and just as in the $U(N)$ case,
 ${N_f}^2+1$ of these remain light.
We must also impose \massConstrSU, which has the effect of removing the FI-term.
We see that ${\cal Z}_A$ and ${\cal Z}_B$  have the same scaling with $s$,
so as before we can remove the prefactor and take the strict $s \rightarrow \infty$ limit to obtain the identity:
\eqn\SUnoeta{
\eqalign{
&\frac{1}{N_c!} \int \prod_{j=1}^{N_c} \frac{d\sigma_j}{\sqrt{-\omega_1\omega_2}}\,\delta \left(\sum_{j=1}^{N_c} \sigma_j\right)\,
 \frac{\prod_{j=1}^{N_c} \prod_{a=1}^{N_f} \Gamma_h(
 {\hat \mu}_a \pm (\sigma_j + m_a+ \frac{\beta}{N_c}))}{\prod_{i<j} \Gamma_h(\pm (\sigma_i - \sigma_j))}= \cr
&\qquad \bigg( \prod_{a,b}^{N_f} \Gamma_h( m_a + {\hat \mu}_a -m_b + {\hat \mu}_b) \bigg) \Gamma_h(2 \omega (N_f+1-N_c) - 2 \sum_{a=1}^{N_f} {\hat \mu}_a ) \times  \cr
&\qquad \frac{1}{(N_f-N_c)!}  \int \prod_{j=1}^{N_f-N_c} \frac{d\ts_j}{\sqrt{-\omega_1\omega_2}}
 \frac{
 \prod_{a=1}^{N_f} \Gamma_h(\omega  - {\hat \mu}_a\pm (\ts_j - m_a + \frac{\beta}{N_f-N_c}))}{\prod_{i<j} \Gamma_h(\pm (\ts_i - \ts_j))} \times   \cr
&\qquad\Gamma_h( -\omega (N_f+1-N_c) \pm \sum_j \ts_j + \sum_{a=1}^{N_f} {\hat \mu}_{a} )\,.}}
This is the partition function of the duality we discussed in section \pureduality. In particular the two hyperbolic Gamma functions appearing in the last line above are the partition functions of the chiral fields $b$ and $\tilde b$ discussed there. The last hyperbolic Gamma function in the second line corresponds to the singlet $Y$.

As discussed in section \gaugetopo, the same duality can be obtained from the $U(N)$ duality of the previous section by gauging the topological $U(1)_J$ symmetry (see also~\KapustinSim). We can see this immediately at the level
of the partition functions. Gauging $U(1)_J$ in~\UNdualitynoeta\ amounts to integrating both sides of the identity with $\int \frac{d\xi}{\sqrt{-\omega_1\omega_2}}$.
In theory A this has the effect of imposing the $SU(N)$ constraint $\sum \sigma_j = 0$,
giving us the left-hand side of~\SUnoeta. In theory B one obtains the theory with gauge group $U(N_f-N_c)\times U(1)$ that we found in ~\UNdualitynoeta, and an application of mirror symmetry to the $U(1)$ factor, using \XYZident\ as above, gives the right-hand side of~\SUnoeta. In particular that proves \SUnoeta\ mathematically, since we argued
above that~\UNdualitynoeta\ is a known identity.
Conversely, by gauging the baryonic $U(1)$ we can obtain the partition functions of the $U(N)$
duality from the $SU(N)$ dualities.
 One can also consider the supersymmetric index (see appendix~A) of the $U(N)$ duality and gauge the $U(1)_J$ at the level
of the index to obtain the index of this $SU(N)$ duality. For this index computation one has to consider the generalized index of~\KapustinJM.

One can test also dualities with Chern-Simons terms using the partition function, by turning on appropriate real masses in the expressions above. In the Chern-Simons cases the partition functions match across dualities up to certain contact terms, which will be
automatically generated by the real masses.

\newsec{Dualities for $USp(2N_c)$ theories}

The dualities of $USp(2N_c)$ gauge theories, both in $4d$ \refs{\SeibergPQ,\IntriligatorNE} and in $3d$ \refs{\KarchUX,\AharonyGP},
are perhaps the simplest ones, due to the absence of baryons in this case. The analysis of the relation between these dualities is similar
to our analysis in section \susection, but there is one new feature. While for $SU(N)$ theories the dual gauge group was the same (up to possible $U(1)$ factors)
in four and in three dimensions, for $USp$ and $SO$ groups there is a shift in the rank of the dual group. This can easily be seen from the brane construction
realization of the dualities (where it is related to the different charges for orientifold planes of different dimensions).  Here we will show how it appears when we derive the $3d$ duality from $4d$, as in our general analysis above.

In four dimensions, the $USp(2N_c)$ SQCD theory (of rank $N_c$) with $2N_f$ chiral multiplets $Q_i$ in the fundamental representation (this number must be even), which we will call theory A, is dual at low energies to a $USp(2N_f-2N_c-4)$ SQCD theory, with $2N_f$ fundamentals $q^i$, $N_f(2N_f-1)$ extra singlet fields $M$, and a superpotential $W = M_{ij} q^i q^j$ ($i,j=1,\cdots,2N_f$), which we will call theory B. When these theories are put on a circle, an extra superpotential $W_A = \eta Y$ ($W_B = {\tilde \eta} {\tilde Y}$ in theory B with $\tilde \eta \propto 1 / \eta$) is generated by instantons wrapping the circle, similar to the $SU(N)$ case discussed above. Here $Y$ ($\tilde Y$) is the natural Coulomb branch coordinate in the $3d$ $USp(2N_c)$ theory \refs{\KarchUX,\AharonyGP}, semi-classically going as
\eqn\spy{Y \simeq \exp({2 \sigma_1\over {\hat g}_3^2}+2ia_1)~,}
where we fix the Weyl symmetry by choosing the eigenvalues of the adjoint matrix $\sigma$ to obey
\eqn\spncoulomb{\sigma_1 \geq \sigma_2 \geq \cdots \geq \sigma_{N_c} \geq -\sigma_{N_c} \geq \cdots \geq -\sigma_1.}
$Y$ labels the part of the Coulomb branch that is not lifted by Affleck-Harvey-Witten instantons. As in our previous discussions, it arises as the low-energy limit of the monopole operator of minimal charge.

Our first claim is then that the $3d$ $USp(2N_c)$ SQCD theory with $2N_f$ fundamentals and $W_A = \eta Y$ is dual at low energies to the $3d$ $USp(2N_f-2N_c-4)$ theory with $2N_f$ fundamentals and $W_B = M q q + {\tilde \eta} {\tilde Y}$. The superpotential breaks precisely the $U(1)$ global symmetry that is anomalous in the $4d$ theory, so the matching of the $\S^3$ partition functions on both sides is again guaranteed by the $4d$ duality (and can be explicitly verified).

As in the $SU(N)$ case, the superpotential lifts the Coulomb branch on both sides. To compare the moduli spaces, we recall that the $3d$ SQCD theory with $N_f=N_c$ has a quantum modified moduli space $Y {\rm Pf}(M) = 1$ (where $M_{ij} = Q_i Q_j$), while the theory with $N_f=N_c+1$ has an effective superpotential $W = - Y {\rm Pf}(M)$ \KarchUX. In theory A the Coulomb branch is lifted, and we have (as in $4d$) the classical Higgs branch, where the rank of $M$ goes up to $2N_c$. In theory B we cannot give a VEV to $qq$ because of the superpotential. Suppose that we try to give in theory B a VEV to $M$ of rank $2(N_c+1)$. This reduces the number of massless fundamentals to $2(N_f-N_c-1)$, which means that the effective description of the remaining light fields in theory B becomes
\eqn\weffbsp{W = - {\tilde Y} {\rm Pf}(q q) + M q q + {\tilde \eta} {\tilde Y}.}
The F-term equations of ${\tilde Y}$ and $M$ cannot be satisfied at the same
time, so the theory with this VEV has no supersymmetric vacuum, in agreement with theory A. As in section \susection, we can easily compare also the complex mass deformations of the two sides.

In order to find a duality for the standard SQCD theory without $\eta$, we can follow the same route as in section \pureduality, starting in theory A from the $USp(2N_c)$ theory with $2(N_f+1)$ fundamentals, and turning on real masses $\pm {\hat m}$ (of opposite signs) for two of the fundamentals. This is a background field in the $SU(2(N_f+1))$ global symmetry group, which maps to a similar mass term also in theory B. The analysis in theory A is quite similar to the $SU(N)$ case of section \susection. There is a supersymmetric vacuum at the origin of the Coulomb branch, and in the low-energy theory the superpotential $W = \eta Y_{high}$ vanishes, so we get the standard SQCD theory.

Unlike in the $SU(N)$ case, here theory B (whose gauge group is $USp(2N_f-2N_c-2)$)
also has a vacuum at the origin of its moduli space. One may think that both theories could also have extra supersymmetric vacua, where we turn on an eigenvalue $\sigma_{1} = {\hat m}$ and connect to the Higgs branch of the flavor that got a real mass. In theory A near this vacuum we break $USp(2N_c) \to USp(2N_c-2)\times U(1)$, and we could have an extra Higgs branch labeled by $M_{2N_f+1,2N_f+2}$. But the effective $U(1)$ theory with one flavor that describes this extra eigenvalue would have a superpotential $W = \eta V_+$ from the high-energy superpotential, in addition to the usual term $W = -V_+ V_- M_{2N_f+1,2N_f+2}$ coming from the dynamics of the $U(1)$ theory with one flavor. This means that in this theory we would have a constraint $V_- M_{2N_f+1,2N_f+2} \sim \eta$ that tells us that when we turn on $M_{2N_f+1,2N_f+2}$ we must go in the Coulomb branch in the direction of decreasing $\sigma_1$, such that this branch can precisely be identified with the Coulomb branch of the $USp(2N_c)$ theory near the origin (with the extra meson $M_{2N_f+1,2N_f+2}$ proportional to its $Y$ coordinate). In theory B we can perform a similar analysis, but we have an extra $M q q$ superpotential that leads to no solutions to the resulting F-term equations. Thus, in both theories the only supersymmetric vacua are at the origin of the moduli space (and the branches emanating from it).

When we analyze theory B near the origin of its moduli space, in addition to the $2N_f$ fundamentals which got no real mass, we still have also a singlet meson $M_{2N_f+1,2N_f+2}$. Giving an expectation value to this field gives a complex mass to the extra flavor, so that the relation between the high-energy and low-energy ${\tilde Y}$ fields is (as in the $SU(N)$ case discussed in section \background) ${\tilde Y}_{low} = {\tilde Y}_{high} / M_{2N_f+1,2N_f+2}$. This means that at low energies we are still left in this theory with an extra singlet field, and with a superpotential $W = {\tilde \eta} M_{2N_f+1,2N_f+2} {\tilde Y}_{low}$. Renaming the extra singlet field $Y \equiv {\tilde \eta} M_{2N_f+1,2N_f+2}$, the duality that we obtain is then precisely the same as the $USp$ duality suggested in \AharonyGP, and we identify the Coulomb branch of theory A with turning on the extra singlet field in theory B (consistently with the discussion of the previous paragraph).

Thus, as in the $U(N_c)$ case, we can derive the known $3d$ duality in this case, mapping the $USp(2N_c)$ theory to a $USp(2N_f-2N_c-2)$ theory, from $4d$.
As discussed already in \AharonyGP, deforming this $3d$ duality by $W = \eta Y$ requires (together with the $W = Y {\tilde Y}$ superpotential) going on the Coulomb branch of theory B (setting ${\tilde Y} = -\eta$) so that the gauge group is broken to $USp(2N_f-2N_c-4)\times U(1)$. Without the superpotential we could dualize the $U(1)$ vector multiplet on this Coulomb branch to the chiral superfield ${\tilde Y}$, but the superpotential gives ${\tilde Y}$ a mass so we can integrate it out.
  However, if we try to move on the Coulomb branch of the remaining $USp(2N_f-2N_c-4)$ theory, there would now be an Affleck-Harvey-Witten superpotential proportional to ${\tilde Y}$ coming from the monopole of $USp(2N_f-2N_c-2)$ that would prevent this. Thus, the low-energy theory we get in theory B is precisely the dual of the theory with $\eta$ that we discussed above.

As in the previous sections, we can also flow to $USp(2N_c)$ theories with Chern-Simons couplings. The simplest way to do this is by starting from the $USp(2N_c)$ theory with $2(N_f+k)$ fundamentals ($k > 0$) and turning on positive real masses for $2k$ of them, inducing at low energies a Chern-Simons level $k$. The real masses are now background fields both in the $SU(2N_f)$ global symmetry and in $U(1)_A$, so in theory B the field $Y$ also receives a real mass (together with $2k$ fundamentals). The dual theory is now a $Up(2N_f+2k-2N_c-2)$ theory at level $(-k)$, still with a $W = M q q$ superpotential, so that we reproduce the known duality of \WillettGP.

\subsec{The partition function}

Let us discuss the partition functions in this case. Reducing the SQCD theory from $4d$,
 theory A is $USp(2N_c)$ with $2N_f$ fundamental chiral multiplets.  The $3d$ partition function is
\eqn\USPetaE{
 {\cal Z}_A=\frac{1}{2^{N_c} N_c!} \int \prod_{j=1}^{N_c} \frac{d\sigma_j}{\sqrt{-\omega_1\omega_2}}\frac{\prod_{j=1}^{N_c} \prod_{a=1}^{2 N_f}
 \Gamma_h(\pm \sigma_j + m_a)}{\prod_{i<j}^{N_c} \Gamma_h(\pm(\sigma_i + \sigma_j)) \Gamma_h(\pm (\sigma_i - \sigma_j)) \prod_{j=1}^{N_c} \Gamma_h(\pm 2 \sigma_j)}\,.
}
Here and in what follows we have absorbed the R-charges into the real masses (as in section \partexample).
There is also a superpotential $W_A = \eta\, Y$, which imposes
\eqn\massConstrSP{
\sum_{a=1}^{2 N_f} m_a = 2\omega( N_f- N_c-1) \,.
}
Theory B has gauge group $USp(2(N_f-N_c-2))$ with $2N_f$
fundamental chiral multiplets, $N_f(2N_f-1)$ uncharged chiral multiplets $M^{ab}$, and a superpotential \eqn\forwb{W_B=\sum_{a<b} M^{ab} q_a q_b + \tilde{\eta} \tilde{Y}.}
The partition function is given by
\eqn\USPetaM{
\eqalign{
& {\cal Z}_B=\frac{1}{2^{N_f-N_c-2} (N_f-N_c-2)!}\prod_{a<b} \Gamma_h(m_a+m_b)  \times\cr
&\qquad  \int \prod_{j=1}^{N_f-N_c-2} \frac{d\ts_j}{\sqrt{-\omega_1\omega_2}} \frac{\prod_{j=1}^{N_f-N_c-2} \prod_{a=1}^{2 N_f}
\Gamma_h(\pm \ts_j + m_a)}{\prod_{i<j}^{N_f-N_c-2} \Gamma_h(\pm(\ts_i + \ts_j)) \Gamma_h(\pm (\ts_i - \ts_j)) \prod_{j=1}^{N_f-N_c-2} \Gamma_h(\pm 2 \ts_j)}\,.
}
}
To remove the $\eta$ superpotentials we start with the theory with $2N_f+2$ fundamentals, and take a large real mass for a pair of them.  Let us set
\eqn\defUSP{ m_{2N_f+1} = s + \alpha, \qquad\qquad m_{2N_f+2} = -s + \alpha\,.}
Here $\alpha$ parameterizes a combination of the two masses that is kept finite in the limit.
As we argued above, the only supersymmetric vacuum is  at the origin of moduli space.
When we take the vacuum $\vev{\sigma}=0$ on both sides of the duality, we obtain the following partition functions
\eqn\PSUnoeta{
\eqalign{
 &{\cal Z}^{(1)}_A \rightarrow \exp \bigg( \frac{2\pi i}{\omega_1 \omega_2}2 N_c s (-\omega + \alpha) \bigg) \times \frac{1}{2^{N_c} N_c!}\,\times\cr
 &\qquad \qquad \int \prod_{j=1}^{N_c} \frac{d\sigma_j}{\sqrt{-\omega_1\omega_2}} \frac{\prod_{j=1}^{N_c} \prod_{a=1}^{2 N_f} \Gamma_h(\pm \sigma_j + m_a)}{\prod_{i<j}^{N_c}
 \Gamma_h(\pm(\sigma_i + \sigma_j)) \Gamma_h(\pm (\sigma_i - \sigma_j)) \prod_{j=1}^{N_c} \Gamma_h(\pm 2 \sigma_j)}\,,\cr
&{\cal Z}^{(1)}_B \rightarrow \exp \bigg( \frac{2\pi i}{\omega_1 \omega_2}2 N_c s (-\omega + \alpha) \bigg) \bigg( \prod_{a<b} \Gamma_h(m_a + m_b) \bigg) \Gamma_h(2 \alpha)\,
\frac{1}{2^{N_f-N_c-1} (N_f-N_c-1)!} \times \cr
& \qquad \qquad \int \prod_{j=1}^{N_f-N_c-1} \frac{d\ts_j}{\sqrt{-\omega_1\omega_2}}
 \frac{\prod_j \prod_{a=1}^{2 N_f} \Gamma_h(\pm \ts_j + m_a)}{\prod_{i<j}^{N_c} \Gamma_h(\pm(\ts_i + \ts_j)) \Gamma_h(\pm (\ts_i - \ts_j)) \prod_{j=1}^{N_c} \Gamma_h(\pm 2 \ts_j)\,.
}
}
}
The two partition functions have the same scaling with $s$.
Removing the $s$-dependent factor, using~\massConstrSP\ to solve for $\alpha$,
and equating the expressions, we find the following identity
\eqn\SPnoeta{
\eqalign{
 &\frac{1}{2^{N_c} N_c!}
\int \prod_{j=1}^{N_c} \frac{d\sigma_j}{\sqrt{-\omega_1\omega_2}} \frac{\prod_{j=1}^{N_c} \prod_{a=1}^{2 N_f} \Gamma_h(\pm \sigma_j + m_a)}{\prod_{i<j}^{N_c}
 \Gamma_h(\pm(\sigma_i + \sigma_j)) \Gamma_h(\pm (\sigma_i - \sigma_j)) \prod_{j=1}^{N_c} \Gamma_h(\pm 2 \sigma_j)}
=\cr
 &\qquad   \bigg( \prod_{a<b} \Gamma_h(m_a + m_b) \bigg) \Gamma_h(-2\omega(N_f-N_c-1) +2 \sum_{a=1}^{N_f} m_a)\,\frac{1}{2^{N_f-N_c-1} (N_f-N_c-1)!}  \times  \cr
 &\qquad
\int \prod_{j=1}^{N_f-N_c-1} \frac{d\ts_j}{\sqrt{-\omega_1\omega_2}}
 \frac{\prod_{j=1}^{N_f-N_c-1} \prod_{a=1}^{2 N_f} \Gamma_h(\pm \ts_j + m_a)}{\prod_{i<j}^{N_f-N_c-1} \Gamma_h(\pm(\ts_i + \ts_j)) \Gamma_h(\pm (\ts_i - \ts_j)) \prod_{j=1}^{N_f-N_c-1} \Gamma_h(\pm 2 \ts_j)
}\,.
}
}
This identity is consistent with the duality we discussed above:
the $USp(2N_c)$ theory with $2N_f$ chiral multiplets and no superpotential is dual to the $USp(2(N_f-N_c-1))$ theory with $2N_f$ fundamental
chiral multiplets, singlet mesons $M_{ab}$, a singlet chiral superfield $Y$, and the  superpotential
 \eqn\forwbn{W_B = Y \tilde{Y}+\sum_{a<b}M_{ab}q^a\,q^b.}
This is precisely the duality of \AharonyGP.
The last hyperbolic Gamma function in the second line in~\SPnoeta\ is the partition function of $Y$, and the other Gamma functions in that line are partition functions of the mesons. Turning on now the superpotential $W = \eta Y$, namely restricting the charges as in~\massConstrSP,
the contribution of  $Y$ vanishes in theory B due to the fact that $\Gamma_h(2\omega)=0$. However, one finds that the gauge integral diverges for this choice of parameters, giving in the end~\USPetaM. This is the  Higgs mechanism discussed above, as seen by the partition function in this particular case.

\vskip 1cm

\noindent {\bf Acknowledgments:}

We would like to thank T.~Dimofte, T.~Dumitrescu, G.~Festuccia,  K.~Intriligator, I.~Klebanov, Z.~Komargodski, D.~Kutasov, L.~Rastelli, B.~Safdi, E.~Witten, and I.~Yaakov for useful discussions. OA is the Samuel Sebba Professorial Chair of Pure and Applied Physics, and he is supported in part by a grant from the Rosa and Emilio Segre Research Award, by an Israel Science Foundation center for excellence grant, by the German-Israeli Foundation (GIF) for Scientific Research and Development, by the Minerva foundation with funding from the Federal German Ministry for Education and Research, and by the I-CORE program of the Planning and Budgeting Committee and the Israel Science Foundation (grant number 1937/12). OA gratefully acknowledges support from an IBM Einstein Fellowship at the Institute for Advanced Study.
 SSR gratefully acknowledges support from the Martin~A.~Chooljian and Helen Chooljian membership
 at the Institute for Advanced Study. The research of SSR was also partially supported by
NSF grant number PHY-0969448. The work of NS was supported in part by DOE grant DE-FG02-90ER40542 and by the United States-Israel Binational Science Foundation (BSF) under grant number~2010/629. The research of BW was supported in part by DOE Grant DE-FG02-90ER40542.

\appendix{A}{The supersymmetric index in $3d$}

In this appendix we briefly discuss the supersymmetric index in $3d$, namely the supersymmetric partition function on $\S^2\times \S^1$.
The $3d$ index is defined
as~\BhattacharyaZY
\eqn\indexDefthree{
\II(x;\,\{u_a\})=\Tr \left[(-1)^{2J_3}\,x^{\Delta+J_3}\,\prod_{a}u_a^{\mu_a}\right]\,.
}
Here $\Delta$ is the energy of the state on $\S^2$ times the radius of the $\S^2$ (for a superconformal theory this is the same as the conformal dimension), $J_3$ is the Cartan generator of the $SO(3)$ isometry of $\S^2$, and $\mu_a$ are charges under
global symmetries (except the R-symmetry). The states that contribute to this index satisfy $\Delta-R-J_3=0$,
where $R$ is the R-charge.

A prescription for computing the index in $3d$ was
given in \refs{\KimWB,\ImamuraSU,\KapustinJM}.  There, the factor $(-1)^{2J_3}$ is written as $(-1)^F$, with $F$ interpreted as the ``naive'' fermion number that can be
read from the fields appearing in the action. However, in the presence of magnetic monopoles (fluxes of the gauge field on $\S^2$) the spin of states can be shifted by a half-integer amount\foot{See~\DimoftePY\
for a recent discussion of this issue in the context of the $3d$ index. See also~\refs{\BeemMB,\HwangJH,\KrattenthalerDA}. For sign subtleties in the case of a $3d$ partition function on a Lens space see~\ImamuraRQ.}
\eqn\FtoJ{
(-1)^{2J_3}=(-1)^{F+e \cdot m}\,,}
where the vector $e$ is the vector of electric charges of a state, and $m$ is the vector of magnetic GNO charges of the monopole background.
The correct definition of the index is as stated in~\indexDefthree. In particular~\indexDefthree\  is
the same  for two dual theories, while in general the index with $(-1)^F$ is not. We will comment on this in what follows.

Taking~\FtoJ\ into account, the $3d$ index
in the presence of magnetic fluxes $n_a$ on $\S^2$ for background fields coupled to the flavor symmetries is
\eqn\threeDindexCorrect{
\eqalign{
&{\cal I}_{\{n_a\}}(x;\,\{(-1)^{n_a}u_a\}) =\sum_{\{m_1,\dots,m_{r_G}\}}\frac{1}{|W|_{\{m_\ell\}}}
(-1)^{\sum_{\ell=1}^{r_G} k^{\ell\ell}m^2_\ell+\sum_{a=1}^{r_F} k^{aa}n^2_a}\,\times\cr
 &\qquad
(-1)^{\sum_\Phi\left (\sum_{\ell=1}^{r_G} \nu^\ell(\Phi)m_\ell+\sum_{a=1}^{r_F} \mu^a(\Phi)n_a\right)\left|\sum_{\ell=1}^{r_G} \nu^\ell(\Phi)m_\ell+\sum_{a=1}^{r_F} \mu^a(\Phi)n_a \right|/2}\,
\oint \prod_{\ell=1}^{r_G}\frac{dz_\ell}{2\pi i\,z_\ell}\;\times\cr
&\qquad\qquad
\prod_{\ell=1}^{r_G} z_\ell^{k^{\ell n}\,m_n+k^{\ell a}\,n_a}\, \prod_{a=1}^{r_F} u_a^{k^{a b}\,n_b+k^{a\ell}\,m_\ell}\,{\cal I}_V(\{z_\ell,\,m_\ell\})\,{\cal I}_M(\{z_\ell,\,m_\ell\},\, \{u_a,\,n_a\})
\times \cr
& \qquad\qquad x^{-\frac{1}{2}\sum_G|\alpha(m)|}\,
\prod_{\Phi}\left(
x^{1-R_\Phi}\,\prod_{\ell=1}^{r_G}
z_\ell^{-\nu^\ell(\Phi)}\,\prod_{a=1}^{r_F}u_a^{-\mu^a(\Phi)}
\right)^{|\nu^\ell(\Phi)\,m_\ell+\mu^a(\Phi)\,n_a|/2}
\,.
}} The signs in the first two lines come from the shifts due to~\FtoJ, which are obtained
by shifting $z_\ell \to (-1)^{m_\ell}\,z_\ell$ and $u_a \to (-1)^{n_a}\,u_a$ in the expressions
for indices of vector and chiral multiplets
in the prescription
of \refs{\KimWB,\ImamuraSU,\KapustinJM}.  The exponent on the second line is always an integer provided there is no parity anomaly.
In the third line the first two products give the classical contribution from the CS terms. We allow for arbitrary (mixed)
CS terms consistent with the gauge and flavor symmetry.
The functions ${\cal I}_M$ and ${\cal I}_V$ contain the contributions from one-loop matter and gauge determinants, respectively.
The  fourth line
 has the contribution from the monopole background with fluxes $\{m_\ell\}$ and $\{n_a\}$.
The product over $\Phi$ is over all chiral superfields, with $\nu^\ell(\Phi)$ and $\mu^a(\Phi)$ being the charges
of the chiral field under the corresponding gauge and flavor Cartan subgroups, respectively.  $|W|_{\{m_\ell\}}$ is the symmetry factor in the given monopole sector, and $\alpha(m)$ are the roots of the gauge group $G$.

The main difference between \threeDindexCorrect\ and the prescription discussed in
\refs{\KimWB,\ImamuraSU,\KapustinJM} is that different magnetic sectors are weighed
with different signs. In a generic duality, states contributing to a particular magnetic sector on one side map to different magnetic sectors in the
other side of the duality, and this difference becomes crucial.
Next, we will discuss several examples.

First, let us take theory A to be a $U(N)$ gauge theory with $N_f$ fundamental flavors. For simplicity, we turn on a fugacity for the axial symmetry $y$ with no flux, and a fugacity $w$ and flux $n$ for the $U(1)_J$
global symmetry.  We also turn on a diagonal CS term at level $k>0$ for
the gauge group.  The index is thus given by
\eqn\threeDindexUN{
\eqalign{
&{\cal I}^A_{n}(x;y,\,(-1)^{n}w) =\sum_{\{m_1,\dots,m_{N}\}}\frac{(-1)^{k\sum_{\ell=1}^N m_\ell}}{|W|_{\{m_\ell\}}}
\oint \prod_{\ell=1}^{N}\frac{dz_\ell}{2\pi i\,z_\ell}
w^{\sum_{\ell=1}^Nm_\ell}\,\prod_{\ell=1}^{N} z_\ell^{k\,m_\ell + n}\,
\,\times \cr
& \qquad\qquad  (x/y)^{N_f\sum_{i=1}^N|m_i|}\,x^{-\sum_{i< j}^N|m_i-m_j|}\,
 {\cal I}_V(\{z_\ell,\,m_\ell\})\,\left({\cal I}^{(0)}_\chi(y,\{z^{\pm1}_\ell,\,m_\ell\})\right)^{N_f}\,.
}
} Here ${\cal I}_\chi^{(r)}(\{z_\ell^{-1},\,m_\ell\})$ is the index of a chiral superfield in the fundamental representation of $U(N)$ with R-charge $r$,
\eqn\chiralInthree{
\eqalign{
&{\cal I}_\chi^{(r)}(y)=\frac{(x^{2-r}\,y^{-1}\,;x^2)}{(x^{r}\,y\,;x^2)}\,,
\qquad  {\cal I}_V(\{z_\ell,\,m_\ell\})=\prod_{i\neq j}(1- (z_i/z_j) \, x^{|m_i-m_j|})\,,
\cr
&{\cal I}_\chi^{(r)}(y,\,\{z_\ell^{},\,m_\ell\})=\prod_{\ell=1}^N\frac{(x^{|m_\ell|+2-r}
\,y^{-1}\,z^{-1}_\ell;x^2)}{(x^{|m_\ell|+r}\,y\,z^{}_\ell;x^2)}\,,
\qquad (a;x)\equiv\prod_{j=0}^\infty (1-a\,x^j)\,.
}
}

Let us now consider the dual theory B.  The dual theory is \GiveonZN\ a $U(N_f+|k|-N)_{-k}$ gauge theory with $N_f$ flavors and $N_f^2$ singlets $M$.
Moreover, one has to introduce also a contact term \refs{\ClossetVG,\ClossetVP}, which is a level $(-{\rm sign}(k))$ CS term for the $U(1)_J$ symmetry. The index of theory B is thus
\eqn\threeDindexUNM{
\eqalign{
&{\cal I}^B_{n}(x;y,\,(-1)^{n}w) =\cr
&\qquad (-w)^{-n}\sum_{\{m_1,\dots,m_{N_f+k-N}\}}\frac{(-1)^{k\sum_{\ell=1}
^{N_f+k-N}m_\ell}}{|W|_{\{m_\ell\}}}
\oint \prod_{\ell=1}^{N_f+k-N}\frac{d\tz_\ell\,\tz_\ell^{-k\,m_\ell + n}}{2\pi i\,\tz_\ell}
w^{\sum_{\ell=1}^{N_f+k-N} m_\ell}\,
\,\times \cr
& \qquad\qquad\qquad y^{N_f\sum_{i=1}^{N_f+k-N}|m_i|}\, x^{-\sum_{i< j}^{N_f+k-N}\,|m_i-m_j|}\,\times\,\cr
&\qquad\qquad\qquad  {\cal I}_V(\{\tz_\ell,\,m_\ell\})\,\left({\cal I}^{(1)}_\chi(y^{-1},\{\tz^{\pm1}_\ell,\,m_\ell\})\right)^{N_f}\;\left({\cal I}_\chi^{(0)}(y^2)\right)^{N_f^2}\,.
}
} One can check that ${\cal I}^B_{n}$ and ${\cal I}^A_{n}$ match for any value of $n$.
Using the prescription of \refs{\KimWB,\ImamuraSU,\KapustinJM}, the two partition functions  match only for even $n$.
In the case of $U(N)$ dualities without fluxes for global symmetries, the extra minus signs of prescription~\threeDindexCorrect\ as
opposed to the one in \refs{\KimWB,\ImamuraSU,\KapustinJM}
can be interpreted as a special FI-term in the latter (namely, the insertion of a fugacity $w_0=(-1)^k$ for the
$U(1)_J$ symmetry). This explains why the indices matched across dualities using the latter prescription (for instance in \HwangQT).

Let us consider now the $SU(N)$
duality. As we discussed in section \gaugetopo, in this case we can obtain it by simply gauging the $U(1)_J$ symmetry. At the level of the index this is achieved simply by integrating and summing, respectively, over the fugacity and flux for the $U(1)_J$ symmetry. The index
of theory A above becomes the $SU(N)$ index
\eqn\threeDindexSUN{
\eqalign{
&{\cal I}^A(x,y) =\sum_{\{m_1,\dots,m_{N}\},\,n}\frac{(-1)^{k\sum_{\ell=1}^Nm_\ell}}{|W|_{\{m_\ell\}}}
\oint \frac{dw}{2\pi i w}\oint \prod_{\ell=1}^{N}\frac{dz_\ell}{2\pi i}
w^{\sum_{\ell=1}^Nm_\ell}\,\prod_{\ell=1}^{N} z_\ell^{k\,m_\ell + n}\,
\,\times \cr
& \qquad\qquad  (x/y)^{N_f\sum_{i=1}^N|m_i|}\,x^{-\sum_{i< j}^N|m_i-m_j|}\,
 {\cal I}_V(\{z_\ell,\,m_\ell\})\,\left({\cal I}^{(0)}_\chi(y,\{z^{\pm1}_\ell,\,m_\ell\})\right)^{N_f}=\cr
&\sum_{\{m_1,\dots,m_{N}\}}\frac{1}{|W|_{\{m_\ell\}}}
\oint \prod_{\ell=1}^{N-1}\frac{dz_\ell}{2\pi i}
\,\prod_{\ell=1}^{N} z_\ell^{k\,m_\ell }\,
 \;(x/y)^{N_f\sum_{i=1}^N|m_i|}\,x^{-\sum_{i< j}^N|m_i-m_j|}\,\times \cr
&\qquad\qquad \qquad  \left.{\cal I}_V(\{z_\ell,\,m_\ell\})\,\left({\cal I}^{(0)}_\chi(y,\{z^{\pm1}_\ell,\,m_\ell\})\right)^{N_f}\right|_{\prod_
 {\ell=1}^Nz_\ell=1,
\sum_{\ell=1}^Nm_\ell=0}
\,.
}
} On the other side
of the duality, due to the contact term,
 we obtain:
\eqn\threeDindexSUNM{
\eqalign{
&{\cal I}^B(x,y) =\sum_{\{m_1,\dots,m_{N_f+k-N}\},n}\,(-1)^{n+k\sum_{\ell=1}^{N_f+k-N}m_\ell}\, y^{N_f\sum_{i=1}^{N_f+k-N}|m_i|}\,\,\times\cr
& \qquad \qquad x^{-\sum_{i< j}^{N_f+k-N}\,|m_i-m_j|}\,\frac{1}{|W|_{\{m_\ell\}}}\oint \frac{dw}{2\pi i w}
\oint \prod_{\ell=1}^{N_f+k-N}\frac{d\tz_\ell\,\tz_\ell^{-k\,m_\ell + n}}{2\pi i\,\tz_\ell}
w^{\sum_{\ell=1}^{N_f+k-N} m_\ell-n}\,
\,\times \cr
&\qquad \qquad
 {\cal I}_V(\{\tz_\ell,\,m_\ell\})\,\left({\cal I}^{(1)}_\chi(y^{-1},\{\tz^{\pm1}_\ell,\,m_\ell\})\right)^{N_f}\;\left({\cal I}_\chi^{(0)}(y^2)\right)^{N_f^2}=\cr
&\sum_{\{m_1,\dots,m_{N_f+k-N}\}}\frac{(-1)^{(k+1)\sum_{\ell=1}^{N_f+k-N}m_\ell}}{|W|_{\{m_\ell\}}}
\oint \prod_{\ell=1}^{N_f+k-N}\frac{d\tz_\ell\,\tz_\ell^{-k\,m_\ell+\sum_{j=1}^{N_f+k-N}m_j}}{2\pi i\,\tz_\ell}
\,\times \cr
&  \qquad \qquad y^{N_f\sum_{i=1}^{N_f+k-N}|m_i|}\, x^{-\sum_{i< j}^{N_f+k-N}\,|m_i-m_j|}\,\times\cr
&\qquad\qquad\qquad  {\cal I}_V(\{\tz_\ell,\,m_\ell\})\,\left({\cal I}^{(1)}_\chi(y^{-1},\{\tz^{\pm1}_\ell,\,m_\ell\})\right)^{N_f}\;\left({\cal I}_\chi^{(0)}(y^2)\right)^{N_f^2}
\,.
}
}
The right-hand side may be interpreted as an index for a $U(N_f+k-N)$ theory, as described in section \csmdualsec.
We have checked that ${\cal I}^B$ and ${\cal I}^A$ exactly agree for a variety of values of $N$ and $N_f$. An example of physical information that one can extract from the index is how the baryons map. For theory A we have $\frac{N_f!}{N!(N_f-N)!}$ baryons $B_A= Q^N$:
these contribute $\frac{N_f!}{N!(N_f-N)!}\,y^N$ to the index.
The  gauge group of theory B has a $U(1)$ factor, and does not
have baryons in the zero monopole sector. We can trace where the contribution $\frac{N_f!}{N!(N_f-N)!}\,y^N$ comes from for theory B  and deduce that, as described in section \csmdualsec, the baryon here is given by
\eqn\baryonb{B_{\rm B}={\tilde X}_+ \, \left(\prod_{i=2}^k \lambda_{1i}\right)\,\prod_{j=1}^{N_f-N}q^{(a(j))}_{k+j}.}
Here $\lambda_{ij}$ are gauginos, ${\tilde X}_+$ is the basic monopole with GNO charges $m=(1,0,\cdots,0)$,
$q$ is the quark of theory B, and $\{a(j)\}_{j=1}^{N_f-N}$ is a subset of the indices $\{1,\cdots,N_f\}$.
The gluino $\lambda_{1i}$ contributes $-x \tz_1/\tz_i$; the quark $q^{(a)}_{i\neq1}$
contributes  $x\,y^{-1}\,\tz_i^{-1}$; the magnetic monopole $m=(1,0,0,\cdots)$ contributes
$(-1)^{k+1}y^{N_f}x^{N-N_f-k+1}\tz_1^{-k}\prod_{i=1}^{N_f+k-N}\tz_i$. Combining all these
factors we get that $B_{\rm B}$ contributes as $y^N$, and the combinatorics also works out correctly.

Next we discuss a chiral example: the $U(N)$ gauge theory with $N_f^+$ fundamental quarks and $N_f^-$ anti-fundamental quarks (choosing the integer $\frac{N_f^+-N_f^-}{2}$ to be positive for concreteness), without a bare Chern-Simons term.
The dual is given \BeniniMF\ by a $U(N_f^+-N)$ gauge theory with $N_f^-$ fundamental quarks and $N_f^+$ anti-fundamental quarks, and $N_f^+\,N_f^-$ singlet mesonic fields.
We will turn on non-trivial flux for the axial symmetry, $n$, and refine the index with the axial fugacity $y$ and $U(1)_J$ fugacity $w$. The index of theory A is given by
\eqn\threeDindexUNCh{
\eqalign{
&{\cal I}^A_{n}(x;\,(-1)^{n}\,y,w) =\sum_{\{m_1,\dots,m_{N}\}}\frac{1}{|W|_{\{m_\ell\}}}
\oint \prod_{\ell=1}^{N}\frac{dz_\ell}{2\pi i\,z_\ell}
w^{\sum_{\ell=1}^N m_\ell}\times\cr
&\qquad \prod_{\ell=1}^{N} ((-1)^{m_\ell}z_\ell)^{\frac{N_f^-}2|m_\ell-n|-\frac{N_f^+}2|m_\ell+n|}\,
((-1)^{n}x/y)^{\frac{N_f^-}2|m_\ell-n|+\frac{N_f^+}2|m_\ell+n|}
\times x^{-\sum_{i< j}^N|m_i-m_j|} \cr
&\qquad
 {\cal I}_V(\{z_\ell,\,m_\ell\})\,\left({\cal I}^{(0)}_\chi(\{y,n\},\{z^{}_\ell,\,m_\ell\})\right)^{N_f^+}
\left({\cal I}^{(0)}_\chi(\{y,n\},\{z^{-1}_\ell,\,m_\ell\})\right)^{N_f^-}\,.
}}
For the dual theory we obtain
\eqn\threeDindexUNMCh{
\eqalign{
&{\cal I}^B_{n}(x;\,(-1)^{n}\,y,w) =\sum_{\{m_1,\dots,m_{N_f^+-N}\}}\frac{w^{-N_f^+\,n}y^{-2N_f^+N_f^-|n|}}
{|W|_{\{m_\ell\}}}
\oint \prod_{\ell=1}^{N_f^+-N}\frac{d\tz_\ell}{2\pi i\,\tz_\ell}
w^{-\sum_{\ell=1}^{N_f^+-N} m_\ell}\,\times\cr
&\qquad \prod_{\ell=1}^{N_f^+-N} ((-1)^{m_\ell}\tz_\ell)^{-\frac{N_f^-}2|m_\ell-n|+\frac{N_f^+}2|m_\ell+n|}\,
((-1)^{n}y)^{\frac{N_f^-}2|m_\ell-n|+\frac{N_f^+}2|m_\ell+n|}
\,\times \cr
& \qquad \qquad   x^{-\sum_{i< j}^{N_f^+-N}|m_i-m_j|}\,
 {\cal I}_V(\{\tz_\ell,\,m_\ell\})\,\left({\cal I}^{(1)}_\chi(\{y^{-1},n\},\{\tz^{}_\ell,\,m_\ell\})\right)^{N_f^-}\,
 \times\cr
&\qquad \qquad \left({\cal I}^{(1)}_\chi(\{y^{-1},n\},\{\tz^{-1}_\ell,\,m_\ell\})\right)^{N_f^+}
 \left({\cal I}^{(0)}_\chi(\{y^{2},n\})\right)^{N_f^+N_f^-}\,.
}
} Here the term $w^{-N_f^+\,n}$ comes from a contact term~\BeniniMF\ between the axial  $U(1)_A$ and the topological $U(1)_J$,
and the rest follows directly from~\threeDindexCorrect. We have checked that the index of theory A  matches the index of theory B for a variety
of parameters.

\appendix{B}{A comment on the index of a chiral superfield}

In this appendix we give a physical derivation of  equation~\threeDlim.
As we mentioned in section 5, the $4d$ index defined in~\indexDef\ can be thought
of as a partition function on $\S^3\times \S^1$, of radii $r_3$ and $r_1$, twisted by fugacities for various global symmetries.  Equivalently, after a change of variables it can be thought of as a partition function on $\S_b^3\times{\widetilde \S}^1$ with the fugacities responsible for the geometric twisting absorbed in the geometry~\ImamuraWG.

It should be possible to compute
the index by first reducing the theory on ${\widetilde \S}^1$ of finite radius, and then computing the $3d$ partition function of the resulting $3d$ theory, including all the KK modes on the ${\widetilde \S}^1$.  The fugacities corresponding to flavor symmetries can be thought of as a coupling to background gauge fields along the $\S^1$ direction, and these simply reduce to real mass parameters in $3d$, as discussed above.  In addition, as we go once around the $\S^1$, we should rotate the $\S^3$ along the Hopf fiber by an angle depending on the fugacities $p$ and $q$.  This has the effect of changing the geometry.  As discussed in \ImamuraWG, there is a change of coordinates, where the metric becomes that of an $\S^3_b \times {\widetilde \S}^1$, where the ${\widetilde \S}^1$ factor is rotated on the $\S^3_b$ base.  The parameters are related by
\eqn\pqrel{p=e^{-2\pi\,b \frac{2r_1}{r_3(b+b^{-1})}} \qquad ; \qquad q=e^{-2\pi\,b^{-1} \frac{2r_1}{r_3(b+b^{-1})}}~.}
This procedure leads to the action used in \ImamuraWG\ to compute the supersymmetric partition function on $\S^3_b$.
Then, we can write the $4d$ index as coming from a theory on $\S^3_b$, with an infinite tower of KK modes.

Let us review this in more detail. We start from a round metric for $\S^3\times \S^1$,\foot{
We follow the notations of~\ClossetRU.
}
\eqn\metricone{
ds^2=\frac{r_3^2}{4}\left[
\left(d\psi+2\sin^2\left(\frac\theta2\right)\,d\phi\right)^2+d\theta^2+\sin^2(\theta)\,d\phi^2
\right]+dx_4^2\,.
} The coordinate $x_4$ parameterizes the $\S^1$, and the angle $\psi\sim \psi+4\pi$ parameterizes the Hopf fiber of $\S^3$, rotated by the $SU(2)$ generator $j_2$
(see discussion around~\indexDef). Twisting the partition function
with fugacities $p$ and $q$ implies that we have to identify,
\eqn\twist{
(x_4,\,\psi)\quad \sim\quad (x_4+2\pi r_1,\, \psi+4\pi\,i\,\frac{b-b^{-1}}{b+b^{-1}}\frac{r_1}{r_3})\,.
} It is thus more convenient to define new coordinates,
\eqn\newcoord{
(x_4,\,\psi)\quad \to \quad(x_4,\,\hat \psi)\equiv(x_4,\,\psi-2\,i\,\frac{b-b^{-1}}{b+b^{-1}}\frac{x_4}{r_3})\,,
} for which the identification above is
\eqn\twisttwo{
(x_4,\,\hat\psi)\quad \sim\quad (x_4+2\pi r_1,\, \hat\psi)\,.
} The metric in the new coordinates becomes,
\eqn\metrictwo{\eqalign{
&ds^2=\frac{r_3^2}{4}\left[
\left(\frac{b+b^{-1}}{2}\right)^2\left(d\hat \psi+2\sin^2\left(\frac\theta2\right)\,d\phi\right)^2+d\theta^2+\sin^2(\theta)\,d\phi^2
\right]+\cr
&\qquad\qquad\qquad \left(\frac{2}{b+b^{-1}}\right)^2\left(dx_4-\frac{i}{8}(b^2-b^{-2})\,r_3
\,\left(d\hat \psi+2\sin^2\left(\frac\theta2\right)\,d\phi\right)\right)^2\,.
}
} On the right-hand side of the first line we see the metric of $\S^3_b$~\refs{\ImamuraWG,\ClossetRU}. In the second line we have
the metric for ${\widetilde \S}^1$. The radius
of ${\widetilde \S}^1$, $\tilde r_1$, is related to the radius of ${\S}^1$, $r_1$, as
\eqn\radii{
\tilde r_1=\frac{2}{b+b^{-1}}\,r_1\,.
} Now we can reduce the partition function along ${\widetilde \S}^1$ to obtain a partition function
on $\S^3_b$~\ImamuraWG.

Next we discuss the reduction in the simplest case of a free field.
For a free chiral field (of R-charge $R$ and charged under a $U(1)_u$ symmetry) this physical statement translates into
\eqn\chiralrefn{
{\cal I}^{(R)}(p,\,q,\,u)\propto \prod_{n=-\infty}^\infty {\cal Z}^{(R)}(\omega_1,\,\omega_2,\,m+\frac{n}
{{\tilde r}_1})\,,
}
with the parameters on the two sides related as in \fugmass.
On the left-hand side we have the $4d$ index of a chiral superfield, and
on the right-hand side the product over $3d$ $\S^3_b$ partition functions of the KK modes on ${\widetilde \S}^1$.
The inverse radius of ${\widetilde \S}^1$, $1/{{\tilde r}_1}$, plays the role of a real mass coupled to the KK momentum.

When one computes the partition function
certain divergences coming from determinants of the modes should be properly regularized, and there are different natural choices for their finite part.  One choice, used in the index computation \indexChi\ and \indexDef, is to normalize it such that the vacuum contributes $+1$ to the partition function.  Another normalization is more natural when we try to relate it to the $3d$ partition function.

Concretely, the twisted partition function of the chiral field on $\S^3\times \S^1$ can be written as\foot{
We set the R-charge to vanish for brevity. General R-charge can be easily reintroduced in the following expressions.
}
\eqn\chirfieldphase{
{\cal I}^{(0)}(p,\,q,\,u)=e^{{\cal I}_0}\;\Gamma( u\,;p,\,q)\,.
} The factor $e^{{\cal I}_0}$ relates the two different natural
 normalizations. It is computed in~\KimAVA:
\eqn\casimir{{{\tilde r}_1}^{-1}{\cal I}_0=\frac14
\left.\left({r}^{-1}\frac{d}{dr}\left(r\,\Gamma_0(
e^{2\pi r  i m};e^{2\pi r i \omega_1},e^{2\pi r i\omega_2})\right)\right)\right|_{r=0}\, ,}
where $\Gamma_0(z;p,q)$ is the so called single particle index, defined by
\eqn\pleth{
\Gamma( u\,;p,\,q)=
\exp\left[\sum_{n=1}^\infty\frac1n \; \Gamma_0(z^n;p^n,q^n)\right]\quad\to\quad
\Gamma_0(z;p,q)=\frac{z-pqz^{-1}}{(1-p)(1-q)}\,.
}
Using the fact that $\Gamma_0$ has a simple pole at $r=0$ and a vanishing constant term in the expansion around $r=0$, equation
\casimir\ leads to
\eqn\casimirtwo{{\cal I}_0=
\frac{\pi\,i\,{\tilde r}_1\, (m-\omega) \left(2\, m\,(m - 2 \omega)+ \omega _1\,\omega _2\,\right)}
{6\,\omega _1\, \omega _2}
\,.
}

Next we compute the right-hand side of \chiralrefn. Using the  $\S^3_b$ partition function~\partChi\
 we can write
\eqn\rhs{\eqalign{
&\prod_{n=-\infty}^\infty {\cal Z}^{(0)}(\omega_1,\,\omega_2,\,m+\frac{n}{{\tilde r}_1})=
\prod_{n=-\infty}^\infty \Gamma_h(m+\frac{n}{{\tilde r}_1};\omega_1,\omega_2)\,.
}
}
The infinite product over $n$ here  diverges, since for large $n$ the hyperbolic Gamma functions
approach a divergent exponential behavior (see~\gammaAsymp).
We can regularize this divergence using zeta-function
regularization ($\sum_{n=1}^\infty n^s=\zeta(-s)$)\foot{
Here we defined ${\rm sign}(n=0)=-1$.
}
\eqn\zetaregprod{\eqalign{
\prod_{n=-\infty}^\infty\,
&e^{-{\rm sign}(n)\,\frac{\pi i}{2\omega_1\omega_2}\left((m+\frac{n}{{{\tilde r}_1}}-\omega)^2-\frac{\omega_1^2+\omega_2^2}{12}\right)}
\longrightarrow  \cr
& \exp\left(\Delta\right)\equiv
\exp\left(
\frac{i \pi \,  \left(2 m (3 m\,{{\tilde r}_1}+1)-2\,(1-6m\,{{\tilde r}_1})\,\omega+{{\tilde r}_1}\,(\omega _1^2+\omega _2^2+3\omega_1\omega_2)\right)}{12\,{{\tilde r}_1}\, \omega _1 \omega _2}
\right)\,.
}
} The precise statement of \chiralrefn\ is then the following equality
\eqn\regrhs{
e^{{\cal I}_0}\;\Gamma( u\,;p,\,q)=
e^{-\Delta} \;\prod_{n=-\infty}^\infty\,
\left[e^{-{\rm sign}(n)\,\frac{\pi i}{2\omega_1\omega_2}\left((m+\frac{n}{{\tilde r}_1}-\omega)^2-\frac{\omega_1^2+\omega_2^2}{12}\right)}
\Gamma_h(m+\frac{n}{{{\tilde r}_1}};\omega_1,\omega_2)\right]\,.
}  The infinite product on the right-hand side is now well-defined, and in fact by using \hypG\ and
 \ellG\
it can be written as a product of two elliptic Gamma functions,
\eqn\slthree{
\Gamma(u;p,\,q)=e^{-\Delta-{\cal I}_0}\;
\frac{\Gamma(e^{2\pi i\frac{m}{\omega_1}};
e^{2\pi i\frac{\omega_2}{\omega_1}},e^{-2\pi i\frac{1}{{{\tilde r}_1}\,\omega_1}})}{\Gamma(e^{2\pi i\frac{m-\omega_1}{\omega_2}};e^{-2\pi i\frac{1}{{{\tilde r}_1}\,\omega_2}},e^{-2\pi i\frac{\omega_1}{\omega_2}})}\,.
}
This equality is discussed in \slthreeZ. It is sometimes viewed as an indication of an $SL(3,\ZZ)$ structure.

Taking the $3d$ limit by sending ${\tilde r}_1$ to zero, we decouple the massive KK modes on the ${\widetilde \S}^1$.  The only term surviving the limit on the right-hand side of \regrhs\ has $n=0$,
and we obtain precisely~\threeDlim.

\listrefs
\end